# Impact of metal nanoparticles on cell survival predicted by the local effect model for cells in suspension and tissue. Part 1: Theoretical framework

**Hans Rabus[1], Leo Thomas[1,2]**

[1] Physikalisch-Technische Bundesanstalt, Berlin, Germany
[2] Technische Universität Berlin, Berlin, Germany

## Abstract

This work investigates the change in cell survival predicted by the local effect model (LEM) for an irradiated cell containing metal nanoparticles (MNPs) depending on the distribution of neighboring cells and the uptake of MNPs into the cells. In this first part of the paper, the theoretical framework is described, which is based on analytical weighting functions for the energy deposition around a single metal nanoparticle and radially symmetric distributions of MNPs. The weighting functions allow calculation of the radial profile of the absorbed dose in the cell nucleus as well as the mean dose and the mean square of the dose in the nucleus. The latter two quantities determine cell survival according to the LEM. The weighting functions are applied to isolated cells in a localized MNP distribution, cells in solution, and densely packed cells in tissue. It is shown that only for the idealistic case of complete uptake of MNPs it is sufficient to consider an isolated cell, as this otherwise leads to a significant underestimation in more realistic situations. In the case of cells in tissue, the MNP concentration within the range of secondary particles around the cell must be taken into account. Different packing densities of the cells may lead to values differing by up to 30% for the mean dose in the cell nucleus, depending on the conceived scenario for the uptake of MNPs. The weighting function offers a versatile method for assessing cell survival under irradiation in the presence of MNPs by the LEM, which is more general than previously reported approaches.

---

## 1. Introduction

Since Hainfeld et al. (Hainfeld *et al* 2004) showed in their pioneering work that the presence of metal nanoparticles (MNPs) in biological cells exposed to ionizing radiation results in decreased cell survival, MNPs have been explored as potential agents for improving radiation therapy of cancer (Kim *et al* 2012, Jain *et al* 2012, Dorsey *et al* 2013, Schuemann *et al* 2016, Cui *et al* 2017, Her *et al* 2017, Lacombe *et al* 2017, Kuncic and Lacombe 2018, Schuemann *et al* 2020, Gerken *et al* 2023, Kumarasamy *et al* 2024). Numerous simulation studies have tried to explain this dose enhancement or radiosensitizing effect of MNPs (Chow *et al* 2012a, 2012b, Mesbahi *et al* 2013, Zygmanski *et al* 2013b, Gao and Zheng 2014, Wälzlein *et al* 2014, Kakade and Sharma 2015, Xie *et al* 2015, Zygmanski and Sajo 2016, Martinov and Thomson 2017, Chow 2018, Zabihzadeh *et al* 2018, Peukert *et al* 2020, Vlastou *et al* 2020, 2022, Xu *et al* 2022). Many simulations have considered the enhanced energy deposition around a single MNP as an initial step (McMahon *et al* 2011a, 2011b, Lin *et al* 2014, 2015, Velten and Tomé 2023, 2024, Li *et al* 2020, 2024, Kim *et al* 2024). Analytical models are used to achieve the transition from the dose distribution around a single MNP to the dosimetric effects at the cellular level (Lin *et* *al* 2015, Melo-Bernal *et al* 2021, Velten and Tomé 2023, 2024, Rabus 2024b). In many of these studies, simplified concentric spherical geometries were considered (Lin *et al* 2015, Xie *et al* 2015, Dressel *et al* 2019, Velten and Tomé 2023, 2024). Some of these studies used the local effect model (LEM) of (Krämer and Scholz 2000) to predict cell survival (McMahon *et al* 2011b, Lin *et al* 2015, Velten and Tomé 2023).

Pitfalls related to simplifications for computational feasibility of application of the LEM to MNPs may lead to severe overestimation of the reduction in cell survival (Rabus 2024b, Velten and Tomé 2024). Methodologically more sound approaches have been presented by (Melo-Bernal *et al* 2021, Rabus 2024b), assuming a simple power law describing the radial dose profile around an MNP.

In the present work, these earlier approaches are developed and used to assess to what extent investigations considering MNPs in and around a single cell (McMahon *et al* 2011b, Lin *et al* 2015, Velten and Tomé 2023) are representative of realistic scenarios of cells in suspension or densely packed in tissue. Similar to preceding work, the analysis is based on idealizing assumptions such as that the dose distribution around an MNP is radially symmetric and that MNPs are present in spatially uniform distributions in parts of the considered geometry. While the emission of secondary particles and the resulting spatial distribution of energy imparted may not be exactly spherically symmetric, anisotropy only occurs in the vicinity of the MNP and, even there, the anisotropy is in the range of a few percent (Derrien *et al* 2023, Rabus 2024a). Similarly, MNPs tend to be non-uniformly distributed in and between cells, which leads to statistical variations in cell survival, as discussed by (Zygmanski *et al* 2013a). This implies caveats to the present approach, which should still be sufficient for addressing questions such as whether predictions of cell survival based on simulations considering isolated cells (Lin *et al* 2015) are representative of the expected outcome for cells in suspension in radiobiological experiments or for survival of cells in tissue.

The approach used in this work is based on geometrical weighting functions for the radial distribution of energy imparted around a single spherical MNP. These weighting functions allow calculation of the radial dose profile in a spherical cell nucleus as well as of the mean dose and the variance of the dose in a spherical cell nucleus from the radial distribution of the dose around a single MNP. This approach is more general than those presented previously (Melo-Bernal *et al* 2021, Rabus 2024b), in that no assumption is made regarding the functional dependence of the dose distribution around an MNP. The weighting functions can be applied to MNPs of any material and also to the often-considered water nanoparticle of the same size.

Such weighting functions are determined for the case of a uniform MNP distribution in a sphere surrounding the cell nucleus as well as for simple extensions such as MNPs only in a spherical shell or MPNs at different concentration in a sphere and one or two spherical shells around. The generalization to a continuously varying radially symmetric concentration profile of MNPs is also discussed. General considerations are made regarding the suitability of approaches found in the literature that consider a localized MNP distribution around a single cell. In addition, the case of cells in tissue is considered, and the applicability of a simplified surrogate model of the MNP concentration. Application of the model with dose distributions around MNPs from several recent publications will be discussed in the second part of the paper.

## 2. Materials and methods

### *2.1 The local effect model*

The LEM formulated by (Krämer and Scholz 2000) assumes that the survival probability of a cell exposed to an ion beam producing a largely inhomogenous dose distribution is given by Eq. (1):

$$S_i(D) = e^{-N_i} \tag{1}$$

In Eq. (1), $N_i$ is the expected number of lethal lesions produced in the cell nucleus, which is given by Eq. (2):

$$N_i\left(\overline{D}, \overline{D^2}; \alpha, \beta, D_t\right) = \langle N_u \rangle = \iiint\limits_{V_n} \frac{N_u(D(\vec{r}); \alpha, \beta, D_t)}{V_n} dV \tag{2}$$

Herein, $D(\vec{r})$ is the local dose at point $\vec{r}$; $V_n$ is the volume of the cell nucleus; and $\alpha, \beta, D_t$ are the coefficients of the linear quadratic linear (LQL) model of cell survival applying to irradiation with photons at a uniform dose distribution (over the dimensions of a cell). $\overline{D}$ and $\overline{D^2}$ are the mean dose and mean square of the dose in the nucleus, respectively (Eq. (3)):

$$\overline{D} = \frac{1}{V_n} \int\limits_{V_n} D(\vec{r}) dV \qquad\qquad \overline{D^2} = \frac{1}{V_n} \int\limits_{V_n} D^2(\vec{r}) dV_n \tag{3}$$

The expected number $N_u(D; \alpha, \beta, D_t)$ of lesions in the nucleus for a uniform dose distribution in the LQL model is given by Eq. (4):

$$N_u(D; \alpha, \beta, D_t) = \begin{cases} \alpha D + \beta D^2 & D \leq D_t \\ (\alpha + 2\beta D_t) D - \beta {D_t}^2 & D > D_t \end{cases} \tag{4}$$

In Eq. (4), $D_t$ is the threshold dose for transition between the linear quadratic dose dependence to the high-dose linear dose dependence. The quantity $N_i\left(\overline{D}, \overline{D^2}; \alpha, \beta, D_t\right)$ is thus the integral of the spatial density of the expected lesions and can be expressed by Eq. (5)[1]:

$$N_i\left(\overline{D}, \overline{D^2}; \alpha, \beta, D_t\right) = \alpha \overline{D} + \beta \overline{D}^2 + \beta \overline{(D - \overline{D})^2} - \beta \overline{(D - D_t)^2}' \tag{5}$$

The third term on the right-hand side of Eq. (5) is the variance of the dose. The quantity $\overline{(D - D_t)^2}'$ represents the mean squared excess dose over the regions in the cell nucleus in which the dose $D$ exceeds $D_t$, as indicated by the integration domain $V_t$ in Eq. (6):

$$\overline{(D - D_t)^2}' = \frac{1}{V_n} \int\limits_{V_t} (D(\vec{r}) - D_t)^2 dV_t \tag{6}$$

The term defined by Eq. (6) leads to a reduction in the number of predicted lesions as per Eq. (5). Neglecting this term, as we also do following the example of (Melo-Bernal *et al* 2021), therefore gives an upper bound for the expected number of lesions. It must be emphasized, however, that if the mean dose exceeds the threshold dose, then the volume $V_t$ comprises the whole nucleus, and the following linear expression is obtained:

[1] By virtue of $2D_t D - {D_t}^2 = D^2 - (D - D_t)^2$.

$$N_i\left(\bar{D},\overline{D^2};\alpha,\beta,D_t\right) = (\alpha + 2\beta D_t)\bar{D} - \beta {D_t}^2 \tag{7}$$

*2.2 Weighting functions for dosimetric effects in a spherical cell nucleus from a uniform distribution of MNPs inside a concentric sphere*

In the following it is assumed that the nucleus is a sphere of radius $R_n$ and that $N_m$ MNPs made of material $m$ are uniformly distributed in a concentric sphere of radius $R_c \geq R_n$. If $p_m$ is the expected number of ionizing radiation interactions in an MNP (or decays for radioactive MNPs), then the expected number of MNPs emitting electrons or photons is given by $N = N_m p_m$, and the number density $\bar{n}_m$ of emitting MNPs is given by Eq. (8):

$$\bar{n}_m = \frac{3N_m p_m}{4\pi {R_c}^3} \tag{8}$$

For a given geometrical arrangement of the $N$ emitting MNPs at positions $\vec{r}_i$ ($i = 1,2,\ldots,N$), the absorbed dose at a point $\vec{r}$ in the nucleus is given by Eq. (9):

$$D(\vec{r}) = D_w + \sum_{i=1}^{N} D_1(|\vec{r} - \vec{r}_i|) \tag{9}$$

Here, $D_w$ is the dose contribution originating in interactions of the incident radiation that occur outside MNPs (zero for radioactive MNPs). The dose contribution from an emitting MNP, $D_1$, is the expected energy imparted per mass at a given radial distance from the center of the MNP due to the particles emitted from the MNP after ionizing radiation interaction or a decay event.

The mean value of the dose and the square of the dose in the nucleus for a fixed geometrical arrangement of the $N$ emitting MNPs are given by Eqs. (10) and (11), respectively:

$$\bar{D} = \frac{1}{V_n}\int_{V_n} D(\vec{r})dV = D_w + \frac{1}{V_n}\int_{V_n} \sum_{i=1}^{N} D_1(|\vec{r} - \vec{r}_i|)\, dV_n \equiv D_w + \overline{D_N} \tag{10}$$

$$\overline{D^2} = {D_w}^2 + 2D_w\overline{D_N} + \overline{D_N^2} + \overline{D_{NN'}^2} \tag{11}$$

The quantities $\overline{D_N^2}$ and $\overline{D_{NN'}^2}$ are the contributions of individual MNPs and of pairs of MNPs to the mean square of the dose given by Eqs. (12) and (13):

$$\overline{D_N^2} = \frac{1}{V_n}\int_{V_n} \sum_{i=1}^{N} [D_1(|\vec{r} - \vec{r}_i|)]^2\, dV_n \tag{12}$$

$$\overline{D_{NN'}^2} = \frac{1}{V_n}\int_{V_n} \sum_{i=1}^{N}\sum_{j\neq i} D_1(|\vec{r} - \vec{r}_i|) D_1\left(|\vec{r} - \vec{r}_j|\right) dV_n \tag{13}$$

These two quantities determine the variance of the dose, as shown in Eq. (14):

$$\mathrm{Var}(D) \equiv \overline{(D - \bar{D})^2} = \overline{D_N^2} + \overline{D_{NN'}^2} - \overline{D_N}^2 \tag{14}$$

### 2.2.1 Weighting function for the dose at a point in the cell nucleus

Eq. (9) makes use of the fact that the absorbed dose is an expectation value and gives the absorbed dose conditional on the geometrical arrangement of the $N$ emitting MNPs. To obtain the unconditional expectation value representing the absorbed dose for a uniform distribution of MNPs in a sphere, Eq. (9) must be averaged over all possible arrangements of the $N$ emitting MNPs. This can be achieved by replacing the sum by an integral over the probability density of MNP positions, which is constant and given by Eq. (8). This leads to Eq. (15):

$$D(\vec{r}) = D_w + \int_{V_c} \bar{n}_m \times D_1(|\vec{r} - \vec{r}_i|)dV_c \tag{15}$$

It should be noted that by the definition of absorbed dose, in both Eqs. (9) and (15), $D(\vec{r})$ is an expectation—in the former case for a fixed arrangement of the MNPs, and in the latter case for a uniform distribution of emitting MNPs in a volume. To evaluate the integral on the right-hand side, it is advantageous to choose the point $\vec{r}$ as the center of the coordinate system (blue point in Fig. 1) and introduce spherical coordinates. This transforms Eq. (15) into Eq. (16):

$$D(\vec{r}) = D_w + \bar{n}_m \int_{r_p}^{\infty} D_1(r_i)\, \Omega(r_i, r|R_c) {r_i}^2 dr_i \tag{16}$$

In Eq. (16), $r_p$ denotes the radius of the spherical MNP, and ${r_i}^2\Omega(r_i, r|R_c)$ is the part of the surface of a sphere of radius $r_i$ around a point at distance *r* from the center of the nucleus which falls inside the sphere where MNPs are present. In Fig. 1, this region is indicated by the yellow shaded area, and a portion of the surface inside the sphere containing the MNPs is indicated by the blue circle and arc, respectively. The solid angle subtended by the part of the surface inside the region with MNPs is given by Eq. (17):

$$\Omega(r_i, r|R_c) = \begin{cases} 0 & r < r_i - R_c \\ 4\pi & r < R_c - r_i \\ \pi \dfrac{{R_c}^2 - (r - r_i)^2}{r_i r} & \text{else} \end{cases} \tag{17}$$

The first of the three cases in Eq. (17) means that the distance $r_i$ from the considered point is so large that all points at this distance from a point inside the nucleus are outside the region containing MNPs. The red dashed circle in Fig. 1(a) corresponds to the smaller value of $r_i$ for which this is the case. The second case in Eq. (17) applies to all points in the nucleus for which the whole sphere of radius $r_i$ is within the region containing MNPs. The solid blue circle in Fig. 1(b) indicates the largest sphere for which this applies. The third case corresponds to distances $r_i$ from the chosen point for which some points are inside (blue arc in Fig. 1(c)) and some are outside (red dashed arc in Fig. 1(c)).

Supplementary Fig. 1 shows sample weighting functions according to Eq. (17) for different values of $\frac{R_c}{R_n}$.

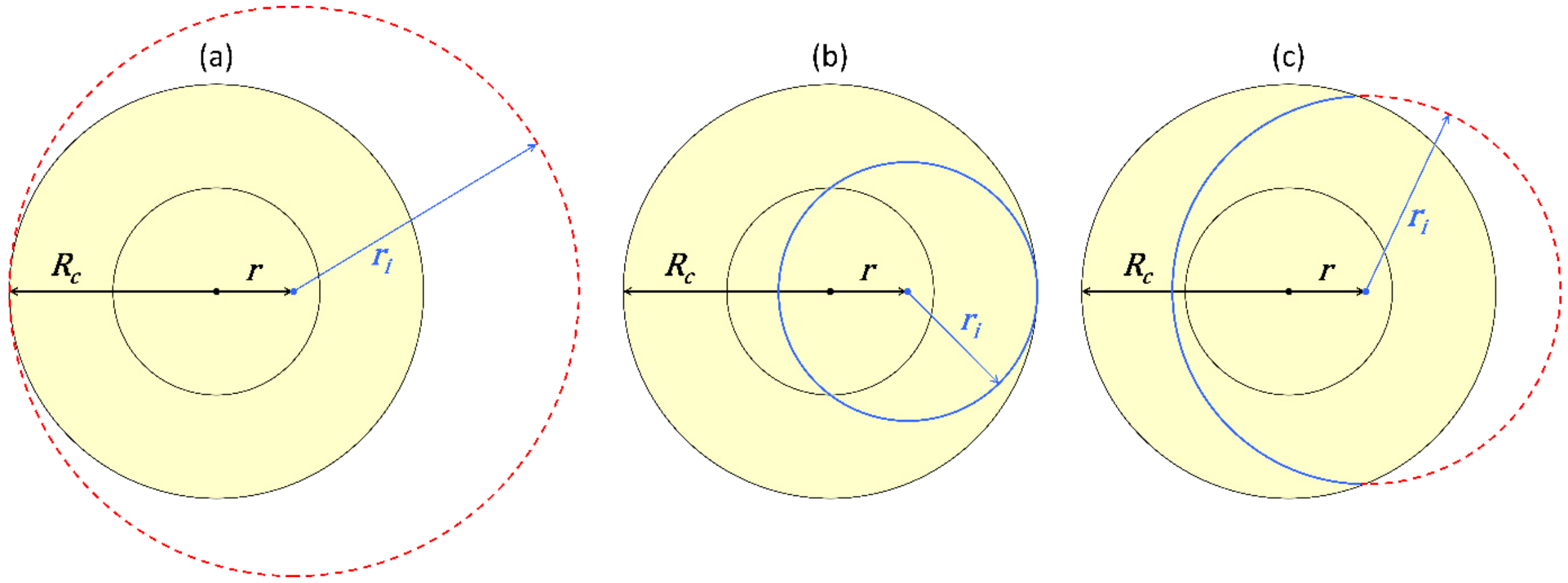

Fig. 1: Schematic illustration of geometrical weighting. In each graph, the yellow shaded circle represents a cut through the sphere containing nanoparticles by a plane passing through its center and a point at a radial distance $r$. The small black circle represents the cell nucleus. The blue circle and blue arc indicate the part of a sphere surface around the chosen point inside the sphere containing nanoparticles, the red dashed arc and circle indicate the portion of the sphere surface outside the region with nanoparticles.

### 2.2.2 Mean dose in a cell nucleus from a uniform distribution of MNPs inside a sphere

As shown in Section S1.1 of Supplement 1, the average excess dose $\overline{D_N}$ produced in a spherical nucleus (radius $R_n$) from a uniform distribution of MNPs in a concentric spherical volume (radius $R_c$, volume $V_c$) can be written as follows:

$$\overline{D_N} = \bar{n}_m \overline{d_1}(R_c) \tag{18}$$

In Eq. (18), $\bar{n}_m$ is the number density of emitting MNPs in the volume $V_c$, and $\overline{d_1}$ is the ratio of the dose contribution from a single emitting MNP to the number density of emitting MNPs and can be obtained from the radial dose distribution around an emitting nanoparticle by the following relation:

$$\overline{d_1}(R_c) = \int_{r_p}^{\infty} D_1(r_i) r_i^2 \times 4\pi \varpi(r_i|R_c) dr_i \tag{19}$$

The dimension of $\overline{d_1}$ is dose times volume. The symbol $r_i$ in Eq. (19) denotes the radial distance from the nanoparticle, and $\varpi(r_i|R_c)$ is a weighting function defined by Eq. (20):

$$\varpi(r_i|R_c) = \frac{3}{{R_n}^3} \int_0^{R_n} \frac{\Omega(r_i, r|R_c)}{4\pi} r^2 dr \tag{20}$$

Explicit expressions for the weighting function are given in Appendix 1. The quantity $\varpi(r_i|R_c)$ is the mean fraction of the surface area of a sphere of radius $r_i$ around an arbitrary point within the nucleus that is within the region loaded with MNPs, averaged over the cell nucleus. Fig. 2 shows sample weighting functions according to Eq. (A.1) for different ratios $R_c/R_n$ between 1 (MNPs only in the cell nucleus) and 2. It can be seen that with increasing radius $R_c$ of the sphere containing MNPs, the dose distribution at larger radial distance from an MNP receives increasing weight.

### 2.2.3 Mean square of the dose in a spherical cell nucleus from a uniform distribution of MNPs inside a concentric sphere

It is evident that the third term in Eq. (11) can be treated in the same way as $\overline{D_N}$. This leads to Eq. (21):

$$\overline{D_N^2} = N\overline{D_1^2} = \bar{n}_m \overline{d_1^2} \tag{21}$$

In Eq. (21), $\overline{D_1^2}$ is the average contribution to the mean square of the dose in the nucleus from a single emitting MNP, and $\overline{d_1^2}$ is the corresponding average contribution per number density of emitting MNPs given by Eq. (22):

$$\overline{d_1^2}(R_c) = \int_{r_p}^{\infty} [D_1(r_i)]^2 \, r_i{}^2 \; \times 4\pi\varpi(r_i|R_c)dr_i \tag{22}$$

The last term in Eq. (11) can be written in an analogous way (see Supplement 1, Section S1.3), where $V_p$ denotes the volume of the MNP:

$$\overline{D_{NN'}^2} = \bar{n}_m{}^2 \times \left[\overline{d_2^2} - 8V_p \times \overline{d_1^2}\right] \tag{23}$$

The quantity $\overline{d_2^2}$ (Eq. (24)) is the ratio of the average synergistic contribution from a pair of MNPs to the number density of emitting MNPs:

$$\overline{d_2^2}(R_c) = \int_{r_p}^{\infty} \int_{r_p}^{\infty} D_1(r_i) r_i{}^2 D_1(r_j) r_j{}^2 \times (4\pi)^2 \bar{\omega}_2(r_i, r_j|R_c) \, dr_j \, dr_i \tag{24}$$

In Eq. (24), $\bar{\omega}_2(r_i, r_j|R_c)$ is a bivariate weighting function defined by Eq. (25):

$$\bar{\omega}_2(r_i, r_j|R_c) = \frac{3}{R_n{}^3} \int_0^{R_n} \frac{\Omega(r_i, r|R_c)}{4\pi} \frac{\Omega(r_j, r|R_c)}{4\pi} \, r^2 dr \tag{25}$$

Explicit expressions for the function $\bar{\omega}_2(r_i, r_j|R_c)$ are given in Appendix 2.

### 2.2.4 Weighting function for the variance

From Eqs. (14), (21), and (23), the variance of the dose can be written as

$$\mathrm{Var}(D) = \bar{n}_m \overline{d_1^2}(1 - \bar{n}_m 8V_p) + \bar{n}_m{}^2 \left(\overline{d_2^2} - \overline{d_1}^2\right) = \bar{n}_m \times \mathrm{var}_1 + \bar{n}_m{}^2 \times \mathrm{var}_2 \tag{26}$$

That is, the variance has one term proportional to the number density of emitting MNPs and one term proportional to the square of the number density. The last term is given by Eq. (27):

$$\mathrm{var}_2 = \overline{d_2^2} - \overline{d_1}^2 = \int_{r_p}^{\infty} \int_{r_p}^{\infty} D_1(r_i) r_i{}^2 D_1(r_j) r_j{}^2 \times (4\pi)^2 \bar{v}_2(r_i, r_j|R_c) \, dr_j \, dr_i \tag{27}$$

The modified weighting function appearing in Eq. (27) is given in Eq. (28):

$$\bar{v}_2(r_i, r_j|R_c) = \bar{\omega}_2(r_i, r_j|R_c) - \varpi(r_i|R_c)\varpi(r_j|R_c) \tag{28}$$

*2.3 Weighting functions for a spherical shell and combinations of sphere and spherical shells*

The weighting functions presented in the previous section are for a uniform distribution of MNPs in a sphere. In potential radiotherapy applications, enhanced uptake of MNPs into cells or even the cell nucleus is desirable. Adversely, the membranes of the cell or nucleus may hinder diffusion of MNPs into the cell nucleus or the cell without completely suppressing it. In both cases, the result may be that the MNPs have different number densities in the nucleus ($\bar{n}_n$), the cytoplasm ($\bar{n}_p$), and the extracellular region ($\bar{n}_x$), while still being uniformly distributed in each of these regions. Examples considered in literature are situations in which the MNPs are only present in the cytoplasm or only outside the cell (Lin *et al* 2015, Velten and Tomé 2023).

Appendix 3, Appendix 4, and Appendix 5 show the derivation of the corresponding relations for the cases of MNPs only in a spherical shell, for MNPs at different concentrations in a sphere and a surrounding spherical shell, and for three different MNP concentrations in a sphere and two enclosing spherical shells. The latter case is suggested as a surrogate for the general case of a continuously varying radially symmetric concentration of emitting MNPs, for which the corresponding expressions can be found in Appendix 6 and Appendix 7.

As can be seen in the appendices, the resulting contributions from single MNPs or pairs of MNPs turn out to be linear or bilinear combinations of the quantities $\overline{d_1}$, $\overline{d_1^2}$, and $\overline{d_2^2}$ described by Eqs. (19), (22), and (24), respectively. In the case of different MNP concentrations in three radial intervals (Appendix 5), the mean contribution to the dose and the square of the dose from MNPs turn out to be given by Eq. (29):

$$\overline{D_N} = \bar{n}_a \overline{d_1}(\theta) \qquad \overline{D_N^2} = \bar{n}_a \overline{d_1^2}(\theta) \qquad \overline{D_{NN'}^2} = \bar{n}_a{}^2 \times \left(\overline{d_2^2}(\theta) - 8V_p \overline{d_{1b}^2}(\theta)\right) \tag{29}$$

Here, $V_n$ is the volume of the nucleus, and $\theta = \{R_n, R_c, R_x, u_n, u_p, u_x\}$ is the set of radii and concentrations relative to the average number density $\bar{n}_a$ in the sphere of radius $R_x$. $\bar{n}_a$ is related to the average number of emitting MNPs per cell, i.e., the ratio of the number density of emitting MNPs and the number density of cells present in the sphere of radius $R_x$, where $R_x$ can be chosen arbitrarily large.

The quantities $\overline{d_1}(\theta)$, $\overline{d_1^2}(\theta)$, $\overline{d_2^2}(\theta)$, and $\overline{d_{1b}^2}(\theta)$ are obtained from the radial dose distribution around a single emitting MNP using Eqs. (30) to (33), with the weighting functions $w_1$, $w_2$, and $w_{1\mathrm{b}}$ given by Eqs. (A.19), (A.20), and (A.21), respectively:

$$\overline{d_1}(\theta) = \int_{r_p}^{\infty} D_1(r_i) r_i{}^2 \times 4\pi w_1(r_i|\theta) dr_i \tag{30}$$

$$\overline{d_1^2}(\theta) = \int_{r_p}^{\infty} [D_1(r_i)]^2 r_i{}^2 \times 4\pi w_1(r_i|\theta) dr_i \tag{31}$$

$$\overline{d_2^2}(\theta) = \int_{r_p}^{\infty} \int_{r_p}^{\infty} D_1(r_i) r_i{}^2 D_1(r_j) r_j{}^2 \times (4\pi)^2 w_2(r_i, r_j|\theta)\, dr_j\, dr_i \tag{32}$$

$$\overline{d_{1b}^2}(\theta) = \int_{r_p}^{\infty} \int_{r_p}^{\infty} [D_1(r_i)]^2 r_i{}^2 \times (4\pi)^2 w_{1\mathrm{b}}(r_i|\theta)\, dr_i \tag{33}$$

### *2.4 Scenarios of MNP distributions*

Many studies in the literature on the dosimetric effects of MNPs at the cellular level have focused on a single cell only. For instance, (Lin *et al* 2015) considered several cases of distributions: (1) MNPs only in the cell; (2) MNPs uniformly distributed only in the cell nucleus; (3) MNPs only in the cytoplasm; (4) MNPs only in a thin spherical shell outside the cell; and (5) MNPs uniformly distributed in the cytoplasm and in a thin spherical shell outside the cell at the same concentration and in the cell nucleus at a different concentration.

The first three correspond to extreme cases. In all these cases, there is complete uptake of all available MNPs into the cell. In the first case, there is even complete uptake by the cell nucleus, which may be considered the ideal case for achieving optimum dose enhancement. The last case may be considered more realistic, in that the nuclear membrane may hinder diffusion of MNPs into the cell nucleus without completely suppressing it, while there is a uniform concentration outside the cell nucleus.

These cases are also among the potential scenarios for MNP uptake by cells considered in the present work (see Table 1). The first five scenarios correspond to those of (Lin *et al* 2015). In scenarios 1 to 3, there is complete uptake of the MNPs into the cells, where the MNPs have a uniform concentration throughout the cell or only in the nucleus or only in the cytoplasm. In scenario 4, there is no uptake, so that the MNPs are only outside the cells. In scenario 5, the average concentration of MNPs in the cell is the same as outside, while only 10% of the MNPs inside the cell are within the nucleus. The additional scenario 6 assumes that the concentration of MNPs in the cell is half that outside the cell, and only 10% of the MNPs in the cell are within the nucleus. In scenario 7, the MNPs are only near the outer surface of the cell nucleus, and in scenario 8, they are only in the middle of the cytoplasm (mimicking MNPs in an endosome). In scenario 4, there is no uptake by the cell, so that the MNPs are only outside.

For all these cases, Eqs. (A.15) to (A.17) can be used to determine the mean dose and mean square of the dose in the cell nucleus by appropriate choice of the radii and MNP concentrations based on the values used in (Lin *et al* 2015).

However, the assumption that MNPs are confined to a region slightly larger than the cell appears not to be very relevant. If the envisaged scenario is a solution of cells in water, as encountered in radiobiologic cell experiments, then it would be more plausible to assume the concentration outside the cell to be uniform everywhere outside cells. Another plausible assumption is then that MNPs are distributed in neighboring cells in the same way as in the considered cell. This will change the density of MNPs contributing to the dose in the considered cell, particularly if cells are densely packed within tissue.

Table 1: Summary of the scenarios considered in this study. The second and third columns specify the intracellular distribution of MNPs by giving the ratios of the nanoparticle concentrations in the nucleus ($\bar{n}_n$) and cytoplasm ($\bar{n}_p$) to the average nanoparticle concentration in a cell ($\bar{n}_c$). The fourth and fifth columns show the ratios of the MNP concentrations in cells and in the extracellular region ($\bar{n}_{xt}$) to the overall average MNP concentration ($\bar{n}_a$). $c_V$ is the packing density of cells or the average fraction of the volume covered by cells. $V_c$ and $V_n$ are the volumes of the cell and nucleus, respectively, where the radii used by (Lin *et al* 2015) were applied (i.e. $R_c$ = 6.75 µm and $R_n$ = 4 µm). $V_s$ is the volume of a sphere of radius slightly larger (100 nm) than the nucleus. $V_a$ and $V_b$ are the volumes of spheres with radius $R_a$ = 5 µm and $R_b$ = 6 µm, respectively.

| Scenario | | $\bar{n}_n/\bar{n}_c$ | $\bar{n}_p/\bar{n}_c$ | $\bar{n}_c/\bar{n}_a$ | $\bar{n}_{xt}/\bar{n}_a$ |
|---|---|---|---|---|---|
| 1 | cell | 1 | 1 | ${c_V}^{-1}$ | 0 |
| 2 | nucleus | $V_c/V_n$ | 0 | ${c_V}^{-1}$ | 0 |
| 3 | cytoplasm | 0 | $V_c/(V_c - V_n)$ | ${c_V}^{-1}$ | 0 |
| 4 | extern | 0 | 0 | 0 | $(1 - c_V)^{-1}$ |
| 5 | n10% | $0.1\, V_c/V_n$ | $0.9V_c/(V_c - V_n)$ | 1 | 1 |
| 6 | &50% | $0.1\, V_c/V_n$ | $0.9V_c/(V_c - V_n)$ | $(2 - c_V)^{-1}$ | $2(2 - c_V)^{-1}$ |
| 7 | surface | 0 | $V_c/(V_s - V_n)$ | ${c_V}^{-1}$ | 0 |
| 8 | endosome | 0 | $V_c/(V_b - V_a)$ | ${c_V}^{-1}$ | 0 |

Therefore, four different cases are considered (Table 2) in this work: isolated cells with MNPs present only in their proximity; cells in solution; and surrogate models for cells in tissue represented by densely packed spheres in simple cubic and face-center cubic lattices.

Table 2: Summary of the cases considered in this study. The packing density given in the third column is the fraction of total volume taken by the spheres representing the cells. $R_n$ and $R_c$ are the radius of the nucleus and the cell, respectively. $R_x$ is the maximum radial distance. The first case is the one considered by (Lin *et al* 2015). The cases sFCC and s_SC are the surrogates for densely packed cells in a face-centered cubic (FCC) and simple cubic (SC) lattice in which the MNP concentration outside a considered cell is replaced by the corresponding average concentration.

| Case number and name | | packing density $c_V$ | $R_x/R_n$ | $\bar{n}_x$ |
|---|---|---|---|---|
| 1 | isolated | $\frac{{R_n}^3}{(2R_c - R_n)^3}$ | $2\frac{R_c}{R_n} - 1$ | $\bar{n}_{xt}$ |
| 2 | solution | $10^{-3}$ | 30 | $n_c c_V + \bar{n}_{xt}(1 - c_V)$ |
| 3 | sFCC | $\frac{\pi}{3\sqrt{2}}$ | 30 | $n_c c_V + \bar{n}_{xt}(1 - c_V)$ |
| 4 | s_SC | $\frac{\pi}{6}$ | 30 | $n_c c_V + \bar{n}_{xt}(1 - c_V)$ |

## 3. Results

### *3.1 Elementary weighting functions for MNPs uniformly distributed in a sphere*

Sample weighting functions that allow calculation of the dose in a cell nucleus for different radial distances from the nucleus center from the dose around a single MNP emitting photons or electrons can be seen in Supplementary Fig. 1. The curves correspond to the cell geometry considered by (Lin *et al* 2015, Velten and Tomé 2023).

Weighting functions for calculating the mean dose to the cell nucleus for different sizes of the sphere filled with MNPs are shown in Fig. 2. It can be seen in Fig. 2 that the weighting function peaks at radial distance 0 and monotonously decreases when the MNPs are only in the cell nucleus. When the radius of the sphere containing MNPs exceeds the radius of the nucleus, an initial plateau is found before the function begins to fall to zero.

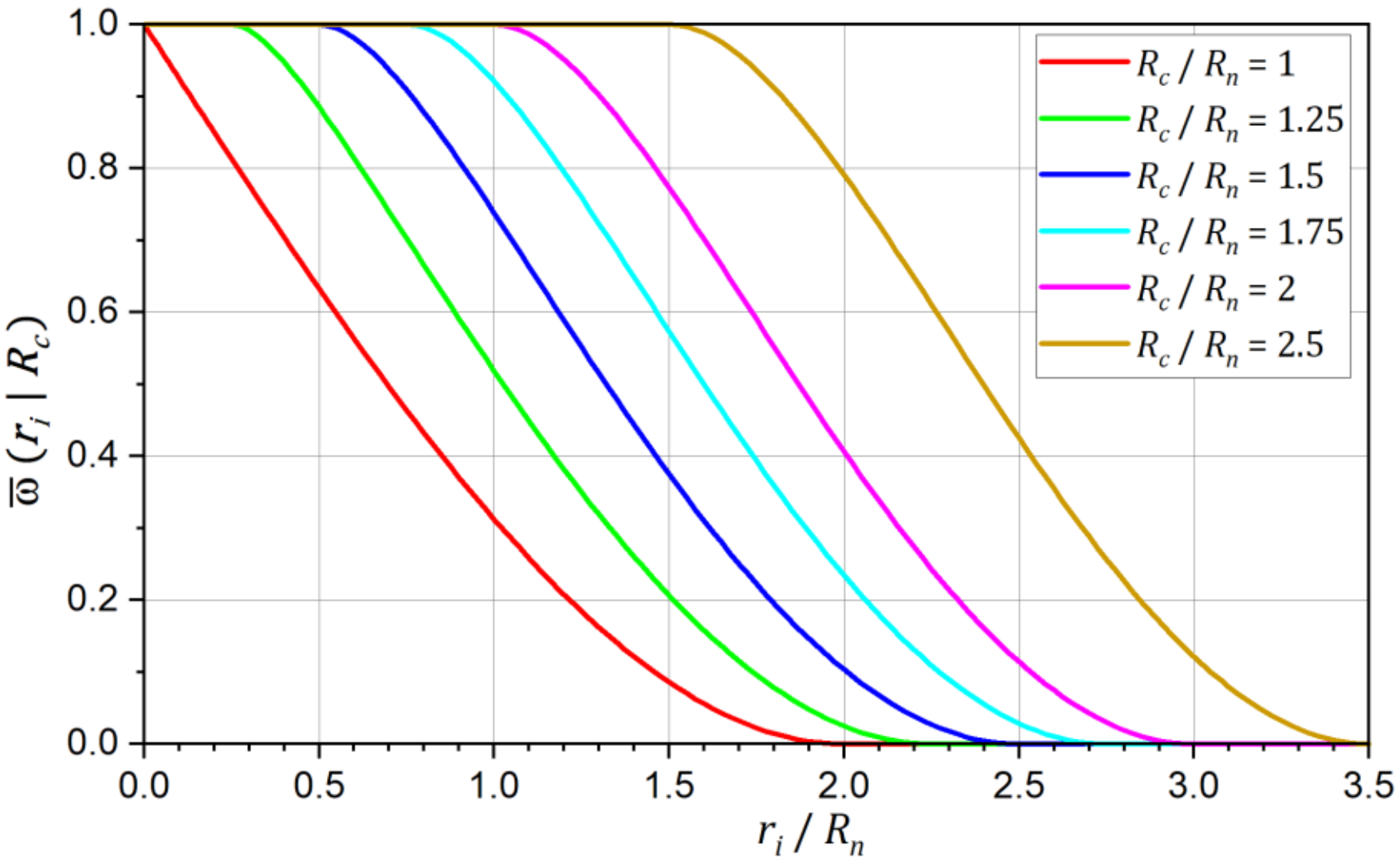

Fig. 2: Weighting function for calculating the mean contribution per emitting MNP to the dose in the cell nucleus (radius $R_n$) from the radial dose distribution around a single MNP. The MNPs are uniformly distributed in a sphere of radius $R_c$ which is concentric to the nucleus. The different curves correspond to different ratios of $R_c/R_n$ (see legend).

An example of a bivariate weighting function (determining the synergistic contribution of pairs of MNPs to the mean square of the dose) is shown in Fig. 3. The value of the parameter $a = R_c/R_n$ determining the distribution corresponds to the cell model used by (Lin *et al* 2015, Velten and Tomé 2023) and the case that the MNPs are uniformly distributed within the whole cell. In this case, the weighting function is unity for radial distances of up to 75% of the nucleus radius and gradually drops to zero while one or both radial distances increase. Weighting functions for some other choices of $R_c/R_n$ can be found in Section S1.3.1. of Supplement 1.

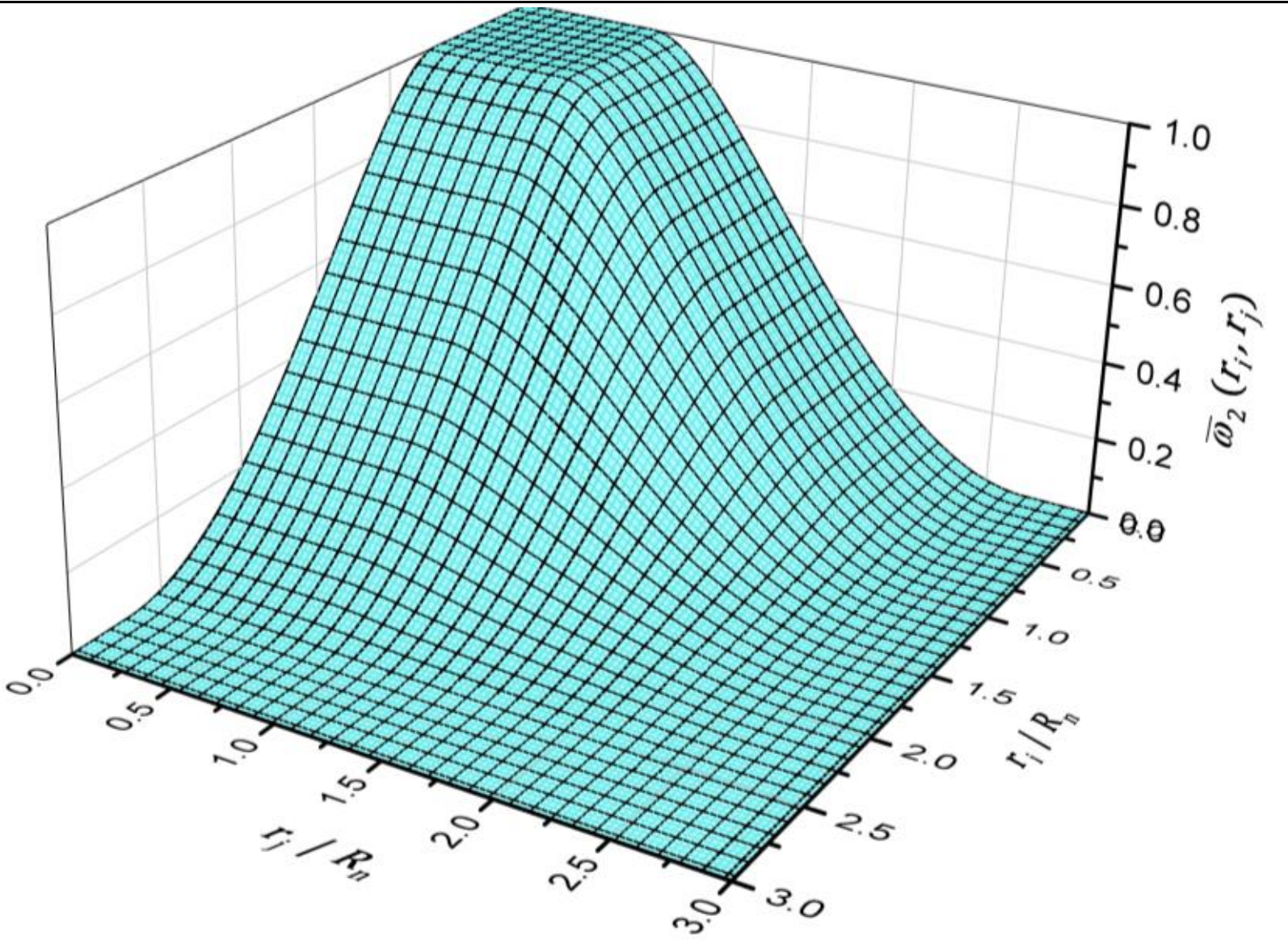

Fig. 3: Functional dependence of the weighting function $\overline{\omega}_2$ (see Eq. (25) and Appendix 2) on the ratios $r_i/R_n$ and $r_j/R_n$ ($x$- and $y$-axes) for the value a = 1.6875 of the ratio $R_c/R_n$.

The functional dependence of the weighting function for the part of the variance depending quadratically on the MNP concentration can be seen in Fig. 4. A general feature of this weighting function is that it vanishes for small values of $r_i/R_n$ and $r_j/R_n$, where a plateau can be seen in Fig. 3. It is worth noting that the peaks and troughs seen in Fig. 4 have only small amplitudes in the order of magnitude of 0.01. This implies that the corresponding contribution to the variance of the dose in the nucleus will only become significant at high number densities of emitting MNPs.

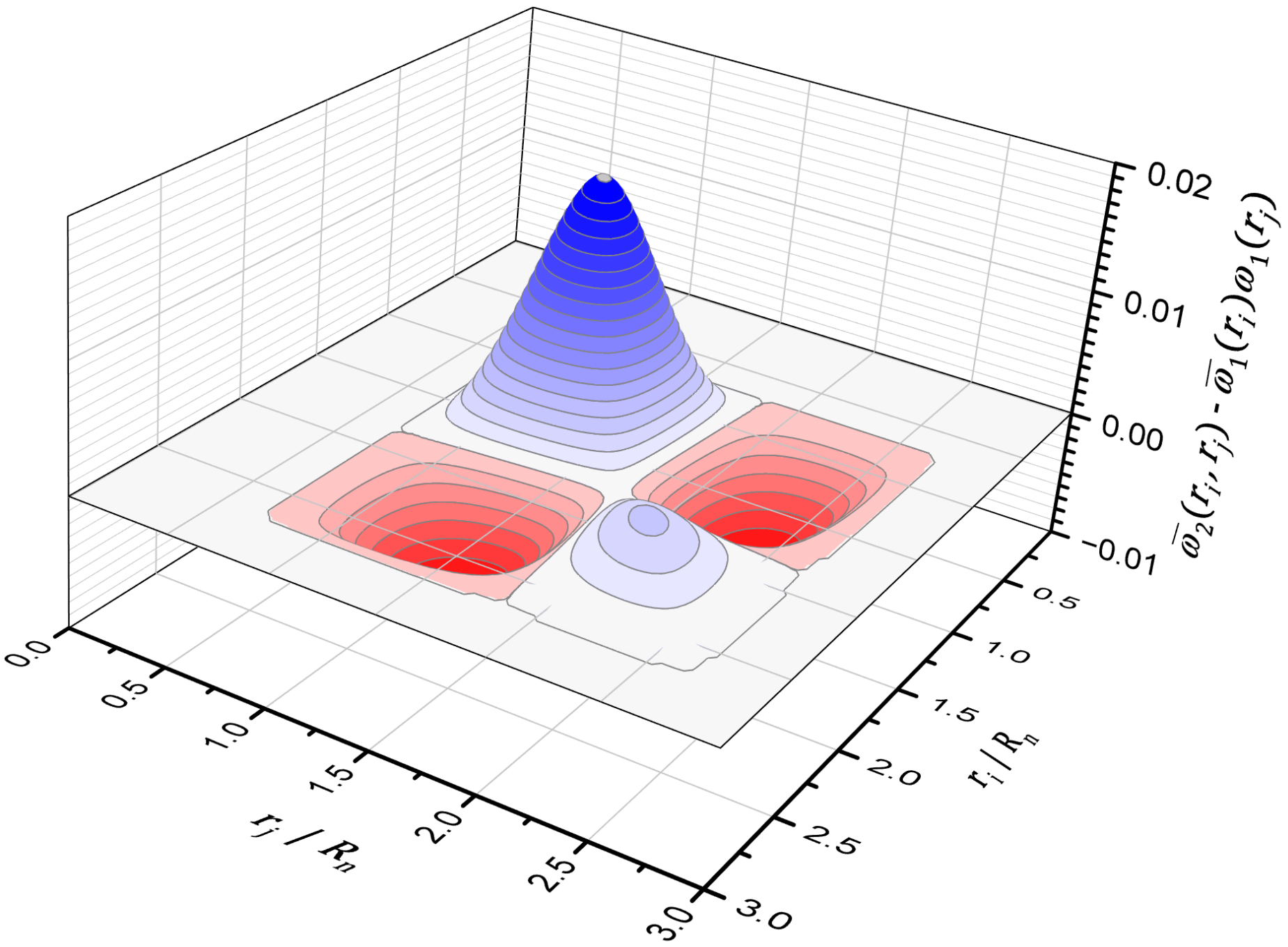

Fig. 4: Functional dependence of the weighting function $\bar{v}_2$ (Eq. (28)) on the ratios $r_i/R_n$ and $r_j/R_n$ ($x$- and $y$-axes) for the value a = 1.6875 of the ratio $R_c/R_n$.

### *3.2 Isolated cells versus cells in solution*

Five of the MNP uptake scenarios listed in Table 1 (see Supplementary Fig. 2) assume a complete uptake of the MNPs by cells. Scenarios 1 to 3 correspond to cases considered by (Lin *et al* 2015, Velten and Tomé 2023) of MNPs uniformly distributed within the entire cell, only in the cell nucleus, or only in the cytoplasm. The other two cases correspond to scenarios discussed in the literature (Douglass *et al* 2012), in which MNPs aggregate at the membrane of the cell nucleus or are clustered in a region of the cytoplasm such as an endosome.

The weighting functions for the radial dose distribution around a single MNP used to calculate the mean contribution to the dose in the spherical cell nucleus per emitting MNP shown in Fig. 5 correspond to these scenarios. They apply to an isolated cell with a ratio $R_c/R_n$ of the cell model used in the work of (Lin *et al* 2015, Velten and Tomé 2023) (about 1.69) with the reference concentration of MNPs pertaining to a sphere of 2.5 nm larger radius than the cell as considered in these papers. The orange curve in Fig. 5 corresponds to MNPs uniformly distributed within the whole cell (scenario 1), whereas the blue curve is for all MNPs in the cell nucleus (scenario 2). The gray line in Fig. 5 corresponds to scenario 3, in which the MNPs are distributed only in a spherical shell of inner radius $R_b$ in Eq. (A.9) identical to the radius $R_n$ of the cell nucleus. The outer radius $R_c$ is equal to the radius of the cell, so that MNPs are only in the cytoplasm. This weighting function prefers the dose distribution around an MNP at distances between $R_n$ and 1.5 $R_n$.

It can be seen from Fig. 5 that the weighting function favors the dose distribution at small radii when the MNPs are only in the nucleus, where the high absolute values of the weights are related to the enhanced concentration of MNPs in the nucleus (for scenario 2). Not surprisingly, the case of the MNPs concentrated on the outer side of the nuclear membrane gives lower weights to very small radial distances which are still significantly higher than what is obtained when the MNPs are uniformly distributed within the entire cell. When the MNPs are uniformly distributed in the cytoplasm or in a layer in the middle of the cytoplasm, the curves show a peak around distances equal to the nuclear radius. Whether the MNPs are spread over the whole cytoplasm or are only in a part that does not neighbor the nucleus does not seem to have a big influence.

It can be seen from Table 1 that the relative concentrations of MNPs listed in the fifth column depend on the fraction of the total volume that is filled by cells. For the geometry considered in the work of (Lin *et al* 2015, Velten and Tomé 2023), the fraction of the total volume is about 40%. However, if such a cell is conceived as being in a solution of cells, a volume fraction of cells of $10^{-3}$ is assumed in the second row of Table 2. For the cell dimensions considered by (Lin *et al* 2015, Velten and Tomé 2023), this corresponds to a number density of cells in the order of 1000 $\mu l^{-1}$, which appears to be a reasonable value. For complete uptake of the MNPs into cells, this implies that the ratio of the MNP concentration in the cells to the average concentration is 1000. Using the average concentration of MNPs in the considered solution may then not be the best option. Using the concentration of MNPs internalized in cells may be preferable in such a situation, as this is also the quantity determined by experimental techniques such as atomic emission spectroscopy (Coulter 2012).

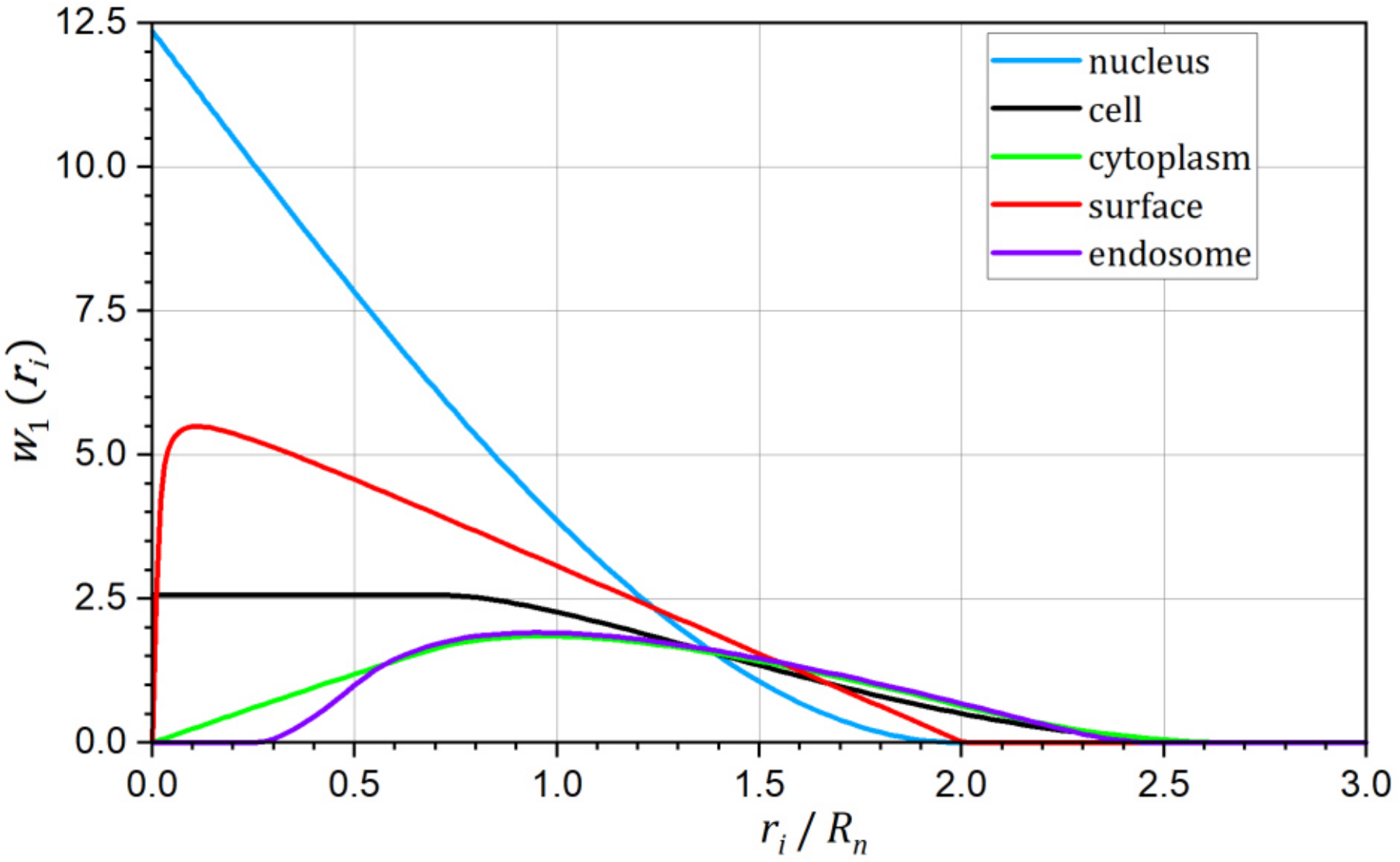

Fig. 5: Weighting function for calculating the mean contribution per emitting MNP to the dose in the spherical cell nucleus (radius $R_n$) from the radial dose distribution around a single MNP. The different curves apply to an isolated cell and different distributions of the MNPs in different compartments of the cell (see legend). The cases are as follows: all MNPs in the nucleus; all in the cell (sphere of radius $R_c$); all in the cytoplasm (spherical shell of inner radius $R_n$ and outer radius $R_c$); all at the nucleus surface (spherical shell of inner radius $R_n$ and outer radius $R_n$+50 nm); and all in a spherical layer in the middle of the cytoplasm (spherical shell of inner radius 1.25$R_n$ and outer radius 1.5$R_n$).

When the cells are in solution, the weighting functions are essentially unchanged, except for a small contribution due to other cells in the order of $10^{-3}$, which would not be discernible in Fig. 5. For the other scenarios, this is different, as illustrated in Fig. 6. Here, the solid lines represent the case of an isolated cell with the considered MNPs present only within a sphere about 40% larger in diameter than the cell. When the MNPs are only outside the cell (red curve), it is the dose around MNPs at distances between one and three times the nuclear radius that contributes to the dose in the nucleus. The blue curve corresponds to the "mixed case" of (Lin *et al* 2015) with the same average concentration in the cell and its environment but only 10% of the intra-cellular MNPs internalized in the nucleus. The solid green curve represents scenario 6 in Table 1, in which the average concentration in the cell is smaller by a factor of 2 than in the extracellular medium. In both scenarios, the presence of MNPs in the cytoplasm and the nucleus results in an increased weighting of the dose distribution at small values, as could be expected.

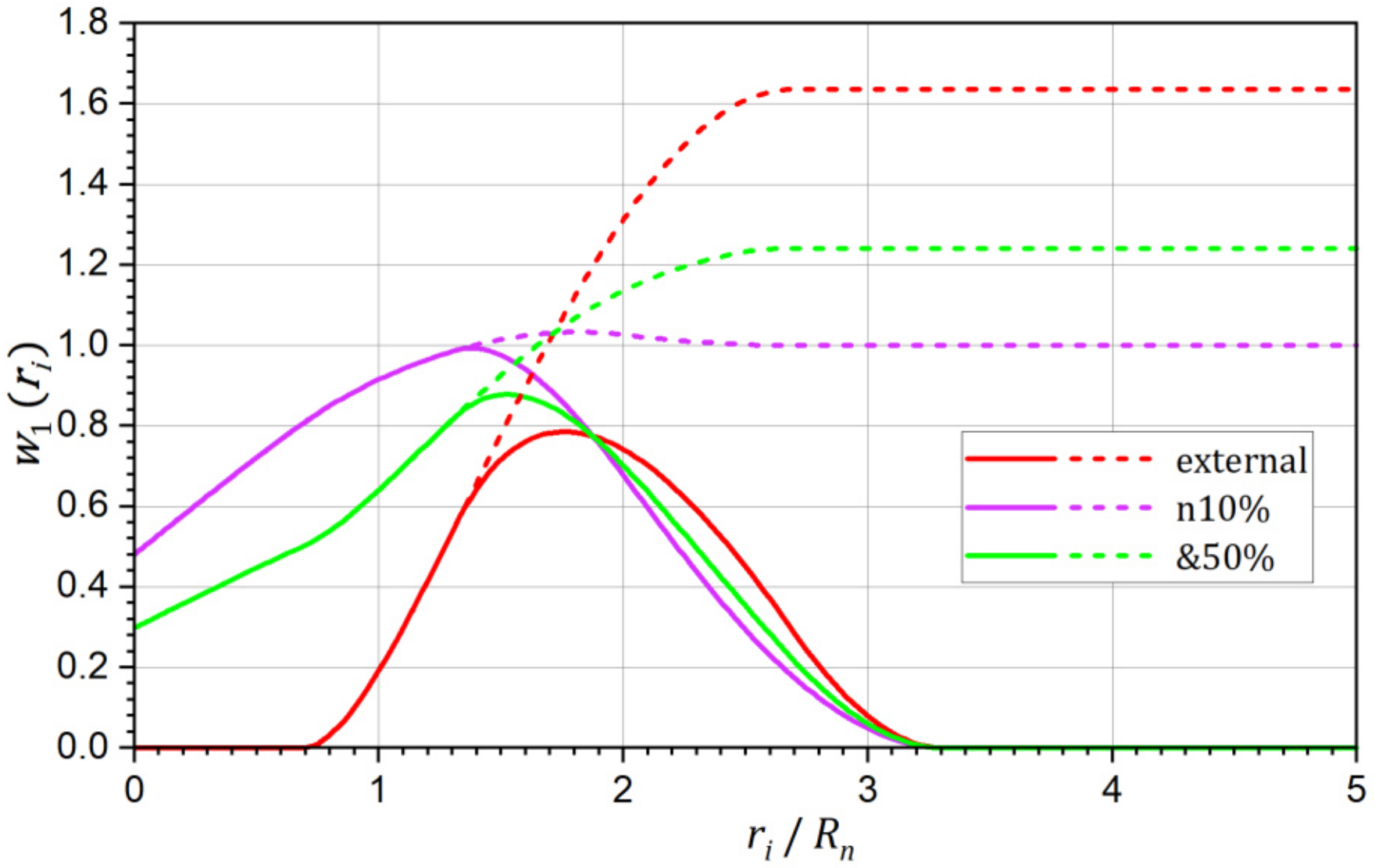

Fig. 6: Weighting function for calculating the mean contribution per emitting MNP to the dose in the spherical cell nucleus (radius $R_n$) from the radial dose distribution around a single MNP. The solid lines correspond to an isolated cell in which MNPs are contained in a sphere of radius 2.5 µm larger than the cell, as in (Lin *et al* 2015). The dashed lines correspond to cells in solution with the same concentration in the extracellular medium as for the solid lines. The cases considered are as follows: MNPs only outside the cells (red and orange); same average concentration of MNPs in the cells as in the extracellular medium, but only 10% of the MNPs internalized in cells are inside the nucleus (blue); the concentration in the extracellular medium is twice the average concentration inside the cells, and 10% of the internalized MNPs are in the nucleus (green).

When the cell is in solution, the assumption that the MNPs are only present in its vicinity is implausible. In this case, it appears more reasonable to assume a constant concentration of MNPs outside cells. The resulting weighting functions for the three scenarios are shown as dashed lines in Fig. 6. For scenario 5 (blue), where the average concentration in the cell was assumed to be the same as outside, the weighting function essentially stays at unity instead of dropping to zero. For the other two cases, a similar plateau is obtained but with a value higher than unity. This is related to the fact that if the concentration outside cells is assumed to be constant and the average concentration is lower inside cells, then the average concentration is higher for cells in solution than for the isolated cell.

The resulting plateaus extend to infinity. This means that the complete dose profile around an MNP contributes equally to the dose in the nucleus for these uptake scenarios and cells in solution. The resulting average dose in the nucleus will thus be significantly higher than what is obtained for a single isolated cell. This is illustrated in Fig. 7, which shows how the estimated dose contribution per density of emitting MNPs varies with the outer radius of the region around the MNP considered in evaluating the weighting function $w_1$ in Eq. (A.19).

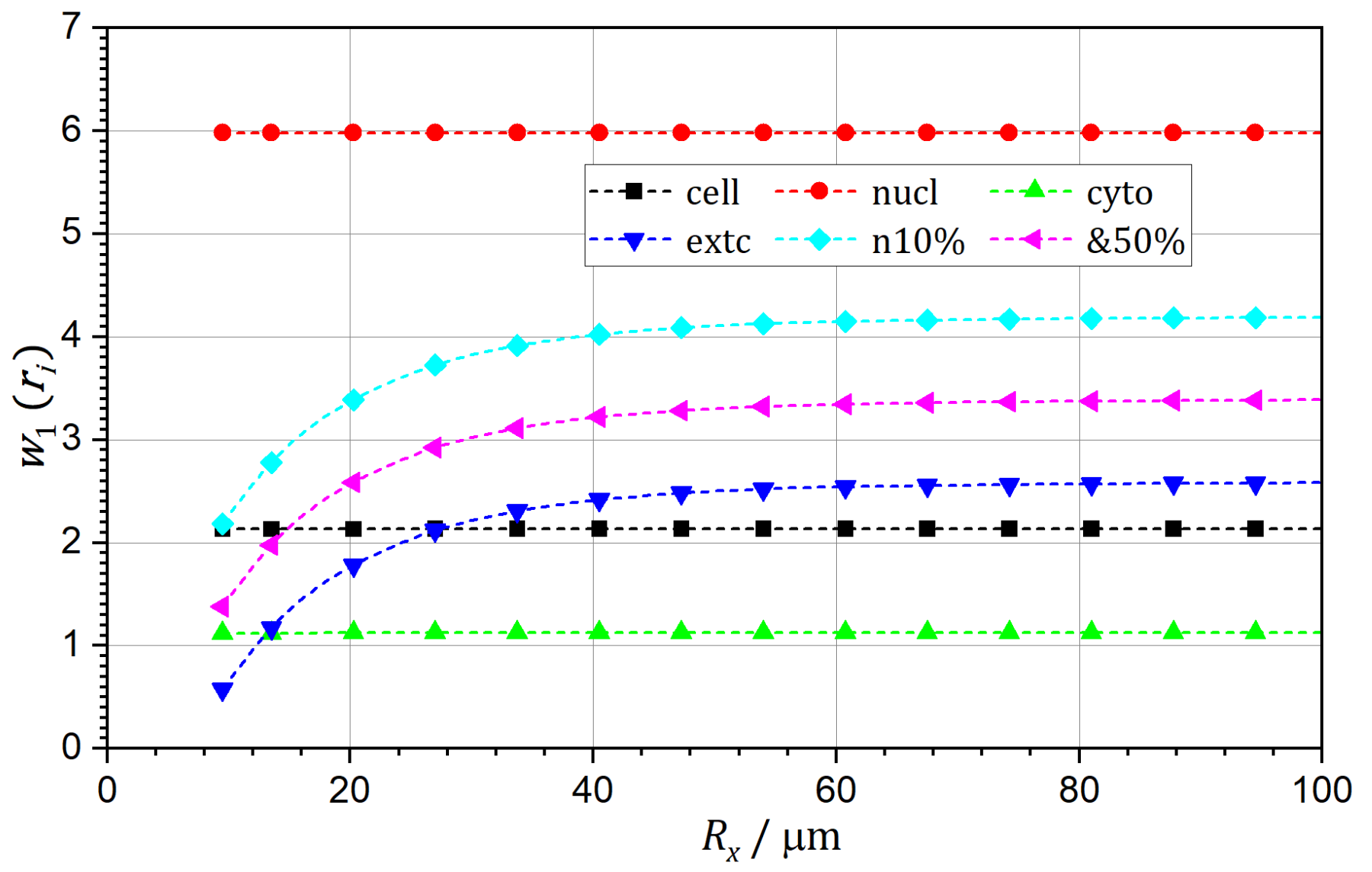

Fig. 7: Dependence of the dose contribution per density of emitting MNPs on the maximum radius considered in its calculation for the first six scenarios (Table 1) of MNP distribution in cells in solution. The curves were produced from the data of (Thomas *et al* 2024) for a 50 nm gold MNP irradiated at 100 µm depth in water by a mixed photon and electron fluence spectrum originating from a 100 kVp X-ray spectrum.

In the scenarios with complete uptake of MNPs into the cells, there is only a small indiscernible increase in the dose contribution per MNP density. In the scenario with all MNPs outside cells (dark blue data in Fig. 7), there is an increase by a factor of almost 5 compared to the starting value at a radius corresponding to the region containing MNPs in the isolated cell case. When the MNP concentration outside cells is the same as the average concentration in cells (magenta symbols), the increase amounts to a factor of 2. When the concentration outside is twice the average concentration in cells, the increase is by more than a factor of 3. This shows that with these uptake scenarios, the dose in the cell nucleus is heavily underestimated when isolated cells are considered.

### 3.3 Cells in tissue

If the cell is part of a tissue, it appears reasonable to assume that neighboring cells also contain MNPs at the same concentrations. (This was also implicitly assumed in the preceding for cells in solution.) For simplicity, two different regular arrangements of the cells are considered, which correspond to different packing densities. Both arrangements correspond to a cubic Bravais lattice. The less densely packed version is the simple cubic (SC) one in which the cells' centers are located on the vertices of the cubic unit cell. In the second arrangement, the face-centered cubic (FCC) one, there are cells at the vertices and the centers of the faces of the cube.

The resulting variation in the average radial concentration profile of MNPs around the center of a particular cell is shown in Fig. 8. All data in Fig. 8 correspond to the cell dimensions used in the works of (Lin *et al* 2015, Velten and Tomé 2023). The black and green lines in Fig. 8(a) correspond to the FCC and SC arrangements of cells, respectively, and relate to the case that the MNPs are uniformly distributed within the cells with no MNPs in interstitial regions. The blue and red lines are the corresponding average concentrations when the MNPs are only in the cell nuclei. The horizontal dashed lines show the surrogate concentration profiles, with constant values corresponding to the packing densities of the spheres as listed in Table 2. These constant concentrations outside the considered cell will be used below to estimate the overall dosimetric effect from a concentration of MNPs.

Fig. 8(a) shows the relative concentrations of MNPs for the four cases up to a value of $r_i/R_n$ of 30 as solid lines. The concentrations have been normalized to the density in the cell or in the nucleus of the considered cell, depending on whether the MNPs are uniformly distributed in the entire cell or only in the nucleus. The dashed lines represent the packing density (volume fraction) of cells or nuclei and correspond to the large-scale average concentration of MNPs. These constant values also apply to all larger radial distances (up to potential boundaries of the region loaded with MNPs). The data suggest that for complete uptake of the MNPs into the cells at constant intracellular concentration, the average concentration of MNPs outside the cell would be 52% or 74% of the concentration in the cells, depending on the cell packing density. If the MNPs are only present in the nuclei, the effective concentration outside the considered cell decreases to 10% and 14% of the concentration inside for SC and FCC packing of the cells, respectively.

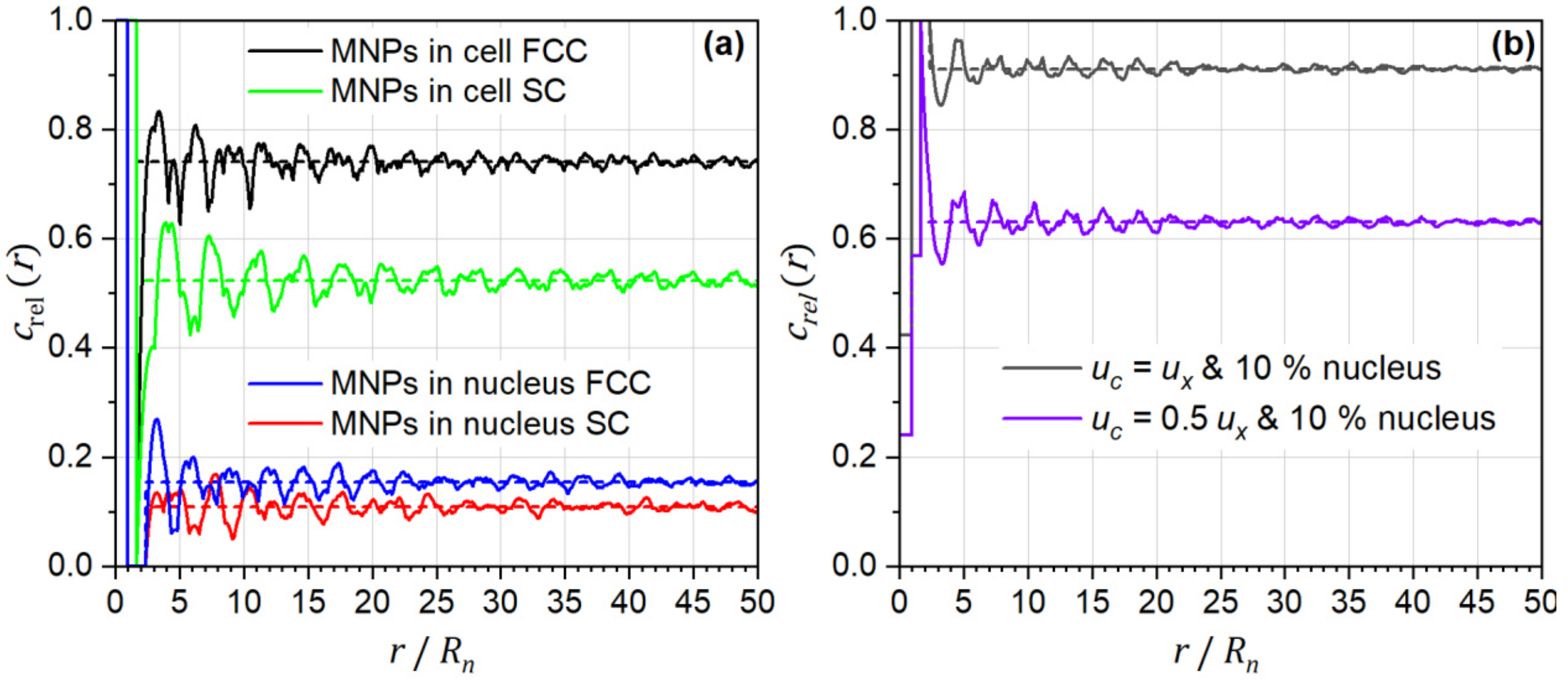

Fig. 8: Average concentrations of nanoparticles around a cell in tissue for (a) complete and (b) partial uptake of the MNPs by the cells. Further assumptions in (a) are a uniform distribution within the entire cell (black and green) or only in the cell nucleus (blue and red curves). The black and blue curves correspond to cells densely packed in a face-centered cubic (FCC) three-dimensional lattice, and the green and red curves to a simple cubic arrangement. In (b) the scenarios 5 and 6 from Table 1 are considered, where only 10% of the MNPs in a cell are in the nucleus. The black line corresponds to equal average concentration inside and outside cells, the blue line to an average cellular concentration of 50% of the concentration in the extracellular region. The dashed lines represent the surrogate density distribution of constant value equal to the average number density at radial distances exceeding the cell radius or twice the cell's radius minus the nuclear radius (for MNPs only in the nuclei). The data correspond to the cell geometry considered by Lin et al. (2015) and are normalized to the concentration in (a) the cell nucleus and (b) the extracellular region. Note that at small radii in (a), the red line is hidden by the blue line and the black line by the green one. In (b) the dashed blue line at small radii is hidden by the solid blue line and the dashed black and solid blue lines are hidden by the solid black line.

Cases 5 and 6 (see Supplementary Fig. 3(c) and Table 1) assume that the nuclear membrane hinders diffusion of MNPs, so that there is a lower concentration of MNPs in the nucleus than in the cytoplasm. In case 5, the average cellular concentration is the same as outside the cell, in case 6 it is only 50% of that concentration. The concentration profile that corresponds to case 5 for an FCC arrangement of the cells is shown as the black line in Fig. 8(b); the corresponding concentration profile for scenario 6 is shown as the blue line in Fig. 8(b). It should be noted that in Fig. 8(b), the concentration is normalized to the maximum concentration, which occurs at the cell surface. When the average concentration in the cell is the same as outside and only 10% of internalized MNPs are in the cell nucleus, the global average MNP density is about 10% lower than the maximum density. When the concentration in the cell is only 50% of that in the extracellular medium, the average concentration (blue dashed line in Fig. 8(b)) is only about 62% of the maximum value. In the nucleus of the considered cell, the relative MNP concentration is significantly higher than 0.1. This is because the scenario assumes 10% of the internalized MNPs to be in the nucleus, while the volume of the nucleus is about 20% of the cell volume for the considered geometry. Therefore, the concentration in the nucleus is about 50% of the concentration outside. (It is not exactly 50% because the concentration in the cytoplasm must be slightly increased to have the required average density in the cell.)

The resulting weighting functions for the surrogate concentration profiles pertaining to FCC packing are shown in Fig. 9. The corresponding data for SC packing can be seen in Supplementary Fig. 4. As would be intuitively expected, radial distances exceeding the nuclear radius or the cellular radius receive a significantly higher weight than small radial distances for the concentration profiles shown in Fig. 8(b), i.e., when there is reduced uptake of the MNPs by the cells or the cell nucleus. For enhanced uptake into the cells, there is more weight given to the dose at a small distance from the MNP.

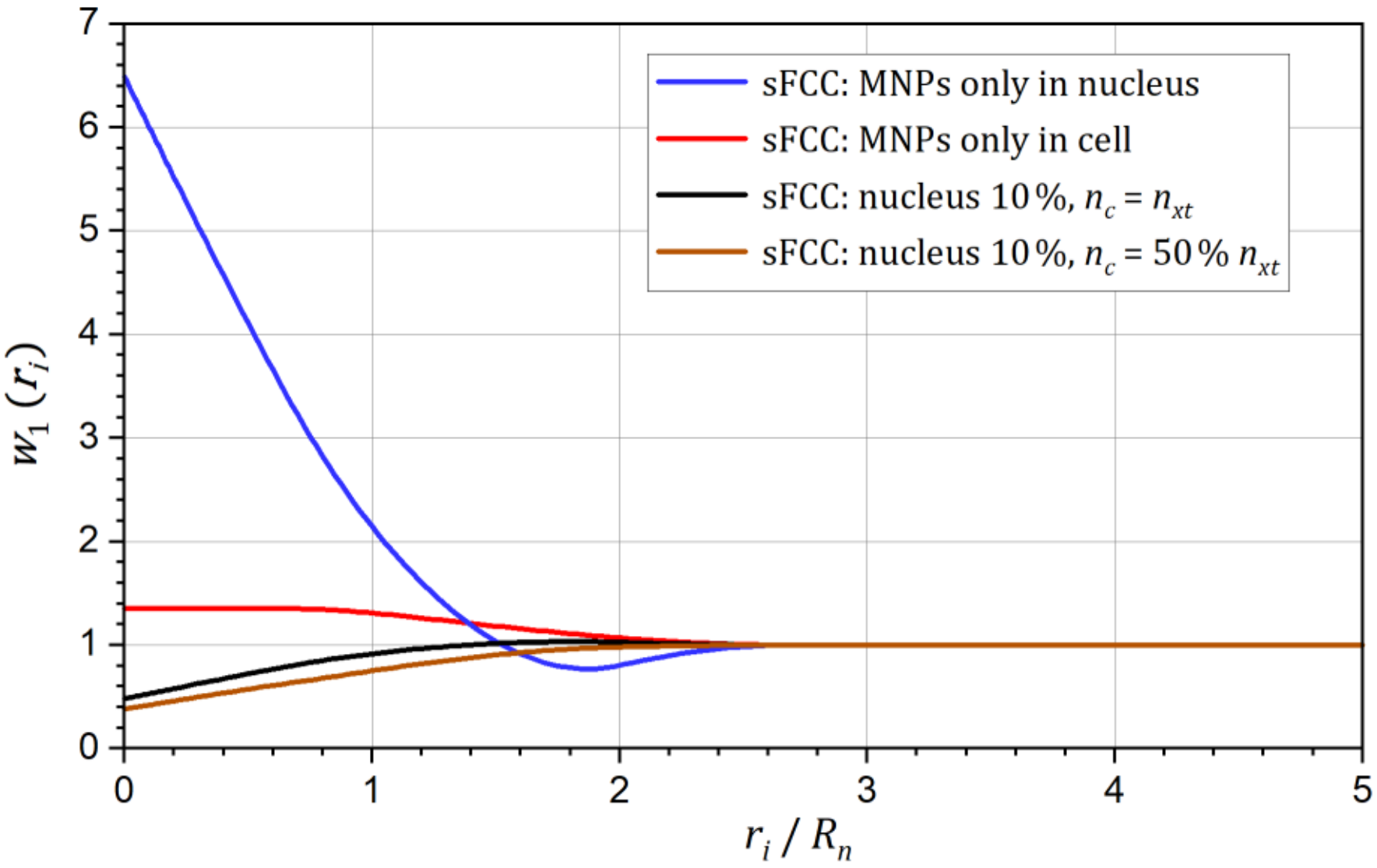

Fig. 9: Weighting function for calculating the mean contribution per emitting MNP to the dose in the spherical cell nucleus (radius $R_n$) from the radial dose distribution around a single MNP for the case of the surrogate MNP distribution for cells that are densely packed in an FCC lattice. The different curves apply to the following situations: MNPs only in the nuclei (red); MNPs only in the cells (orange); same concentration of MNPs in the cells and extracellular medium, but only 10% of the MNPs internalized in cells are inside the nucleus (black); the concentration in the extracellular medium is twice the average concentration inside the cells, and only 10 % of the internalized MNPs are in the nucleus. The ratio of the radii of the spherical cells to those of the concentric spherical nuclei corresponds to the cell geometry employed by (Lin *et al* 2015).

### *3.4 Comparison of the FCC and SC concentration profiles with the respective surrogates*

The MNP concentration profiles of the FCC and SC arrangements of cells deviate significantly from the dashed lines in Fig. 8. The corresponding weighting functions have been obtained based on the approach presented in Appendix 6. Those for the FCC cell arrangement are displayed in Fig. 10 for the first six scenarios for MNP uptake into the cells (see Table 1).

The weighting functions in Fig. 10 have a much smoother appearance than the density profiles in Fig. 8, but they still show significant oscillations where the crests reflect the coordination shells. The weighting functions oscillate around unity at radial distances from the MNP exceeding the cell radius, because the average MNP concentration is identical to the values indicated by the dashed horizontal lines.

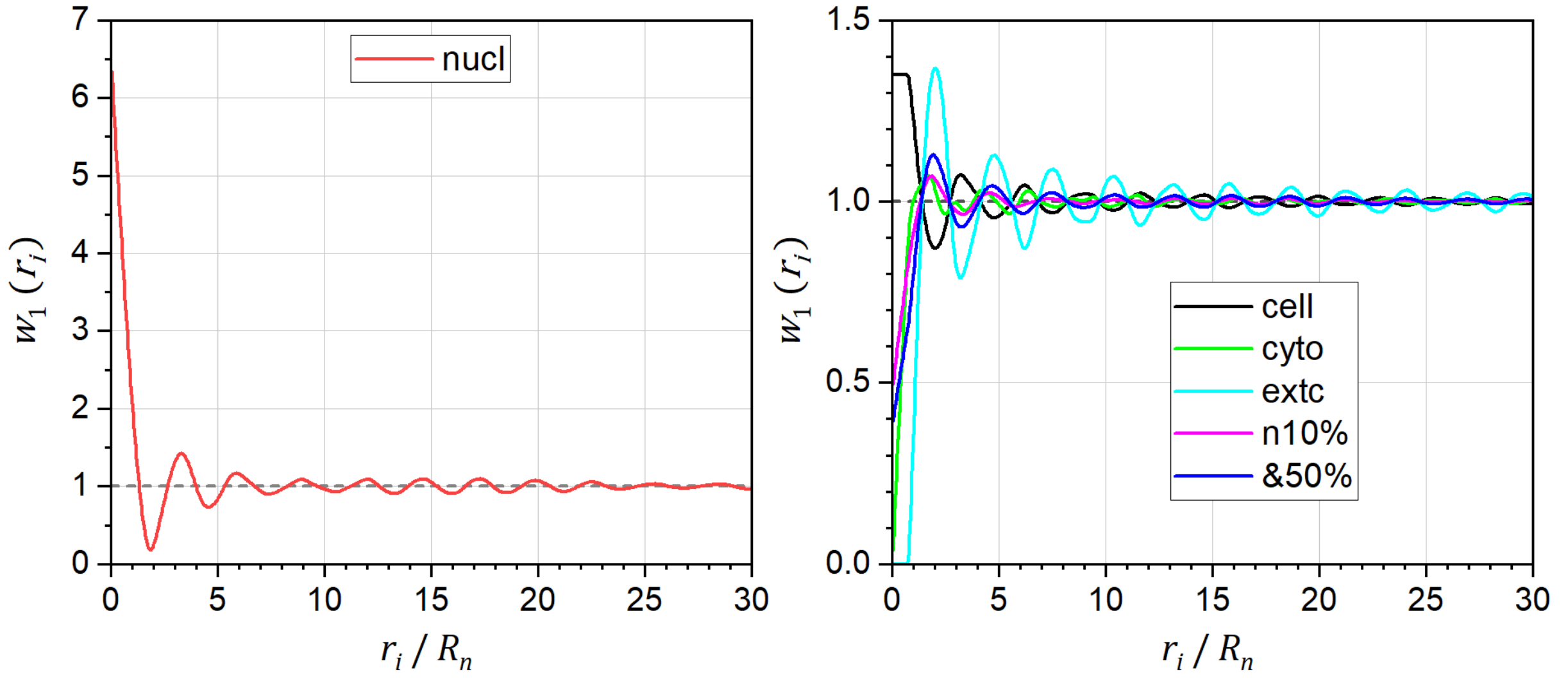

Fig. 10: Weighting function for calculating the contribution per emitting MNP to the mean dose in the spherical cell nucleus (radius $R_n$) from the radial dose distribution around a single MNP for cells that are densely packed in an FCC lattice. The curves apply to the first six uptake scenarios in Table 1. (a) MNPs only in the nuclei; (b) MNPs uniformly distributed in the cells (black); MNPs uniformly distributed in the cytoplasm (green); MNPs only in extracellular medium (cyan); same concentration of MNPs in the cells and extracellular medium, but only 10% of the MNPs internalized in cells are inside the nucleus (violet); the concentration in the extracellular medium is twice the average concentration inside the cells, and only 10% of the internalized MNPs are in the nucleus (blue). The ratio of the radii of the spherical cells to those of the concentric spherical nuclei corresponds to the cell geometry employed by (Lin *et al* 2015).

The bias introduced by using the surrogate weighting functions instead of those shown in Fig. 10 is illustrated in Fig. 11. This figure shows the contribution to the mean dose in the nucleus per average MNP concentration, $\overline{d_1}(R_n, R_c, R_x, u_n, u_p, u_x)$ (Eq. (A.22)), as a function of the radius of the outermost sphere representing the MNP-containing region around the considered cell for the first six uptake scenarios in Table 1. The colors indicate the uptake scenario (see legends), the symbols represent the results obtained using the surrogate MNP concentration outside the cell shown as dashed lines in Fig. 8. The solid lines are the corresponding results based on the weighting function shown in Fig. 10 and Supplementary Fig. 5, where Fig. 11(a) and (c) relate to the FCC arrangement of cells and Fig. 11(b) and (d) to the SC packing.

It can be seen that there appears to be almost no bias for the case of full MNP uptake into the nuclei in FCC-packed cells (Fig. 11(a)) and only a small bias for SC-packed cells (Fig. 11(b)). Larger offsets can be seen in Fig. 11(c) and (d) for the other scenarios with both packing densities of cells. In the case of no uptake into the cells (dark blue symbols and lines), the discrepancy is in the order of 20%. However, there seems to be an offset rather than a variation of the discrepancy with the value of the upper integration limit. This suggests that the discrepancy is already accumulated in the vicinity of the cell surface where the weighting functions show the highest oscillations (Fig. 10) and the deviations between the actual and surrogate concentrations are largest (see Fig. 8).

This suggests that the deviations seen in Fig. 11(c) and (d) could be reduced by using additional spherical shells to better approximate the concentration profiles shown as solid lines in Fig. 8. On the other hand, cells in tissue do not form a crystal lattice, so that the two cases of FCC and SC packing and their surrogates are only an estimate of the situation that would be found with real cells in real tissue. The difference between the two packing densities amounts to almost 35% in the case of complete uptake into the nucleus (see Fig. 11(a) and (b)). For MNPs uniformly distributed in cells (red curves and symbols in Fig. 11(c) and (d)), the difference is in the order of 10%, and only for the "mixed" cases with reduced MNP uptake in the cell nuclei do the differences appear to be in the range of a few percent.

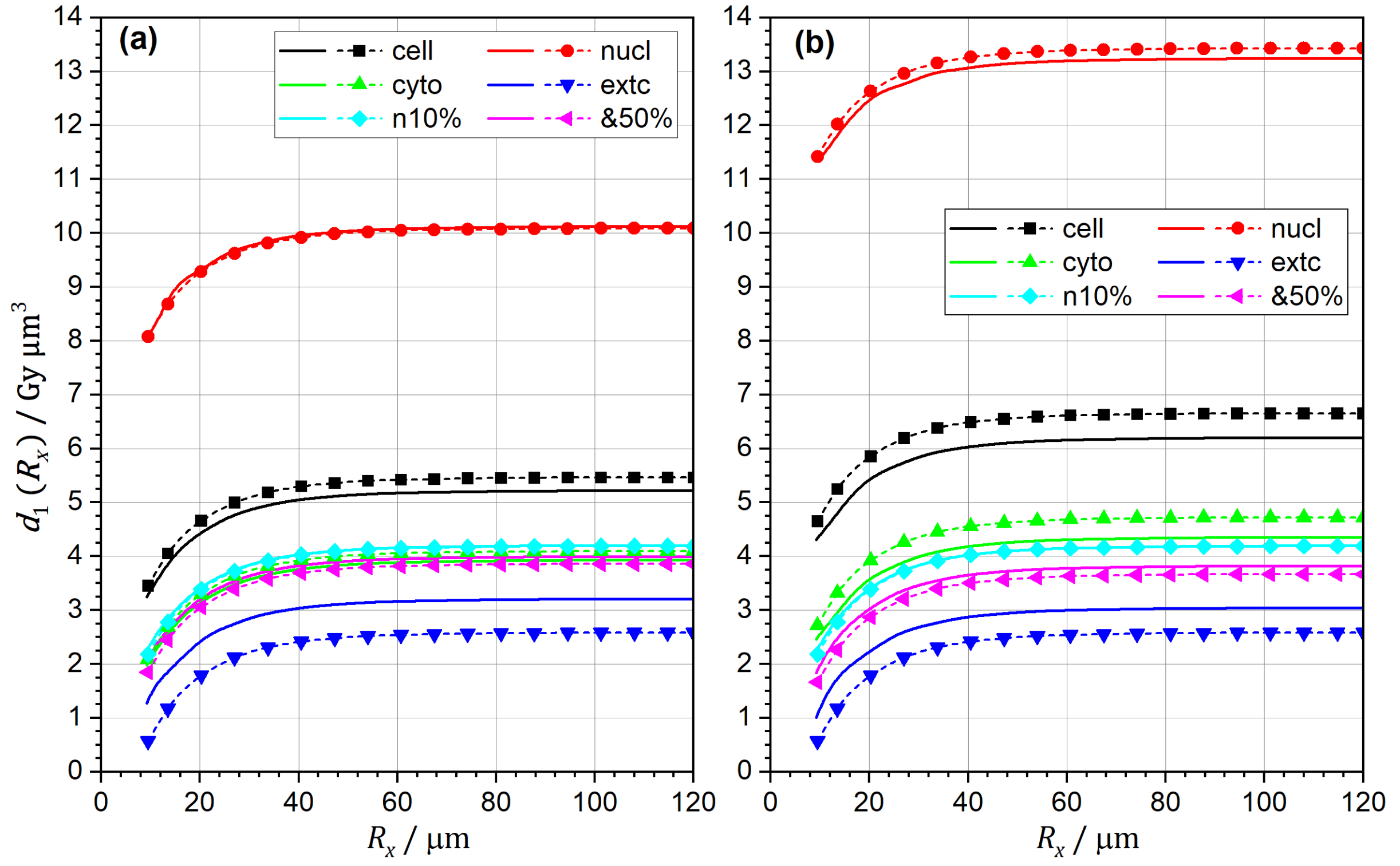

Fig. 11: Contribution per emitting MNP concentration to the mean dose in the cell nucleus for different upper radius values used in the determination of the weighting function for the MNP density distributions (lines) and their surrogates (symbols) as shown in Fig. 8. (a) and (b) pertain to full MNP uptake into the nuclei, (c) and (d) pertain to the uptake scenarios 2 to 6 in Table 1 (see legends). (a) and (c) are for FCC packing of cells, and (b) and (d) are for SC packing. The data apply to the dose distribution (shown in Supplementary Fig. 6) around a 50 nm gold MNP that is irradiated with a mixed photon and electron field originating from a 100 kVp X-ray photon spectrum from (Thomas *et al* 2024).

## 4. Discussion

### *4.1 Key features of the presented approach*

An approach is presented that allows calculation of the mean and variance of the dose in a cell nucleus from the dose resulting from the emission of particles from an MNP undergoing an ionizing radiation interaction or decay. The approach is based on weighting functions derived from the concentration profile of MNPs around the cell under consideration, which are dimensionless and depend parametrically on the geometry and the relative concentration of the MNPs in different regions of the geometry. The actual dosimetric effects produced by a given MNP distribution are the sums of terms scaling with the average number density of emitting MNPs, $\bar{n}_a$, or its square (Eq. (29)). The factors scaled by $\bar{n}_a$ and $\bar{n}_a{}^2$ are weighted averages of the radial distribution of the dose contribution from a single emitting MNP or of a quadratic form (bilinear combination) of this radial dose distribution.

The corresponding results are expectation-true estimates of the quantities considered, and their estimated uncertainties can be determined by using the same weighting functions, following the Guide to the Expression of Uncertainty in Measurements (JCGM 2008). Determining the uncertainties could, in principle, also be achieved by Monte Carlo techniques. However, to the authors' knowledge, all studies determining the dose effects of MNPs based on the LEM have only simulated the dose in the cell nucleus for one configuration of a set number of randomly distributed MNPs. That is, only for a single sample of the distribution of positions. Assessing the contribution to the uncertainty resulting from different sets of MNP positions would require multiple such simulations, which may become challenging in terms of computation time.

#### 4.1.1 Required input data

To apply the weighting functions, the dose distribution from particles emitted upon an ionizing interaction (or a decay) occurring in the MNP is needed. To determine this quantity, several authors, such as (Lin *et al* 2015, Velten and Tomé 2023), first simulated the radiation interaction with an MNP in a vacuum to obtain the fluence of emitted secondary particles. The latter was then used as input for a second simulation step to determine the contribution of the emitted particles to the dose around the MNP. (Li *et al* 2020, 2024) simulated an MNP in water irradiated by a confined beam. The difference between the results obtained with the MNP and without can be used as an estimate of the contribution of an ionizing interaction in the MNP to the dose around it (Rabus *et al* 2019, 2020, 2021).

A similar approach was used by (McMahon *et al* 2011b, Thomas *et al* 2024), with the difference that in a precursor simulation, the fluence of photons and electrons impinging upon the MNP was determined as input for the simulation of the radiation interaction and ensuing dose deposition around the MNP. In (Thomas *et al* 2024), these two further steps were determined in the same simulation, so that again, the difference between the results in the presence and absence of the MNP must be used as surrogate for the dose produced by particles emerging from the MNP as a result of an ionization occurring within it. As will be shown in the second part of the paper, the fact that (Thomas *et al* 2024) determined imparted energy separately for the cases of incident photons and electrons and for emitted electrons and photons allows this drawback to be remedied.

The best dataset for use with the weighting functions is obtained when two conditions are met. First, the fluence impinging on the MNP is determined in a precursor simulation, such as in (Thomas *et al* 2024), and this fluence—and not the fluence of the primary beam—is used in the next step. Second, in the next step, it is assured that the contribution from interactions in the MNP is separated from the contributions due to interactions in water of the incident particles that have not interacted in the MNP. This can be achieved by a two-step approach like the one used by (Lin *et al* 2015, Velten and Tomé 2023). Alternatively, the histories of primary particles leaving the MNP can simply be terminated in the simulation. With data obtained in this way also for a water nanoparticle, it is possible to use the weighting functions to determine the dose contribution from all water nanoparticles present in a geometry without need for obtaining this information by a separate simulation (as in (Thomas *et al* 2024)) or by reference to literature data (as in (Rabus *et al* 2019, 2020, 2021)).

A further condition placed on the dataset is that it covers the complete radial range in which energy is imparted by particles emitted by the MNP or their descendants. For the data of (Thomas *et al* 2024), this is the case, as can be seen in Supplementary Fig. 7. Supplementary Fig. 7(a) shows that the energy imparted per radial interval tends toward zero at the highest radial distances used for scoring. As a consequence, the curves of the cumulative energy imparted in spheres of increasing radius around the emitter show saturation behavior.

### *4.2 Comparison with the approach of Melo-Bernal et al.*

The first and most important difference to the approach taken by Melo-Bernal et al. (Melo-Bernal *et al* 2021) is the different perspective taken as a starting point. Melo-Bernal et al. considered the part of the dose distribution around a given MNP that falls inside a region such as a cell or a cell nucleus. Here, the initial question was, how many MNPs at a given distance from a considered point inside the scoring region contribute to the dose at this point? The latter perspective allows determination of the dose at different radial distances from the center of the nucleus. When the mean dose or mean square of the dose from single a MNP is determined, both approaches are equivalent and give the same results. However, the approach presented here also allows determination of the synergistic contribution from pairs of MNPs to the mean square of the dose in the nucleus.

In the approach of Melo-Bernal et al. (Melo-Bernal *et al* 2021), the synergistic contribution from pairs of MNPs cannot easily be assessed, and Melo-Bernal et al. argued that it is negligible, given the low probability of radiation interaction within a MNP. However, Eq. (23) shows that the synergistic contribution varies with the square of the MNP concentration, like the square of the mean dose. As can be seen in Fig. 4, the weighting function for the difference between the synergistic term and the square of the mean dose has very small values compared to the weighting function for the synergistic term shown in Fig. 3. This means that the variance of the dose in the nucleus is essentially given by the contribution to the mean square of the dose from single MNPs, $\overline{D_N^2}$, which is proportional to the MNP concentration, as is the contribution to the mean of the dose from MNPs, $\overline{D_N}$.

However, this does not imply that the approximation of neglecting the synergistic term made by Melo-Bernal et al. was justified. On the contrary, when the synergistic term is neglected, the mean square of the dose is underestimated. This is because the synergistic term is approximately the same as the square of the mean dose which is responsible for the nonlinear part of the logarithm of survival probability.

A second difference is that the weighting functions determined here parametrize geometry and the uptake scenario and can be readily used with data produced by any studies determining the radial dependence of the dose contribution from an emitting MNP. They can also be easily coupled with parameterizations of the radial dose distribution, such as was done in (Melo-Bernal *et al* 2021). However, the assumption that the radial dependence of the dose distribution around an MNP can be described by a power law is a very rough approximation, as can be seen in Supplementary Fig. 6. In this figure, the data used to produce the results in Fig. 7 and Fig. 11 are displayed as the product of dose and radius squared, i.e., as the quantity to which the weighting functions are applied in Eqs. (19) and (22). It is evident that the curve plotted in Supplementary Fig. 6 (in a log–log representation) is not well represented by a straight line.

Further differences are that the present approach involves a transparent separation of the overall quantitative effect, encoded by the average number density of emitting MNPs[2], the relative effects related to geometry and uptake scenario, and the specific dose distribution around an emitting MNP. The radial dose distribution around a single emitting MNP depends on the MNP radius and radiation quality. The corresponding weighted quantities $\overline{d_1}(R_c)$ (Eq. (19)), $\overline{d_1^2}(R_c)$ (Eq. (22)), and $\overline{d_2^2}(R_c)$ (Eq. (24)) for a uniform MNP distribution in a sphere of radius $R_c$ depend parametrically on the radius $r_p$ of the MNP and the ratio of $R_c$ to the radius $R_n$ of the cell nucleus. They do not depend on the average concentration of emitting MNPs. The resulting quantities for an uptake scenario (Eq. (29)) are linear or bilinear combinations of these elementary weighted dose quantities. They depend on radiation quality as well as MNP radius, geometry, and relative concentrations, but they do not depend on the absolute concentration of emitting MNPs. It must be noted, however, that the density of emitting MNPs is determined by the concentration of the MNPs and the expected number of interactions per MNP (Eq. (8)). For external irradiation, the latter depends on MNP size, the radiation quality, and fluence; for radioactive MNPs, they depend only on the activity per MNP and the exposure time.

### *4.3 Relevance*

The surrogate model of a sphere plus two concentric shells is a generalization of approaches in the literature to assess the impact of MNPs on the survival of cells within the framework of the LEM (e.g., (Lin *et al* 2015)). It was shown to approximate the case of cells in tissue reasonably well. Regarding the effects of MNPs in different uptake scenarios, it was shown that in contrast to cases with complete uptake of MNPs by cells (Fig. 5), the weighting functions for incomplete uptake scenarios show large differences between isolated cells and cells in solution (Fig. 6). For in vitro studies of cells and incomplete uptake of MNPs by the cells, the background concentration of MNPs in the solvent can therefore have a significant effect on the mean dose in the cell nucleus. With the data shown in Fig. 7, this effect translates into a factor of between 2 and 5 for the considered radiation quality and uptake scenarios. This implies that for these cases, there is an underestimation of the lesions resulting from the linear term of the LQ model by a factor of 2 to 5 and of the lesions according to the square term by a factor of 4 to 25! And this is even without considering the LEM-specific additional lesions due to the dose non-uniformity.

For scenarios with complete uptake of the MNPs by cells, Fig. 7 showed no significant influence of the environment when the cells are in solution. However, for cells in tissue there is a significant contribution to the mean dose in a considered cell due to MNPs emitting in neighboring cells. This contribution amounts up to several 10% (Fig. 11). For incomplete uptake scenarios, similar factors are obtained as with cells in solution. Comparison of Fig. 11(a) and Fig. 11(b) shows that different packing densities of cells also result in changes in the order of up to several 10%. It will be interesting to explore how these differences affect the predicted survival curves. This will be investigated in the second part of the paper.

[2] The average number density $\bar{n}_a$ of emitting MNPs may appear to be a somewhat unimaginative concept. In principle, it could be replaced by the product of $\bar{n}_a$ and the cell or nucleus volume. In which case the resulting reference quantity would be the ratio of the number of emitting MNPs to the number of cells or nuclei. Or the mean number of emitting MNPs per cell or cell nucleus but without them necessarily being present in the cell or the nucleus (depending on uptake scenario). While this may be more imaginative, correctly phrasing what this reference quantity is exactly appeared a bit cumbersome.

### 4.4 Impact of the approximations implied

Several approximations were implicitly made in the approach (and in that of (Melo-Bernal *et al* 2021)). One is that the contribution of an MNP to the dose in its vicinity has spherical symmetry. (Derrien *et al* 2023) reported an anisotropy of energy imparted around a spherical MNP amounting to several 10% for monoenergetic photons with a simulation setup that did not assure charged particle equilibrium (CPE). (Rabus 2024a) showed that after correction for CPE, the anisotropy was reduced to below 4% for the extreme cases reported by (Derrien *et al* 2023). (Rabus 2024a) also pointed out that this non-isotropy may, to some extent, be a simulation artefact, since some radiation transport codes model photoemission from all atomic shells according to Sauter's formalism, which strictly only applies to K shells.

A second assumption is that the radial dose distribution around an MNP does not change in the presence of other MNPs. In the study of (Thomas *et al* 2024), it was found that for electrons impinging upon MNPs, the fluence of outgoing electrons is only slightly different from that of incident electrons, with a minor reduction in the percent range at energies higher than 1 keV. At kinetic energies below 100 eV, a reduction by about 50% was found. But these electrons are stopped within some tens of nanometers and can therefore be ignored. Higher-energy electrons (energies of several tens of keV) have ranges of 50 µm or higher and may therefore interact with multiple MNPs.

Since elastic electron scattering occurs predominantly in the forward direction, elastic collisions will not lead to major distortions of the spatial distribution of energy transfer points. However, the small energy losses caused by impact ionization interactions with MNPs will accumulate and may lead to a reduction in the electron range. At the same time, there is also production of bremsstrahlung photons, which will carry their energy over longer distances than electrons would. The mentioned effects may lead to a change in the radial dose profile which can presumably be treated as a small uncertainty. However, this must be confirmed in further studies which could, e.g., simulate the dose distribution around an MNP immersed in uniform mixtures of water and the MNP material at different concentrations of the latter.

Another approximation was the use of the LQ model instead of the LQL model of cell survival. This leads to an overestimation of the number of induced lesions owing to the contribution from emitting MNPs located in the nucleus (Eq. (5)). This effect is not expected to be large. Using the weighting functions for the radial dose profile in the cell nucleus, it is possible to calculate the quantity defined in Eq. (6) for given radial distance from the nucleus center after evaluating the expected dose according to Eq. (16). However, this is more involved, since it requires determination of the dose profile for a specific scenario and the average number density of emitting MNPs. Therefore, addressing this point is postponed until future work.

Finally, the approach assumes that the cell nucleus is a sphere and that the MNPs have a spherically symmetric distribution, thus also implying that cells are spherical. Cells are generally not spheres, and studies have shown that the choice of cellular geometry has an impact on the dose to the cell nucleus in the presence of MNPs (Sung *et al* 2017). Similarly, scenario 8, presented as a surrogate for the situation of MNPs clustered inside an endosome, appears far from realistic, since endosomes are generally not in the form of a spherical shell. Another obvious over-idealization was assuming cells to be arranged in a simple cubic or face-centered cubic lattice like atoms or ions in a crystal. However, for the latter case of cells in tissue, it was shown that the ideal arrangement of cells can be approximated by a surrogate concentration profile for the MNPs with a constant value in the extracellular region. The value of the constant depends on the average density of cells only and not on their exact arrangement. If the cells are randomly distributed within a certain range around the ideal lattice points, the system will be amorphous, and the oscillations seen in Fig. 8 and Fig. 9 with the radial density profile and the resulting weighting function will be smeared out and dampened away.

With the case of MNPs in the endosome, a similar approach of angular averaging of the MNP density can be applied by determining the solid angle of a sphere of radius $r_i$ around a point in the nucleus according to Eq. (16). The key point here is that the weighting functions refer only to radial distances. Leaving aside synergistic effects that have been reported for MNP clusters (Rudek *et al* 2019), it should not matter for the dose to the nucleus whether the MNPs are concentrated in a small volume or are smeared out over the spherical shell defined by the MNPs closest and farthest from the center of the nucleus. In the latter case, the larger solid angle and value of the weighting function is compensated by the reduction in MNP density. Therefore, only the assumption of a uniform density in the spherical shell may be questioned. Given the insight from the case of cells in tissue, it can be expected that a surrogate concentration profile with a constant value may also give a reasonably adequate approximation in this case. This conjecture can be confirmed by using a similar approach as for cells in tissue and is left for future work.

Another caveat to be mentioned is that for the case of cells in tissue, it was implicitly assumed that all cells would be surrounded by cells containing the same concentration of MNPs and that these cells would therefore have the same expectations for the mean and standard deviation of the dose. However, there may be a distribution of the mean and standard deviation of the dose per cell. For instance, in a tumor uniformly loaded with MNPs, the cells near the surface have a lower coordination number of neighbors containing MNPs and, therefore, have a lower mean dose than cells in the inner of the tumor. There also could be a gradient from the surface inward owing to hampered uptake (like as happens with hypoxia).

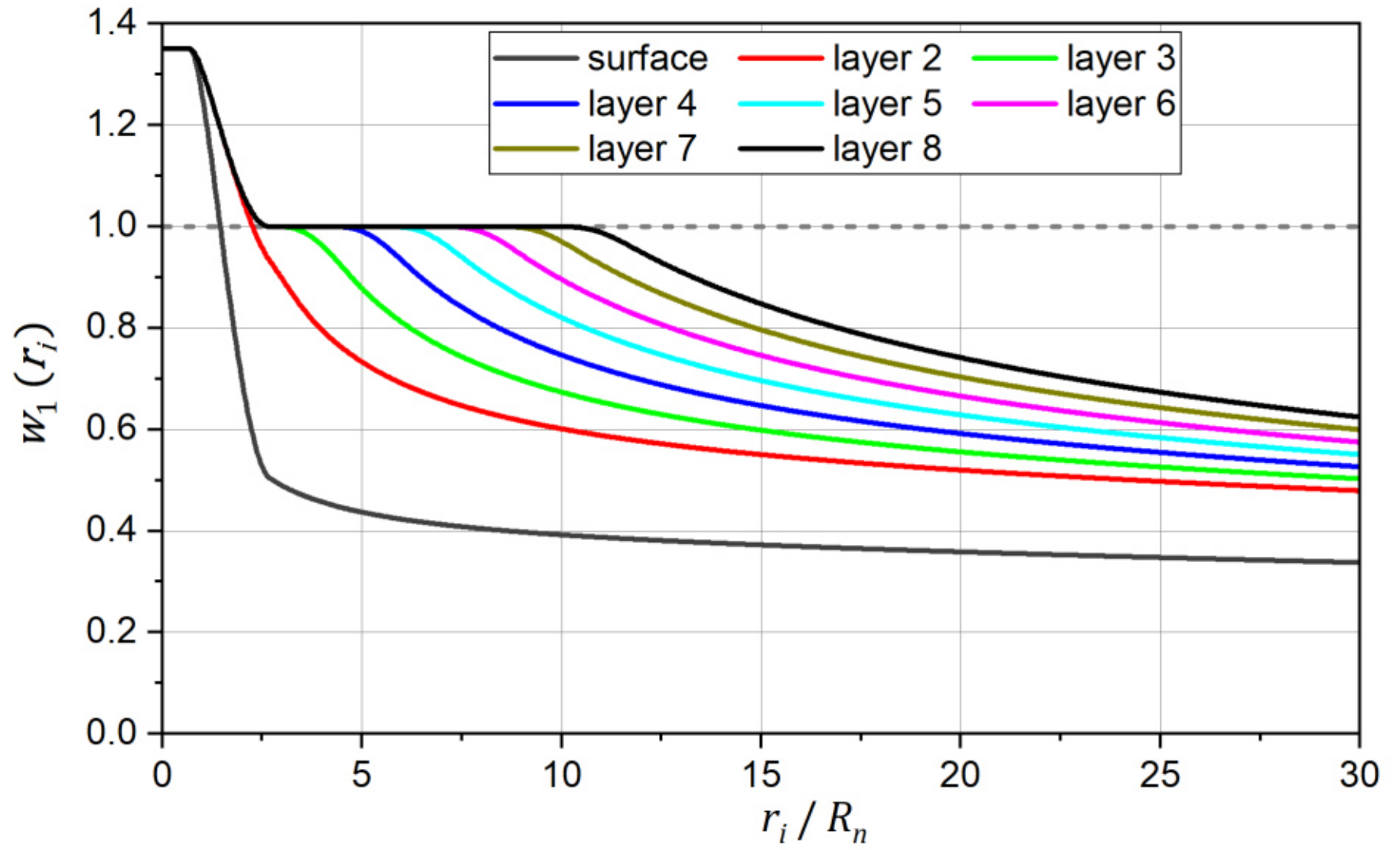

Fig. 12: Modified weighting functions for the sFCC case (cf. Table 2) and complete uptake into the cells for cells located near the surface of a spherical tumor of 1 mm diameter. The dark gray curve corresponds to cells directly at the surface, the colored curves correspond to cells at between two and eight cell layers from the tumor surface (see legend).

As was pointed out by (Zygmanski *et al* 2013a), a non-uniform dose distribution in a cell population will lead to modifications of cell survival fractions. However, these cases can also be approximately addressed by the formalism discussed in this paper. As an illustration, Fig. 12 shows the modified weighting functions for cells located at between one and eight cell layers from the surface of a spherical tumor, assuming the surrogate case of densely packed cells and MNPs uniformly distributed within the cells. For the cells at the surface, the weighting factor outside of the cell is significantly lower than unity for all radial distances larger than the cell radius. The same applies to the second layer. Starting with the third layer, a plateau appears in which the weighting function is unity until it finally starts to drop (when the distance from the center of the considered cell exceeds its distance from the surface). This assessment relies on the generalized weighting functions given in Appendix 6.

## 5. Conclusions

An approach has been presented that allows calculation of the mean dose and variance of the dose in a cell nucleus from the radial distribution of the additional dose around an MNP undergoing an ionizing interaction or radioactive decay. The methodology is based on the probability of finding MNPs at a given distance from a point at which the dose is evaluated. With this different perspective, it also became possible to assess the synergistic contribution of pairs of MNPs to the variance of the dose. It could be shown that this contribution is of similar magnitude to the square of the mean dose and therefore cannot be neglected as was done in the previous study of (Melo-Bernal *et al* 2021). The weighting functions for several surrogate cases have been explicitly stated in the appendix and can be used by other researchers to assess cell survival according to the LEM from their simulations of the dose around single MNPs. Based on the weighting function, conclusions could be drawn regarding the suitability of model cases of isolated cells in a spatially confined distribution of MNPs. It could be shown that for complete uptake scenarios and cells in solution, such simple cases can be used, whereas for partial uptake scenarios or cells in tissue, the presence of MNPs in an extended environment cannot be ignored. The extent of this environment is defined by the radial range within which the particles emitted by an MNP impart their energy. The case of a sphere and two surrounding spherical shells with constant MNP concentration in each but different between these regions has been suggested as a surrogate model for cells in tissue and shown to approximate to within a few 10% the results obtained with an idealized exact placement of the cells in a regular lattice. Similar deviations were found between different packing densities of cells. The approach presented here will be used in the second part of the paper to derive predictions of cell survival based on the LEM for different radiation qualities and uptake scenarios.

## Acknowledgements

Leo Thomas acknowledges support by the Metrology for Artificial Intelligence in Medicine (M4AIM) program which is funded by the German Federal Ministry for Economy and Climate Action within the framework of the QI Digital initiative. The authors thank Christian Velten for his comments on an earlier version of this manuscript and Piotr Zygmanski and Wolfgang Hoegele for an interesting discussion on their approach to handling the stochastic effects seen with nanoparticles. The authors thank Sally Aldenhoven for proof-reading the manuscript for correct English.

## Appendices

### *Appendix 1 Analytical expression for the monovariate weighting function*

The weighting function $\bar{\omega}(r_i|R_c)$ for the dose at a radial distance $r_i$ from the MNP used to obtain the mean dose in the cell nucleus from a uniform distribution of MNPs in a sphere of radius $R_c$ is given by Eqs. (A.1):

$$\bar{\omega}(r_i|R_c) = \bar{\omega}_1(x_i|a) = \begin{cases} 1 & x_i \leq a-1 \\ \bar{\omega}_1{}^{(2)}(x_i|a) & a-1 \leq x_i \leq a+1 \\ 0 & x_i > a+1 \end{cases} \tag{A.1}$$

Here, the reduced quantities given by Eq. (A.2) have been used:

$$x_i = \frac{r_i}{R_n} \; ; \quad a = \frac{R_c}{R_n} \tag{A.2}$$

The function $\bar{\omega}_1{}^{(2)}$ is given by Eq. (A.3):

$$\bar{\omega}_1{}^{(2)}(x_i|a) = \frac{1}{2}(1+a^3) - \frac{3}{8}(1+a^2)x_i + \frac{1}{16}x_i{}^3 - \frac{3}{16}(1-a^2)^2 x_i{}^{-1} \tag{A.3}$$

### *Appendix 2 Analytical expression for the bivariate weighting function $\bar{\omega}_2$*

The analytical expression of the weighting function $\bar{\omega}_2$ (for the derivation see Section S1.3 of Supplement 1) is given by Eq. (A.4) using the reduced quantities defined by Eq. (A.5):

$$\bar{\omega}_2(r_i, r_j|R_c) = \begin{cases} \bar{\omega}_1(x_i|a) & x_j \leq a-1 \leq x_i \leq a+1 \\ \bar{\omega}_2{}^{(3)}(x_i, x_j|a) & a-1 \leq x_j \leq 2a - x_i \;\wedge\; x_j \leq x_i \leq a+1 \\ \bar{\omega}_2{}^{(4)}(x_i, x_j|a) & 2a - x_i \leq x_j \leq x_i \leq a+1 \\ 0 & x_i \geq x_j > a+1 \end{cases} \tag{A.4}$$

$$a = \frac{R_c}{R_n} \; ; \quad x_i = \frac{\max(r_i, r_j)}{R_n}; \quad x_j = \frac{\min(r_i, r_j)}{R_n} \tag{A.5}$$

The functions $\bar{\omega}_1$, $\bar{\omega}_2{}^{(3)}$, and $\bar{\omega}_2{}^{(4)}$ appearing in Eq. (A.4) are given by Eqs. (A.1),(A.6), and (A.7), respectively:

$$\begin{aligned} \bar{\omega}_2{}^{(3)}(x_i, x_j) = \bar{\omega}(x_i) + \frac{1}{2}\bar{\omega}(x_j) - \frac{1}{2} + \frac{x_i}{x_j}P_3{}^{(3)}(x_j|a) + \frac{x_j}{x_i}P_4{}^{(3)}(x_j) \\ + \frac{1}{x_i x_j}P_5{}^{(3)}(x_j|a) \end{aligned} \tag{A.6}$$

$$\bar{\omega}_2{}^{(4)}(x_i, x_j) = \frac{1}{2}\bar{\omega}(x_i) + \frac{x_i}{x_j}P_3{}^{(3)}(x_i) + \frac{x_j}{x_i}P_4{}^{(4)}(x_i|a) + \frac{1}{x_i x_j}P_5{}^{(4)}(x_i|a) \tag{A.7}$$

${P_k}^{(3)}(x_j|a)$ and ${P_k}^{(4)}(x_i|a)$ ($k \in \{3,4,5\}$) are polynomials of third degree in the first argument. The polynomial coefficients depend on the parameter $a$ and are listed in Table 3 and Table 4.

Table 3: Coefficients of the polynomials ${P_3}^{(3)}(x_j|a)$, ${P_4}^{(3)}(x_j)$, and ${P_5}^{(3)}(x_j|a)$ in Eq. (A.6).

| | ${P_3}^{(3)}$ | ${P_4}^{(3)}$ | ${P_5}^{(3)}$ |
|---|---|---|---|
| ${x_j}^0$ | $\frac{(a-1)^2(2a+1)}{16}$ | 0 | $-\frac{(a-1)^3(8a^2+9a+3)}{80}$ |
| ${x_j}^1$ | $\frac{3}{16}(1-a^2)$ | $-\frac{1}{8}$ | $\frac{3}{32}(a^2-1)^2$ |
| ${x_j}^2$ | $\frac{3}{16}$ | 0 | $\frac{1}{16}(a-1)^2(2a+1)$ |
| ${x_j}^3$ | $\frac{1}{16}$ | $\frac{1}{160}$ | 0 |

Table 4: Coefficients of the polynomials ${P_3}^{(4)}(x_i|a)$ and ${P_5}^{(4)}(x_i|a)$ in Eq. (A.7).

| | ${P_3}^{(4)}$ | ${P_4}^{(4)}$ | ${P_5}^{(4)}$ |
|---|---|---|---|
| ${x_i}^0$ | 0 | $\frac{(1-a^2-a^3)}{16}$ | $-\frac{(8a^5+15a^4-a^2+3)}{80}$ |
| ${x_i}^1$ | $-\frac{1}{8}$ | $-\frac{3}{16}(1-a^2)$ | $-\frac{3}{32}(a^2-1)^2$ |
| ${x_i}^2$ | 0 | $\frac{3}{16}$ | $\frac{(1-a^2-a^3)}{16}$ |
| ${x_i}^3$ | $-\frac{1}{160}$ | $-\frac{1}{16}$ | $-\frac{1}{8}$ |

*Appendix 3 MNPs uniformly distributed in a spherical shell only*

Since the dose from different sources is additive, the weighting functions for the radial dose distribution in the nucleus and for the mean dose in the nucleus are easily obtained from Eqs. (16) and (19). The adjusted formulas for MNPs inside a spherical shell of inner radius $R_n$ and outer radius $R_c$ only are given by Eqs. (A.8) and (A.9):

$$D(\vec{r}) = D_w + \bar{n}_m \int_{r_p}^{\infty} D_1(r_i)\,[\Omega(r_i, r|R_c) - \Omega(r_i, r|R_n)]{r_i}^2 dr_i \tag{A.8}$$

$$\bar{D}(R_n, R_c) = D_w + \bar{n}_m\left[\overline{d_1}(R_c) - \overline{d_1}(R_n)\right] \tag{A.9}$$

Here, $\bar{n}_m$ is the number density of MNPs in the spherical shell, and $\overline{d_1}(R_n)$ and $\overline{d_1}(R_c)$ are the dose contributions per number density of emitting MNPs corresponding to spheres of radius $R_n$ and $R_c$, respectively (given by Eq. (18)).

The average contribution of a single MNP to the mean square of the dose can be calculated in the same manner as the last term in Eq. (A.9) by substituting $\overline{d_1^2}$ for $\overline{d_1}$.

However, for the mixed terms in Eq. (24) pertaining to the synergistic contribution from pairs of MNPs to the mean square of the dose, it is not possible to simply subtract the weighting functions for the smaller sphere from those of the larger one. This is because the contribution from two MNPs within the larger sphere is the sum of the cases that (a) both MNPs are in the spherical shell, (b) both are in the smaller shell, and that (c) one of them is in the spherical shell and the other in the inner sphere.

The last contribution is given by the product of the mean doses per number density for one MNP in the spherical shell and the other in the inner sphere. This leads to Eq. (A.10), which can be rewritten as Eq. (A.11):

$$\overline{D_{NN'}^2} = \bar{n}_m{}^2 \times \left[\overline{d_2^2}(R_c) - 8V_p \times \overline{d_1^2}(R_c)\right] - \bar{n}_m{}^2 \times \left[\overline{d_2^2}(R_n) - 8V_p \times \overline{d_1^2}(R_n)\right] \\ -2\bar{n}_m\overline{d_1}(R_n) \times \bar{n}_m\left[\overline{d_1}(R_c) - \overline{d_1}(R_n)\right] \tag{A.10}$$

$$\overline{D_{NN'}^2}(R_n, R_c) = \bar{n}_m{}^2\left(\overline{d_2^2}(R_c) - \overline{d_2^2}(R_n) - 2\overline{d_1}(R_n)\left[\overline{d_1}(R_c) - \overline{d_1}(R_n)\right]\right) \\ - \bar{n}_m{}^2 8V_p\left(\overline{d_1^2}(R_c) - \overline{d_1^2}(R_n)\right) \tag{A.11}$$

### *Appendix 4 MNP distribution with different constant values in a sphere and a surrounding spherical shell*

The case that MNPs are uniformly distributed in a sphere of radius $R_n$ at a number density $\bar{n}_n$ and in a surrounding spherical shell at a different number density $\bar{n}_p$ could, for instance, occur if there is complete uptake of the MNPs into a cell, while the uptake into the nucleus is reduced compared to the cytoplasm. The case is also treated here as a preliminary to the generalization to a continuous variation of the number density with radial distance from the center of the cell nucleus (see Appendix 6 and Appendix 7). For the case of the contribution of MNPs to the mean dose, the resulting expression is simply the sum of Eq. (18) and the second term on the right-hand side of Eq. (A.9), as shown in Eq. (A.12):

$$\bar{D} = D_w + \bar{n}_n\overline{d_1}(R_n) + \bar{n}_p\left[\overline{d_1}(R_c) - \overline{d_1}(R_n)\right] \tag{A.12}$$

The contribution to the mean square of the dose from the same MNP is similarly obtained by Eq. (A.16), with the $\overline{d_1^2}$ given by Eq. (22). This results in Eq. (A.23):

$$\overline{D_N^2} = \bar{n}_n \times \overline{d_1^2}(R_n) + \bar{n}_p \times \left[\overline{d_1^2}(R_c) - \overline{d_1^2}(R_n)\right] \tag{A.13}$$

For the synergistic term originating from pairs of MNPs, the situation is a bit more complex. In addition to the contributions of the sphere (Eq. (24)) and the spherical shell (Eq. (A.10)), there is a contribution due to pairs of MNPs of which one is located in the sphere and the other in the spherical shell. This leads to Eq. (A.14), where $V_n$ and $V_c$ denote the volumes of the spheres of radius $R_n$ and $R_c$, respectively:

$$\overline{D_{NN'}^2} = \bar{n}_n{}^2\overline{d_2^2}(R_n) + \bar{n}_p{}^2\left(\overline{d_2^2}(R_c) - \overline{d_2^2}(R_n) - 2\,\overline{d_1}(R_n)\left[\overline{d_1}(R_c) - \overline{d_1}(R_n)\right]\right) \\ - 8V_p\left(\bar{n}_n{}^2\overline{d_1^2}(R_n) + \bar{n}_p{}^2\left[\overline{d_1^2}(R_c) - \overline{d_1^2}(R_n)\right]\right) \\ + 2\bar{n}_n\bar{n}_p\overline{d_1}(R_n)\left[\overline{d_1}(R_c) - \overline{d_1}(R_n)\right] \tag{A.14}$$

The last term in Eq. (A.14) is the synergistic term for MNP pairs in which one MNP is in the sphere and the other in the spherical shell.

*Appendix 5 MNP distribution with different constant values in a sphere and two concentric spherical shells*

In the case of different MNP concentrations $\bar{n}_n$, $\bar{n}_p$, and $\bar{n}_x$ in the nucleus, cytoplasm, and extracellular region, respectively, the expression for the mean dose contribution from MNPs is given by Eq. (A.15):

$$\overline{D_N} = \bar{n}_n \overline{d_1}(R_n) + \bar{n}_p\left[\overline{d_1}(R_c) - \overline{d_1}(R_n)\right] + \bar{n}_x\left[\overline{d_1}(R_x) - \overline{d_1}(R_c)\right] \tag{A.15}$$

The contribution to the mean square of the dose from the same MNP is similarly obtained by Eq. (A.16), with $\overline{d_1^2}$ given by Eq. (22):

$$\overline{D_N^2} = \bar{n}_n \overline{d_1^2}(R_n) + \bar{n}_p\left[\overline{d_1^2}(R_c) - \overline{d_1^2}(R_n)\right] + \bar{n}_x\left[\overline{d_1^2}(R_x) - \overline{d_1^2}(R_c)\right] \tag{A.16}$$

The synergistic term takes the rather complex form of Eq. (A.17), where $V_n$, $V_c$, and $V_x$ denote the volumes of the nucleus, cell, and extracellular region, respectively:

$$\begin{aligned}\overline{D_{NN'}^2} = {} & \bar{n}_n{}^2\overline{d_2^2}(R_n) + \bar{n}_p{}^2\left(\overline{d_2^2}(R_c) - \overline{d_2^2}(R_n) - 2\,\overline{d_1}(R_n)\left[\overline{d_1}(R_c) - \overline{d_1}(R_n)\right]\right) \\ & + \bar{n}_x{}^2\left(\overline{d_2^2}(R_x) - \overline{d_2^2}(R_c) - 2\,\overline{d_1}(R_c)\left[\overline{d_1}(R_x) - \overline{d_1}(R_c)\right]\right) \\ & - 8V_p\left(\bar{n}_n{}^2\overline{d_1^2}(R_n) + \bar{n}_p{}^2\left[\overline{d_1^2}(R_c) - \overline{d_1^2}(R_n)\right]\right) \\ & - 8V_p\bar{n}_x{}^2\left[\overline{d_1^2}(R_x) - d_1^2(R_c)\right] + 2\bar{n}_n\,\bar{n}_p\,\overline{d_1}(R_n)\left[\overline{d_1}(R_c) - \overline{d_1}(R_n)\right] \\ & + 2\bar{n}_n\bar{n}_x\overline{d_1}(R_n)\left[\overline{d_1}(R_x) - \overline{d_1}(R_c)\right] \\ & + 2\bar{n}_p\,\bar{n}_x\left[\overline{d_1}(R_c) - \overline{d_1}(R_n)\right] \times \left[\overline{d_1}(R_x) - \overline{d_1}(R_c)\right]\end{aligned} \tag{A.17}$$

By normalizing the three number densities appearing in Eqs. (A.15), (A.16), and (A.17) to the average number density $\bar{n}_a$ in the sphere of radius $R_x$, the relative concentrations $u_n$, $u_p$, and $u_x$ are obtained (Eq. (A.18)):

$$u_n = \frac{\bar{n}_n}{\bar{n}_a} \;\; ; \;\; u_p = \frac{\bar{n}_p}{\bar{n}_a} \;; \;\; u_x = \frac{\bar{n}_x}{\bar{n}_a} \tag{A.18}$$

Together with the radii $R_n$, $,R_c$, and $R_x$, these three parameters characterize the uptake scenario. The set $\theta = \{R_n, R_c, R_x, u_n, u_p, u_x\}$ parameterizes new weighting functions $w_1$, $w_2$, and $w_{1\mathrm{b}}$, as given by Eqs. (A.19), (A.20), and (A.21), respectively:

$$w_1(r_i|\theta) = (u_n - u_p)\varpi(r_i|R_n) + (u_p - u_x)\varpi(r_i|R_c) + u_x\varpi(r_i|R_x) \tag{A.19}$$

$$\begin{aligned}w_2(r_i, r_j|\theta) = {} & [u_n{}^2 - u_p{}^2]\bar{\omega}_2(r_i, r_j|R_n) + [u_p{}^2 - u_x{}^2]\bar{\omega}_2(r_i, r_j|R_c) \\ & + u_x{}^2\bar{\omega}_2(r_i, r_j|R_x) \\ & + 2(u_n - u_p)u_p\varpi(r_i|R_n)\left(\varpi(r_j|R_c) - \varpi(r_j|R_n)\right) \\ & + 2(u_p - u_x)u_x\varpi(r_i|R_c)\left(\varpi(r_j|R_x) - \varpi(r_j|R_c)\right) \\ & + 2(u_n - u_p)u_x\varpi(r_i|R_n)\left(\varpi(r_j|R_x) - \varpi(r_j|R_c)\right)\end{aligned} \tag{A.20}$$

$$w_{1b}(r_i|\theta) = [u_n{}^2 - u_p{}^2]\bar{\omega}_1(r_i|R_n) + [u_p{}^2 - u_x{}^2]\bar{\omega}_1(r_i|R_c) + u_x{}^2\bar{\omega}_1(r_i|R_x) \tag{A.21}$$

With these weighting functions, the contributions to the mean dose and mean dose squared specific to the uptake scenario are obtained by Eqs. (A.22), (A.23), and (A.24):

$$\overline{D_N} = \bar{n}_a\overline{d_1}(\theta) = \bar{n}_a\int_{r_p}^{\infty} D_1(r_i)r_i{}^2 \times 4\pi w_1(r_i|\theta)dr_i \tag{A.22}$$

$$\overline{D_N^2} = \bar{n}_a \overline{d_1^2}(\theta) = \bar{n}_a \int_{r_p}^{\infty} [D_1(r_i)]^2 {r_i}^2 \times 4\pi w_1(r_i|\theta) dr_i \tag{A.23}$$

$$\overline{D_{NN'}^2} = {\bar{n}_a}^2 \times \left(\overline{d_2^2}(\theta) - 8V_p \overline{d_{1b}^2}(\theta)\right) \tag{A.24}$$

The synergistic term of MNP pairs contributing to the mean square of the dose in the nucleus is given by Eq. (A.25), and the correction for nonoverlapping of the two MNPs by Eq. (A.26):

$$\overline{d_2^2}(\theta) = \int_{r_p}^{\infty} \int_{r_p}^{\infty} D_1(r_i){r_i}^2 D_1(r_j){r_j}^2 \times (4\pi)^2 w_2(r_i, r_j|\theta)\, dr_j\, dr_i \tag{A.25}$$

$$\overline{d_{1b}^2}(\theta) = \int_{r_p}^{\infty} \int_{r_p}^{\infty} [D_1(r_i)]^2 {r_i}^2 \times (4\pi)^2 w_{1\mathrm{b}}(r_i|\theta)\, dr_i \tag{A.26}$$

*Appendix 6 Generalized weighting function for the contribution to the mean dose*

From Eq. (A.15), it is evident that considering further radial shells would simply add additional terms of the same structure as the last two. If the separation between subsequent radii is an infinitesimal increment and their number increases to infinity, then the sum becomes an integral over the radial density of MNPs, as in Eq. (A.27):

$$\bar{D} = D_w + \bar{n}_n \overline{d_1}(R_n) + \int_{R_n}^{R_x} \bar{n}_m(R)\, \overline{d_1}'(R) dR \tag{A.27}$$

In Eq. (A.27), $\bar{n}_m(R)$ is the number density of emitting MNPs as a function of the radial distance from the center of the cell nucleus, and $\overline{d_1}'(R)$ is the derivative of $\overline{d_1}(R)$ with respect to $R$. From Eq. (19), this derivative is given by Eq. (A.28):

$$\overline{d_1}'(R) = \int_{r_p}^{\infty} D_1(r_i){r_i}^2 \times 4\pi \frac{\partial \bar{\omega}(r_i|R)}{\partial R} dr_i \tag{A.28}$$

The partial derivative of $\bar{\omega}(r_i|R)$ with respect to $R$ is given by Eqs. (A.29) and (A.30), where the reduced quantities $x_i = r_i/R_n$ and $a = R/R_n$ have been used:

$$\frac{\partial \bar{\omega}(r_i|R)}{\partial R} = \begin{cases} \frac{1}{R_n} \frac{\partial \bar{\omega}_1(x_i|a)}{\partial a} & a - 1 \le x_i \le a + 1 \\ 0 & \text{else} \end{cases} \tag{A.29}$$

$$\frac{\partial \bar{\omega}_1(x_i|a)}{\partial a} = \frac{3}{2}a^2 - \frac{3}{4}a[x_i - (1 - a^2){x_i}^{-1}] \tag{A.30}$$

The sequence of integration can be interchanged in the double integral obtained by substituting Eq. (A.28) into Eq. (A.27), so that Eq. (A.31) is obtained with $\tilde{w}_1(r_i)$ defined by Eq. (A.32):

$$\overline{D} = D_w + \bar{n}_n \overline{d_1}(R_n) + \bar{n}_r \int_{r_p}^{\infty} D_1(r_i) {r_i}^2 \times 4\pi \widetilde{w}_1(r_i) dr_i \tag{A.31}$$

$$\widetilde{w}_1(r_i) = \int_{R_n}^{R_x} \frac{\bar{n}_m(R)}{\bar{n}_r} \frac{\partial \varpi(r_i|R)}{\partial R} dR \tag{A.32}$$

The factor $\bar{n}_r$ in front of the integral in Eq. (A.31) is the mean number density of emitting MNPs inside a chosen reference volume $V_r$ (e.g., the cell) and was introduced for convenience to make the weighting function $w_1(r_i)$ dimensionless. The first fraction under the integral in Eq. (A.32) gives the relative profile of the MNP concentration.

*Appendix 7 Generalized weighting function for the contribution to the square of the dose*

As before, the same weighting function applies to the contribution to the square of the dose from single MNPs. For the synergistic term due to pairs of MNPs, the starting point is to consider Eq. (A.17) for the case that $R_x = R_c + dR$ is only infinitesimally larger than $R_c$. This leads to the following expression for the derivative of $\overline{D_{NN'}^2}$ (with $R_c$ replaced with $R$):

$$\begin{aligned} \overline{D_{NN'}^2}'(R) &= \left(\bar{n}_m(R)\right)^2 \times \left[\overline{d_2^2}'(R) - 8V_p \overline{d_1^2}'(R)\right] - 2\left(\bar{n}_m(R)\right)^2 \overline{d_1}(R)\, \overline{d_1}'(R) \\ &\quad + 2\bar{n}_m(R)\ \overline{d_1}'(R) \left[\bar{n}_n \overline{d_1}(R_n) + \int_{R_n}^{R} \bar{n}_m(R')\ \overline{d_1}'(R')\, dR'\right] \end{aligned} \tag{A.33}$$

Integration by parts of the integral in the last term transforms Eq. (A.33) into Eq. (A.34):

$$\begin{aligned} \overline{D_{NN'}^2}'(R) &= \left(\bar{n}_m(R)\right)^2 \times \left[\overline{d_2^2}'(R) - 8V_p \overline{d_1^2}'(R)\right] \\ &\quad + 2\bar{n}_m(R)\, \overline{d_1}'(R) \times [\bar{n}_n - \bar{n}_m(R_n)] \overline{d_1}(R_n) \\ &\quad - 2\bar{n}_m(R)\, \overline{d_1}'(R) \int_{R_n}^{R} \frac{d\bar{n}_m(R')}{dR} \overline{d_1}(R')\, dR' \end{aligned} \tag{A.34}$$

Using the same line of argument as in the previous section, Eq. (A.35) is obtained with the quantities $\widetilde{w}_2(r_i, r_j)$ and $\widetilde{w}_2(r_i, r_j)$ given by Eqs. (A.36) and (A.37):

$$\begin{aligned} \overline{D_{NN'}^2} &= \overline{D_{NN'}^2}(R_n) + {\bar{n}_r}^2 \int_{r_p}^{\infty} \int_{r_p}^{\infty} D_1(r_i) r_i^2 D_1(r_j) r_j^2 \times (4\pi)^2 \widetilde{w}_2(r_i, r_j)\, dr_j\, dr_i \\ &\quad - {\bar{n}_r}^2 8V_p \int_{r_p}^{\infty} [D_1(r_i)]^2 r_i^2 \times 4\pi \widetilde{w}_3(r_i)\, dr_i \\ &\quad + 2[\bar{n}_n - \bar{n}_m(R_n)] \overline{d_1}(R_n) \times \bar{n}_r \int_{r_p}^{\infty} D_1(r_i) {r_i}^2 \times 4\pi \widetilde{w}_1(r_i) dr_i \end{aligned} \tag{A.35}$$

$$\begin{aligned} \widetilde{w}_2(r_i, r_j) &= \int_{R_n}^{R_x} \left(\frac{\bar{n}_m(R)}{\bar{n}_r}\right)^2 \frac{\partial \varpi_2(r_i, r_j|R)}{\partial R} dR \\ &\quad + \int_{R_n}^{R_x} \frac{\bar{n}_m(R)}{\bar{n}_r} \frac{\partial \varpi(r_i|R)}{\partial R} \int_{R_n}^{R} \frac{1}{\bar{n}_r} \frac{d\bar{n}_m(R')}{dR'}\ \varpi(r_j|R') dR' dR \end{aligned} \tag{A.36}$$

$$\widetilde{w}_3(r_i) = \int_{R_n}^{R_x} \left(\frac{\bar{n}_m(R)}{\bar{n}_r}\right)^2 \frac{\partial \bar{\omega}_1(r_i|R)}{\partial R} dR \tag{A.37}$$

## 6. Supplementary Figures and Tables

Supplementary Table 1: Glossary of symbols used.

| Symbol | Meaning |
|---|---|
| $\overline{D}$ | Mean dose in the cell nucleus |
| $\overline{D^2}$ | Mean square dose in the cell nucleus |
| $\bar{n}_\#$ | Mean number of emitting MNPs in the considered large sphere ($\bar{n}_a$), nucleus ($\bar{n}_n$), cytoplasm ($\bar{n}_p$), cell ($\bar{n}_c$), and extracellular medium ($\bar{n}_x$) |

### *6.1 Weighting functions for radial dose dependence*

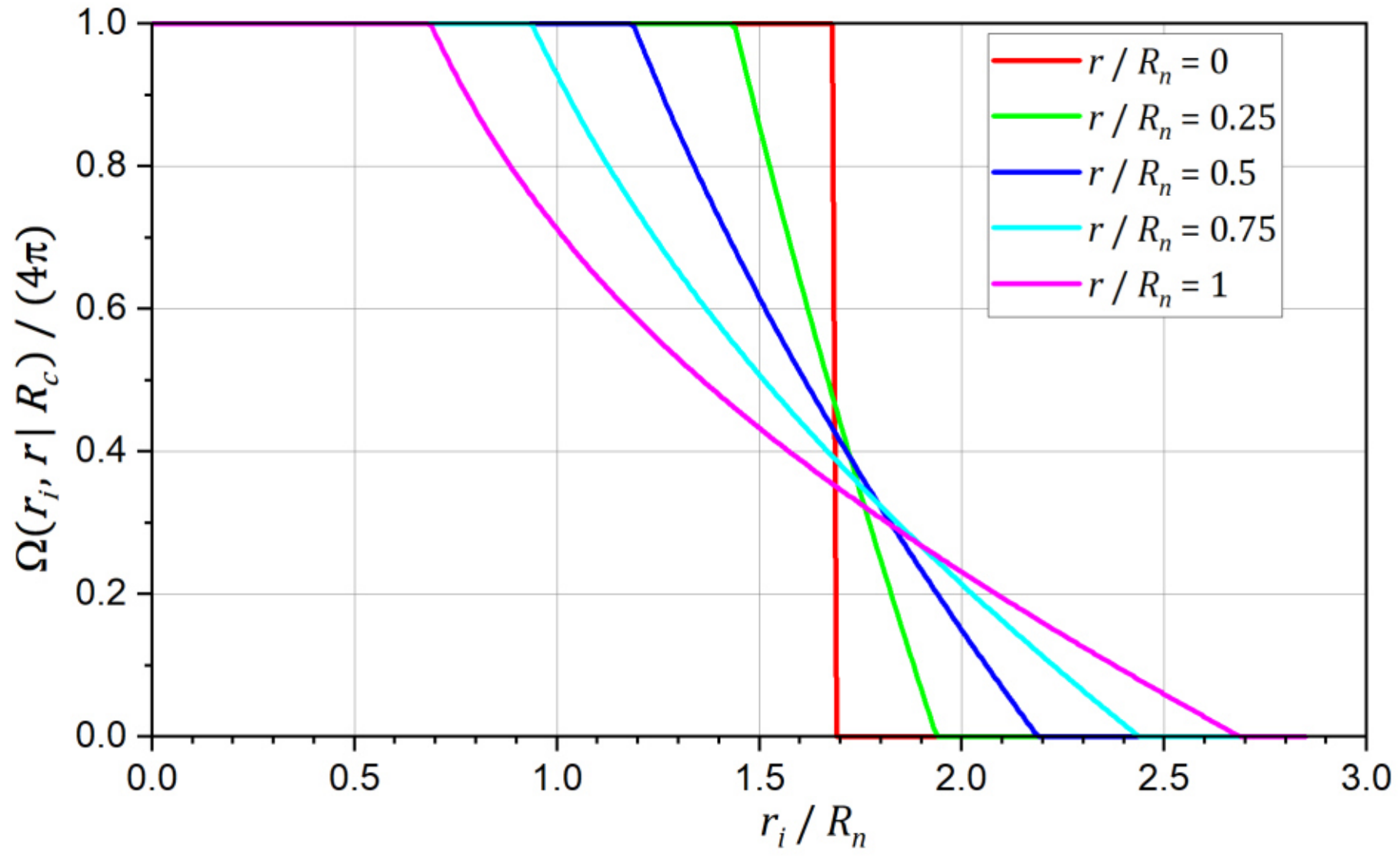

Supplementary Fig. 1: Weighting function (Eq. (17)) for calculating the mean contribution per emitting MNP to the dose at a given radial distance $r$ from the center of the cell nucleus (radius $R_n$) from the radial dose distribution around a single MNP. The MNPs are uniformly distributed in a sphere of radius $R_c$ which is concentric to the nucleus. The different curves correspond to different ratios of $r/R_n$. The data pertain to a ratio $R_c/R_n$ = 1.75, close to the value in the cell geometry employed by (Lin *et al* 2015).

*6.2 Illustration of the cases considered*

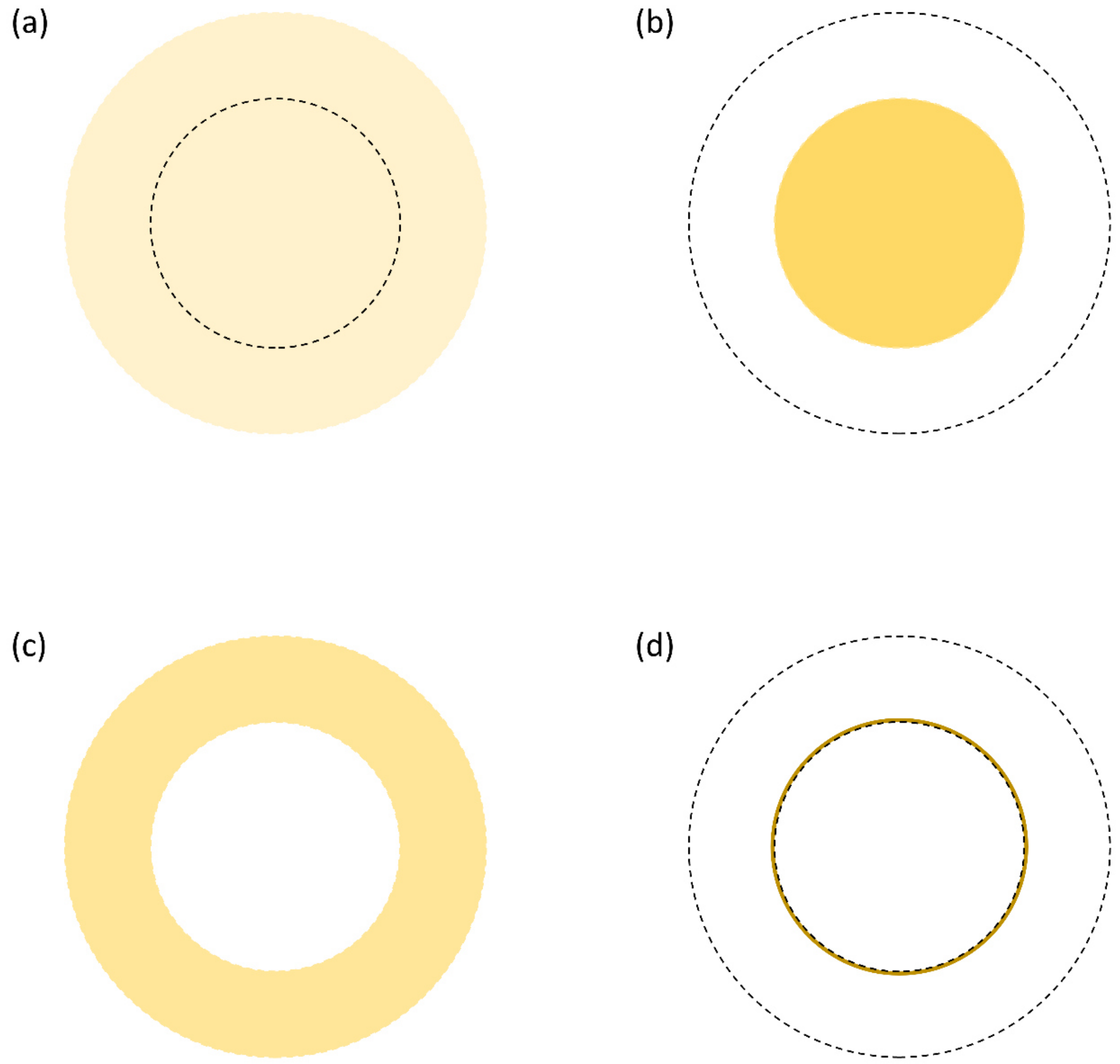

Supplementary Fig. 2: Illustration of the MNP distributions in scenarios 1 to 3 and 7 for an isolated cell. MNPs are uniformly distributed (a) within the whole cell, (b) in the cell nucleus only, (c) in the cytoplasm only, and (d) in a thin layer of the cytoplasm at the nucleus surface only.

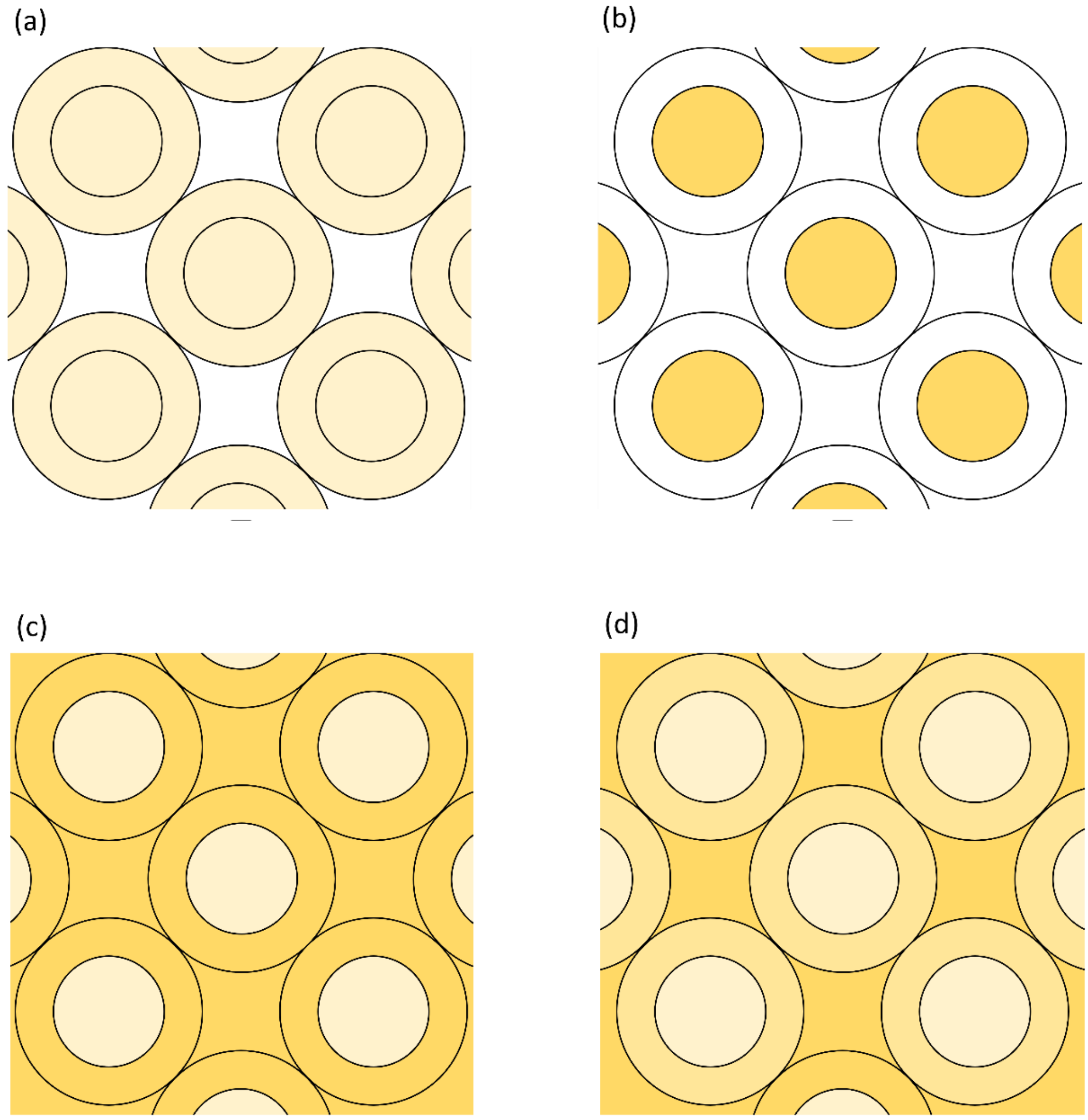

Supplementary Fig. 3: Illustration of the MNP distributions in scenarios 1, 2, 5, and 6 (see Table 1 in the main text) for the case of densely packed cells in tissue (see Table 2). The cells are arranged on a regular face-centered cubic lattice. MNPs are uniformly distributed (a) within the whole cells; (b) in the cell nuclei only; (c) everywhere, with reduced concentration in the nucleus; and (d) everywhere, with reduced concentration in the cell and even lower concentration in the cell nuclei.

## 6.3 Weighting functions for the mean dose of cells in tissue

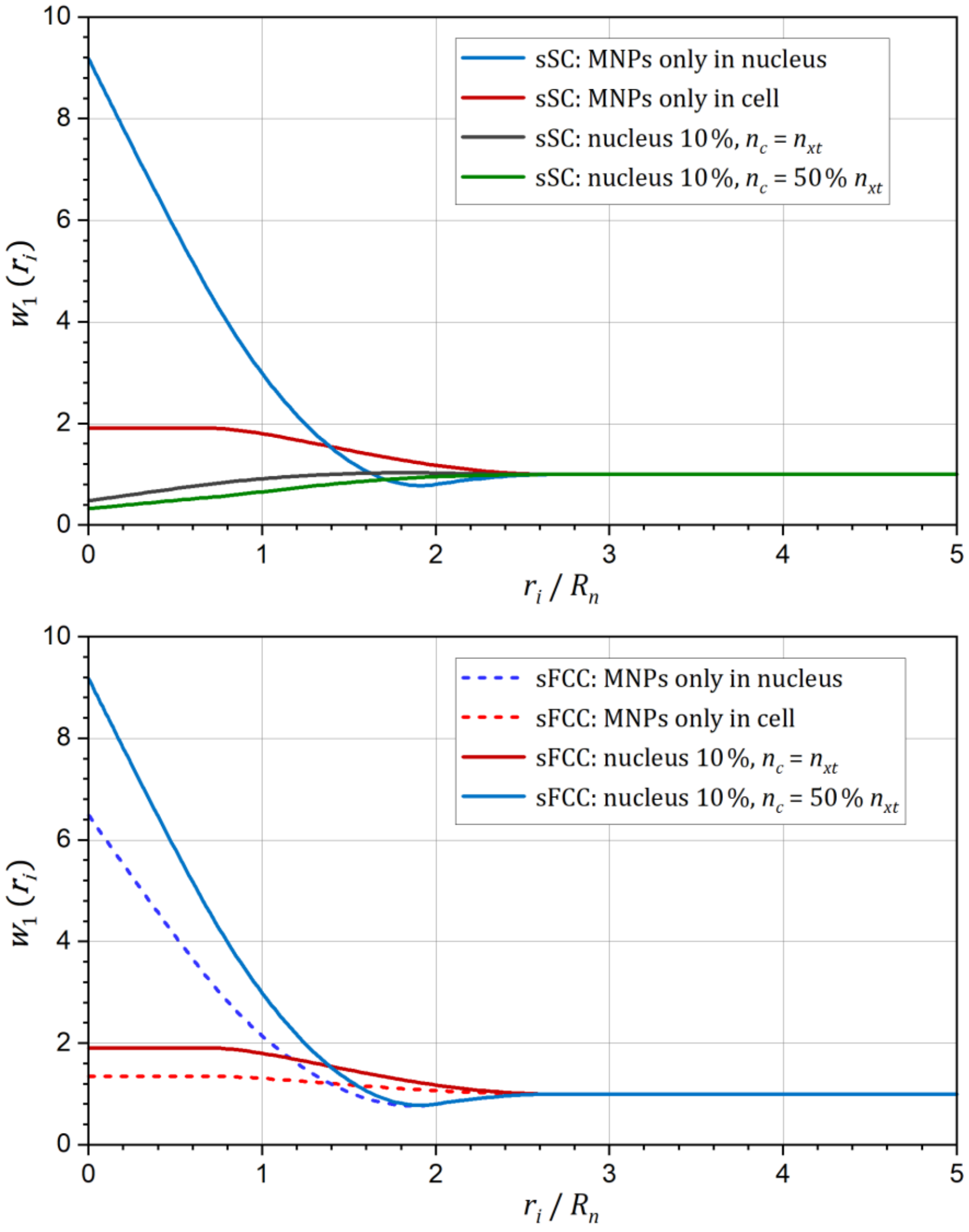

Supplementary Fig. 4: Top: Weighting function (Eq. (A.19)) for calculating the mean contribution per emitting MNP to the dose in the spherical cell nucleus (radius $R_n$) from the radial dose distribution around a single MNP for the case of the surrogate MNP distribution for cells that are densely packed in an SC lattice. The different curves apply to the following situations: MNPs only in the nuclei (red); MNPs only in the cells (orange); same concentration of MNPs in the cells and extracellular medium, but only 10% of the MNPs internalized in cells are inside the nucleus (black); the concentration in the extracellular medium is twice the average concentration inside the cells, and only 10% of the internalized MNPs are in the nucleus. The ratio of the radii of the spherical cells to those of the concentric spherical nuclei corresponds to the cell geometry employed by (Lin *et al* 2015). Bottom: Comparison of the results for FCC and SC packing of cells for the cases of complete uptake into the cell.

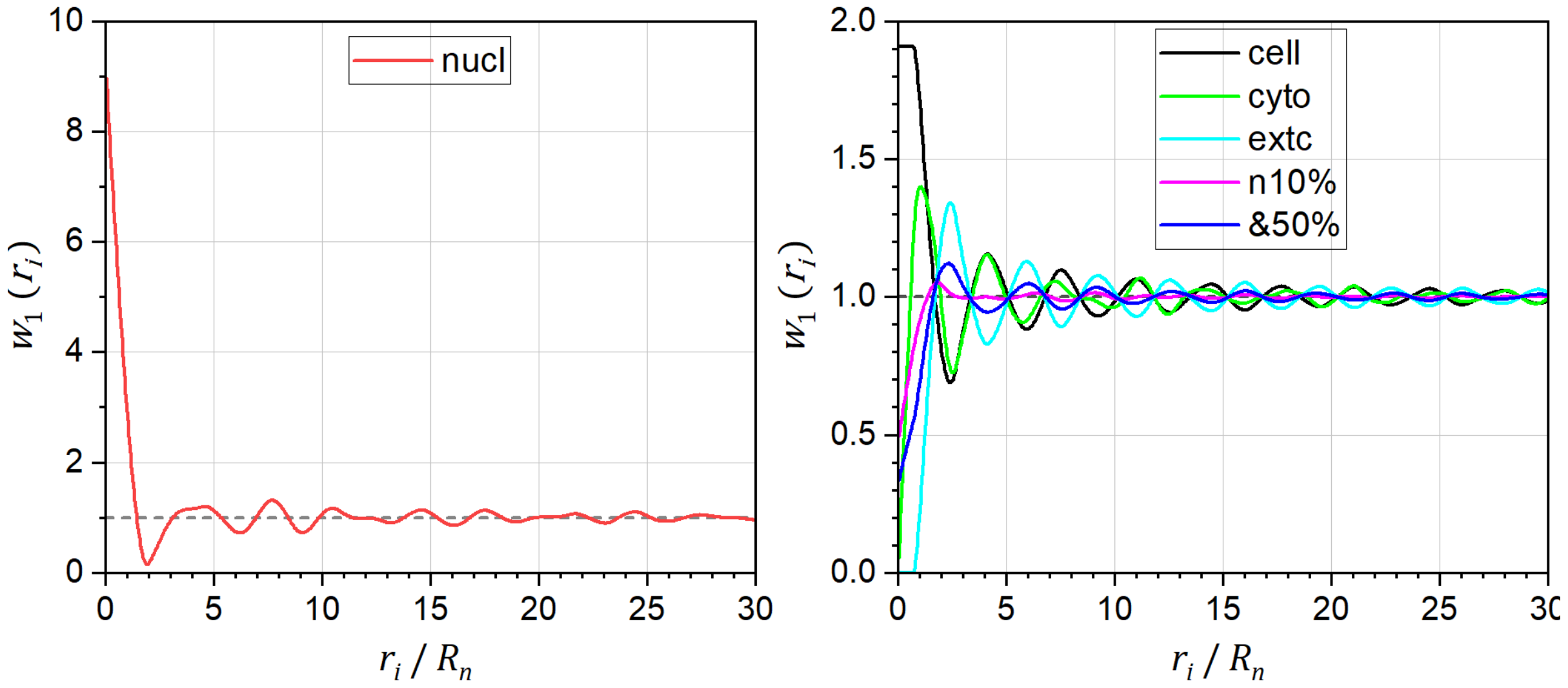

Supplementary Fig. 5: Weighting function for calculating the contribution per emitting MNP to the mean dose in the spherical cell nucleus (radius $R_n$) from the radial dose distribution around a single MNP for cells packed in an SC lattice. The curves apply to the first six uptake scenarios in Table 1. (a) MNPs only in the nuclei; (b) MNPs uniformly distributed in the cells (black); MNPs uniformly distributed in the cytoplasm (green); MNPs only in extracellular medium (cyan); same concentration of MNPs in the cells and extracellular medium, but only 10% of the MNPs internalized in cells are inside the nucleus (violet); the concentration in the extracellular medium is twice the average concentration inside the cells, and only 10% of the internalized MNPs are in the nucleus (blue). The ratio of the radii of the spherical cells to those of the concentric spherical nuclei corresponds to the cell geometry employed by (Lin *et al* 2015).

### *6.4 Sample radial dose distributions around an MNP*

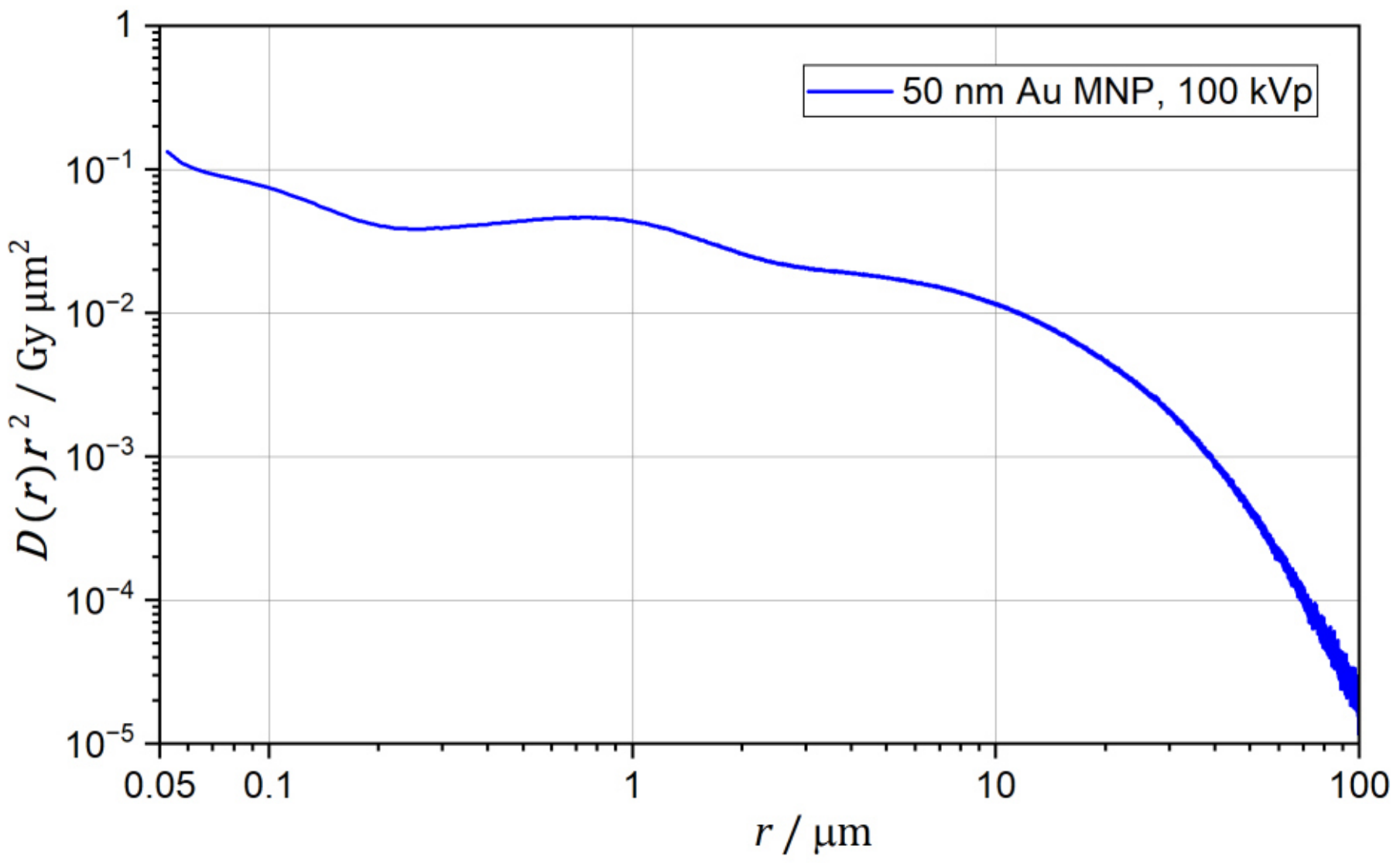

Supplementary Fig. 6: Product of the square of the radius and the radial dose distribution around a single gold MNP of 50 nm radius irradiated at 100 µm depth in water by the mixed photon and electron field resulting from a primary 100 kVp X-ray photon spectrum (from (Thomas *et al* 2024)). This is the quantity to which the weighting functions are applied.

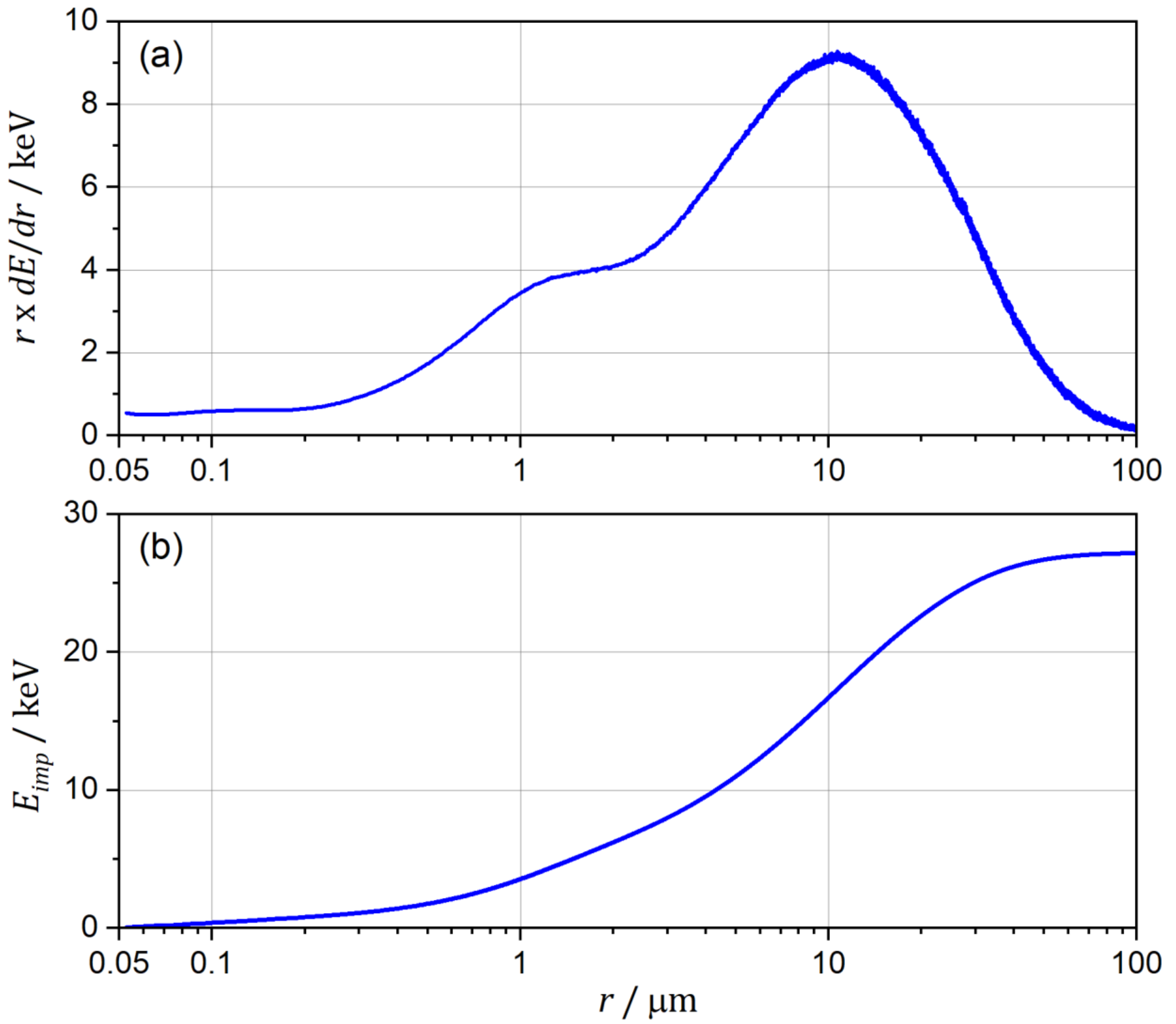

Supplementary Fig. 7: The same data as shown in Supplementary Fig. 6 but now converted to energy imparted per length (left) or cumulative energy imparted within a sphere of radius $r$ (right-hand side). The left-hand side shows the frequency density in a logarithmic radial interval (hence the prefactor $r$). In this presentation, the area under the curve is proportional to the contribution of a radial range to the total integral. The data apply to the radial distribution of energy imparted around a single gold MNP of 50 nm radius irradiated at 100 µm depth in water by the mixed photon and electron field resulting from a primary 100 kVp X-ray photon spectrum (from (Thomas *et al* 2024)).

**Supplement 1: Mathematical derivation of the contributions from a uniform nanoparticle distribution to the dose and its square in a spherical nucleus.**

The derivation is based on the following assumptions:

1. The nanoparticles (NPs) are spheres of radius $r_p$.
2. The cell nucleus is spherical (radius $R_n$).
3. The NPs are uniformly distributed in a sphere of radius $R_c$ which is concentric to the cell nucleus.
4. For a given arrangement of the nanoparticles at fixed points $\vec{r}_i$ ($i$ = 1, 2, .., $N$), the dose at any point $\vec{r}$ in the cell nucleus is given by

$$D(\vec{r}) = D_0 + \sum_{i=1}^{N} D_1(|\vec{r} - \vec{r}_i|) \tag{1}$$

where $D_0$ is the dose from secondary electrons produced in interactions of the incident radiation in the volume outside the NPs. ($D_0$ is approximately the dose in the absence of NPs.)

The average dose in the cell nucleus and the average square of the dose in the cell nucleus are then given by Eqs. (2) and (3), respectively.

$$\overline{D} = \frac{1}{V_n}\int_{V_n} D(\vec{r})dV = D_0 + \frac{1}{V_n}\int_{V_n} \sum_{i=1}^{N} D_1(|\vec{r} - \vec{r}_i|)\, dV_n \equiv D_0 + \overline{D_{1N}}\ , \tag{2}$$

$$\overline{D^2} = {D_0}^2 + 2D_0\frac{1}{V_n}\int_{V_n} \sum_{i=1}^{N} D_1(|\vec{r} - \vec{r}_i|)\, dV_n + \frac{1}{V_n}\int_{V_n} \sum_{i=1}^{N}\sum_{j=1}^{N} D_1(|\vec{r} - \vec{r}_i|)D_1(|\vec{r} - \vec{r}_j|)\, dV_n \tag{3}$$

Using Eq. (2) and separating the last term into contributions from the same NP and from pairs of NPs transforms Eq. (3) into Eq. (4):

$$\overline{D^2} = {D_0}^2 + 2D_0\overline{D_N} + \frac{1}{V_n}\int_{V_n} \sum_{i=1}^{N}[D_1(|\vec{r} - \vec{r}_i|)]^2\, dV_n + \frac{1}{V_n}\int_{V_n} \sum_{i=1}^{N}\sum_{j\neq i} D_1(|\vec{r} - \vec{r}_i|)D_1(|\vec{r} - \vec{r}_j|)\, dV_n \tag{4}$$

The third and fourth terms on the right-hand side of Eq. (4) are designated $\overline{D_N^2}$ and $\overline{D_{NN'}^2}$.

$$\overline{D_N^2} = \frac{1}{V_n}\int_{V_n} \sum_{i=1}^{N}[D_1(|\vec{r} - \vec{r}_i|)]^2\, dV_n\,; \quad \overline{D_{NN'}^2} = \frac{1}{V_n}\int_{V_n} \sum_{i=1}^{N}\sum_{j\neq i} D_1(|\vec{r} - \vec{r}_i|)D_1(|\vec{r} - \vec{r}_j|)\, dV_n \tag{5}$$

## S1.1 Average dose from a uniform NP distribution

The quantity $\overline{D_N}$ is the sum of the contributions to the dose at a point $\vec{r}$ originating from NPs located at positions $\vec{r}_i$ ($i$ = 1, 2, ... $N$) as given by Eq. (6).

$$\overline{D_N} = \frac{1}{V_n}\int_{V_n} \sum_{i=1}^{N} D_1(|\vec{r} - \vec{r}_i|)\, dV_n \tag{6}$$

The integral is over the volume $V_n$ of the cell nucleus. The assumption of a uniform distribution of the NPs means that the number density of NPs is constant. Therefore, the number density $\bar{n}_m$ of NPs undergoing an interaction is given by the ratio of the number $N$ of nanoparticles undergoing an interaction to the volume $V_c$ containing NPs:

$$\bar{n}_m = \frac{N}{V_c} \tag{7}$$

Therefore, one can replace the sum in Eq. (6) with an integral.

$$\frac{1}{V_n}\int_{V_n}\sum_{i=1}^{N} D_1(|\vec{r}-\vec{r}_i|)\,dV_n = \frac{1}{V_n}\int_{V_n}\bar{n}_m\int_{V_c} D_1(|\vec{r}-\vec{r}_i|)dV_c\,dV_n \tag{8}$$

For the evaluation of Eq. (8), it is convenient to introduce spherical coordinates and to use the following trick: For the inner integral (over the NP positions) the coordinate system is chosen such that it is centered on the point $\vec{r}$ at which the dose is evaluated, i.e., the radial coordinate of the NP is equal to its distance from the considered point. Without loss of generality, this point can be chosen to be on the z-axis at a radial distance $r$ from the center of the spherical nucleus. (This is because the geometry has spherical symmetry.)

This transforms Eq.(8) into Eq. (9):

$$\overline{D_N} = \frac{4\pi}{V_n}\int_0^{R_n}\bar{n}_m\int_{r_p}^{R_c+R_n} D_1(r_i)\,\Omega(r_i,r|R_c)r_i^2dr_i\,r^2dr \tag{9}$$

In Eq. (9), $r_i^2\Omega(r_i,r)$ is the part of the surface of a sphere of radius $r_i$ around a point at distance $r$ from the center of the nucleus which falls inside the sphere in which NPs are present. The ratio of the part of the surface inside the sphere containing the NPs to the total surface of the considered sphere is given by

$$\Omega(r_i,r|R_c) = \begin{cases} 0 & r < r_i - R_c \\ 4\pi & r < R_c - r_i \\ \pi\dfrac{R_c^2-(r-r_i)^2}{r_i r} & \text{else} \end{cases} \tag{10}$$

The first of the three cases in Eq. (10) means that the distance $r_i$ from the considered point is so large that all points at this distance from a point inside the nucleus are outside the region containing NPs. The second case applies to all points in the nucleus for which the whole sphere of radius $r_i$ is within the region containing NPs.

It is important to note that despite the fact that the second coordinate system was chosen relative to a point in the first, the integration domains of the two integrals in Eq. (9) are independent so that the sequence of the integrals can be interchanged.

$$\overline{D_N} = \bar{n}_m\int_{r_p}^{R_c+R_n} D_1(r_i)\frac{4\pi}{V_n}\int_0^{R_n}\Omega(r_i,r|R_c)r^2dr\,r_i^2dr_i = \bar{n}_m\overline{d_1}(R_c) \tag{11}$$

$$\overline{d_1}(R_c) = \int_{r_p}^{R_c+R_n} D_1(r_i)\,r_i^2\ \times 4\pi\bar{\omega}(r_i|R_c)dr_i \tag{12}$$

This means that the mean dose contribution from NPs can be derived from the radial dependence of local dose around the NP using the weighting function $\bar{\omega}(r_i|R_c)$ given by Eq. (13):

$$\bar{\omega}(r_i|R_c) = \frac{3}{R_n^3}\int_0^{R_n}\frac{\Omega(r_i,r|R_c)}{4\pi}\,r^2dr \tag{13}$$

The calculation of the analytical expression for $\overline{\omega}(r_i)$ is shown in section S1.4. The result is reproduced here as Eq. (14):

$$\overline{\omega}(r_i) = \begin{cases} 1 & r_i \le R_c - R_n \\ \dfrac{1}{2}\left(1+\dfrac{R_c^3}{R_n^3}\right) - \dfrac{3}{8}\left(1+\dfrac{R_c^2}{R_n^2}\right)\dfrac{r_i}{R_n} + \dfrac{1}{16}\left(\dfrac{r_i}{R_n}\right)^3 - \dfrac{3R_n}{16r_i}\left(1-\dfrac{R_c^2}{R_n^2}\right)^2 & R_c - R_n \le r_i \le R_c + R_n \\ 0 & r_i > R_c + R_n \end{cases} \tag{14}$$

It should be noted that the weighting function $\overline{\omega}$ is a function of the ratio $r_i/R_n$ and depends parametrically on the ratio $R_c/R_n$.

## S1.2 Average square of the dose from the same NP

The contribution to the average square of the dose in the cell nucleus due to the square of the dose contribution from the same NP (first identity in Eq. (5) can be simply obtained by substituting $[D_1(|\vec{r}-\vec{r}_i|)]^2$ for $D_1(|\vec{r}-\vec{r}_i|)$ in Eqs. (6) and (8) and $[D_1(r_i)]^2$ for $D_1(r_i)$ in Eqs. (9) and (11). Therefore,

$$\overline{D_N^2} = N\overline{D_1^2} = \bar{n}_m\overline{d_1^2}; \qquad \overline{d_1^2} = \int_{r_p}^{R_c+R_n} [D_1(r_i)]^2 r_i^2 \times 4\pi\bar{\omega}(r_i|R_c)\, dr_i \tag{15}$$

with $w(r_i)$ given by Eq. (13).

## S1.3 Product of the dose contributions from two NPs from a uniform GNP distribution in a sphere

In analogy to Section S1.1, the sums in the expression for $\overline{D_{NN'}^2}$ in Eq. (5) can be replaced with integrals, and the integrals over the NP positions can be expressed in a spherical coordinate system centered at a point in the nucleus at which the doses are evaluated. Generally, one of the two NPs will be further away from the considered point. This is taken into account by assuming without loss of generality that $r_i \geq r_j$ and multiplying by a factor of 2 in Eq. (16):

$$\overline{D_{NN'}^2} = \frac{4\pi}{V_n}\int_0^{R_n} 2\bar{n}_m \int_{r_p}^{R_c+R_n} D_1(r_i)\Omega(r_i, r|R_c)\bar{n}_m \int_{r_p}^{r_i} D_1(r_j)\Omega^*(r_j, r|R_c, r_i)\, r_j^2 dr_j\; r_i^2 dr_i\; r^2 dr \tag{16}$$

The function $\Omega$ is the same as in Eq. (10). The function $\Omega^*$ is given by Eq. (17):

$$\Omega^*(r_j, r|R_c, r_i) = \Omega(r_j, r|R_c) - \Omega(r_i, r_j|2r_p) \tag{17}$$

The second term on the right-hand side of Eq. (17) takes into account that the two nanoparticles cannot overlap. Therefore the center of the second nanoparticle must have a distance from the center of the first of at least twice the radius of the nanoparticle.

Since the integration domains of the inner integrals do not depend on $r$, the outer and the two inner integrals can be interchanged, which leads to Eq. (18):

$$\overline{D_{NN'}^2} = 2\bar{n}_m{}^2 \int_{r_p}^{R_c+R_n} D_1(r_i) \int_{r_p}^{r_i} D_1(r_j) \frac{4\pi}{V_n}\int_0^{R_n} \Omega(r_i, r|R_c)\Omega^*(r_j, r|R_c, r_i)\; r^2 dr\, r_j^2 dr_j\; r_i^2 dr_i \tag{18}$$

Introducing the bi-variate weighting functions $\bar{\omega}_2$ and $\bar{\omega}_2^*$ by Eqs. (19) and (20), respectively

$$\bar{\omega}_2(r_i, r_j|R_c) = \frac{3}{R_n{}^3}\int_0^{R_n} \frac{\Omega(r_i, r|R_c)}{4\pi}\frac{\Omega(r_j, r|R_c)}{4\pi}\; r^2 dr \tag{19}$$

$$\bar{\omega}_2^*(r_i, r_j|R_c) = \frac{3}{R_n{}^3}\int_0^{R_n} \frac{\Omega(r_i, r|R_c)}{4\pi}\; r^2 dr \times \frac{\Omega(r_i, r_j|2r_p)}{4\pi} \quad , \tag{20}$$

Eq. (18) can be rewritten as $\overline{D_{NN'}^2} = \bar{n}_m{}^2 \times \left[\overline{d_2^2} - \overline{d_2^{2*}}\right]$, where $\overline{d_2^2}$ and $\overline{d_2^{2*}}$ are defined as follows:

$$\overline{d_2^2} = 2\int_{r_p}^{R_c+R_n}\int_{r_p}^{r_i} D_1(r_i)r_i^2 D_1(r_j)r_j^2 \times (4\pi)^2\bar{\omega}_2(r_i, r_j|R_c)\, dr_j\, dr_i \tag{21}$$

$$\overline{d_2^{2*}} = 2\int_{r_p}^{R_c+R_n}\int_{r_i-2r_p}^{r_i} D_1(r_i)r_i^2 D_1(r_j)r_j^2 \times (4\pi)^2\bar{\omega}_2^*(r_i, r_j|R_c)\, dr_j\, dr_i \tag{22}$$

Since in Eq. (22) $r_i$ and $r_j$ are close to each other, one may approximate $D_1(r_j)$ by $D_1(r_i)$ and perform the integral over $r_j$ only for $\bar{\omega}_2^*$. The results can be shown to be given by Eq. (23):

$$2\int_{r_i-2r_p}^{r_i} r_j^2\bar{\omega}_2^*(r_i, r_j|R_c)\, dr_j = \frac{(2r_p)^3}{3} \times \bar{\omega}(r_i|R_c) \tag{23}$$

Therefore, $\overline{d_2^{2*}}$ is approximately given by Eq. (24), where $V_p$ denotes the volume of the NP:

$$\overline{d_2^{2*}} = \frac{4\pi}{3}(2r_p)^3 \int_{r_p}^{R_c+R_n} [D_1(r_i)]^2 r_i^2 \times 4\pi\bar{\omega}(r_i|R_c)\, dr_i = 8V_p \times \overline{d_1^2} \tag{24}$$

This gives the following final expression for $\overline{D_{NN'}^2}$:

$$\overline{D_{NN'}^2} = \bar{n}_m{}^2 \times \left[\overline{d_2^2} - 8V_p \times \overline{d_1^2}\right] \tag{25}$$

The analytical expressions for the bivariate weighting function $\varpi_2(r_i, r_j)$ are derived in section S1.5 and summarized in section S1.5.8. As with the weighting function $\varpi$, $\varpi_2$ is a function of the ratios $r_i/R_n$ and $r_j/R_n$, and depends parametrically on the ratio $R_c/R_n$. Graphical representations of $\varpi_2$ for some values of $R_c/R_n$ are shown in Fig. 1, Fig. 2, and Fig. 3.

### *S1.3.1 Sample graphical representations of the bivariate weighting function $\varpi_2$*

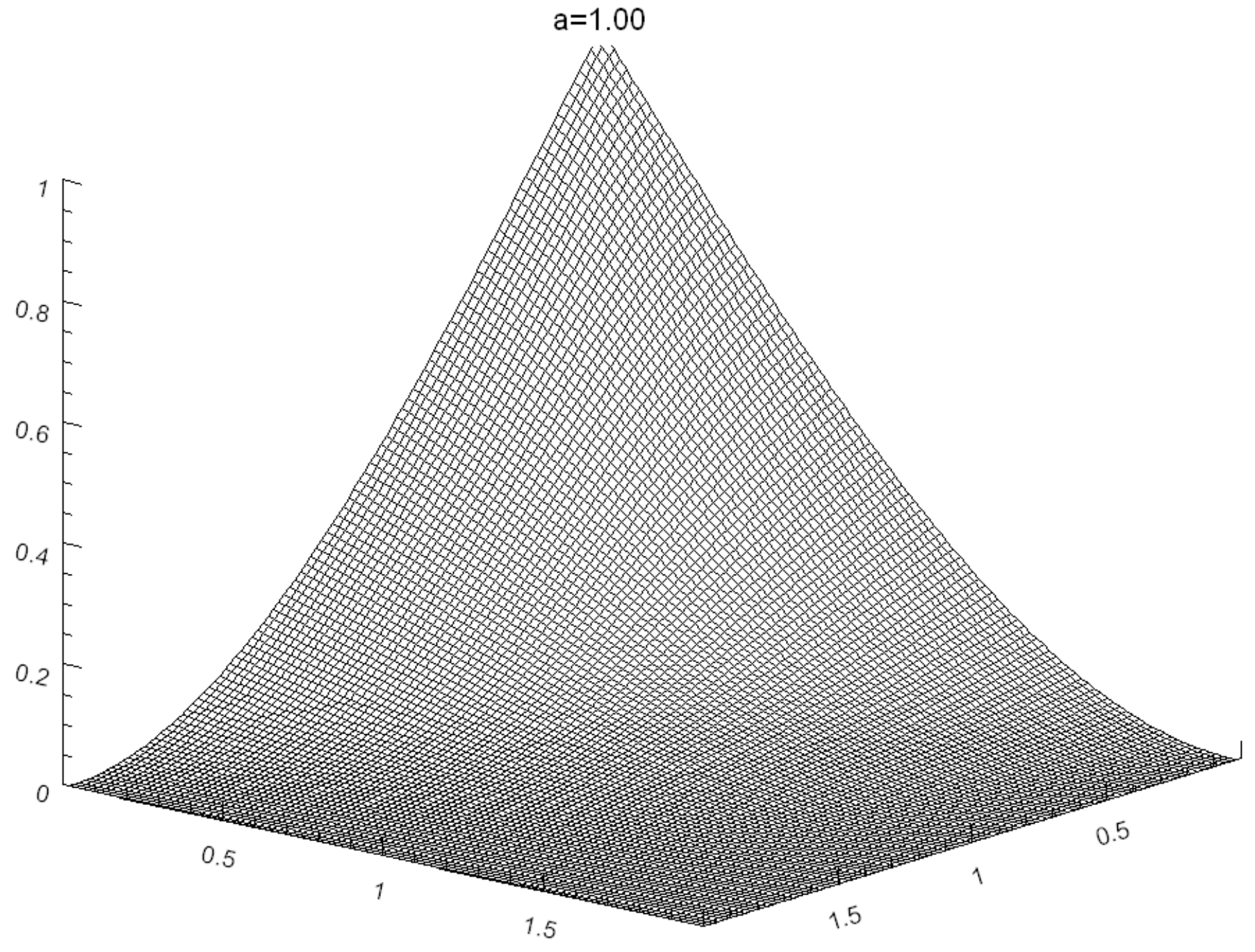

Fig. 1: Functional dependence of the weighting function $\varpi_2$ on the ratios $r_i/R_n$ and $r_j/R_n$ ($x$- and $y$-axes) for the value a = 2 of the ratio $R_c/R_n$.

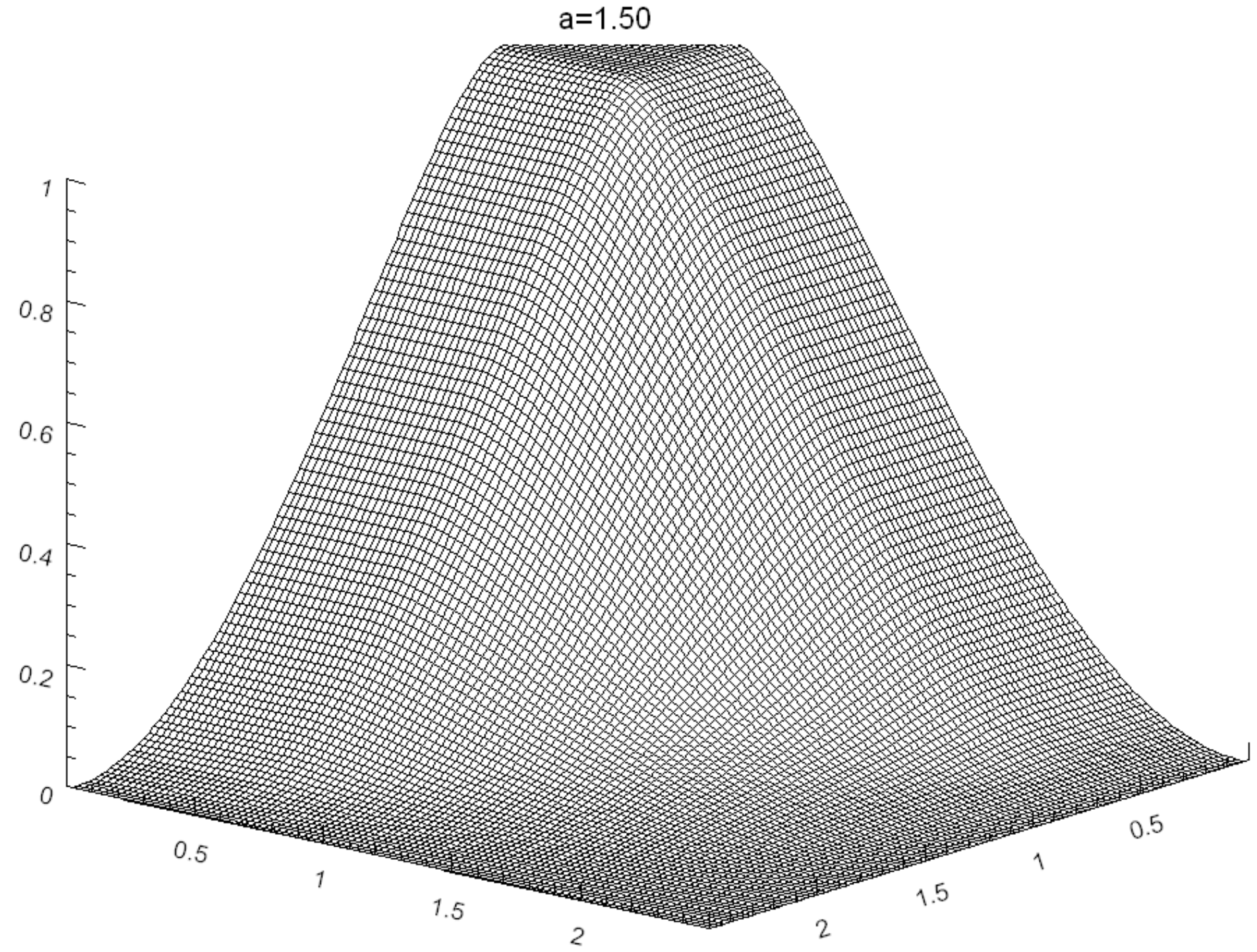

Fig. 2: Functional dependence of the weighting function $\varpi_2$ on the ratios $r_i/R_n$ and $r_j/R_n$ ($x$- and $y$-axes) for the value a = 1.5 of the ratio $R_c/R_n$.

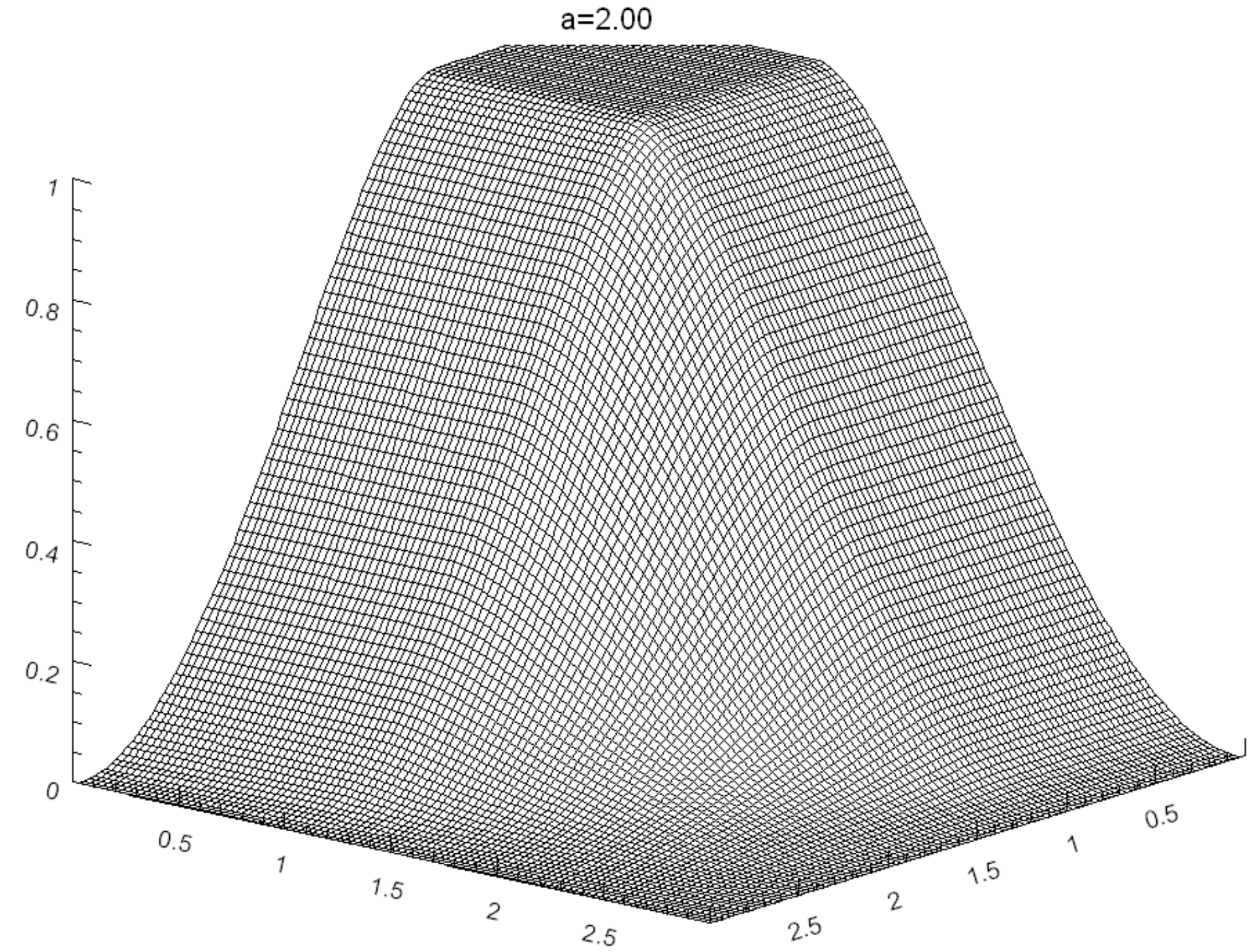

Fig. 3: Functional dependence of the weighting function $\overline{\omega}_2$ on the ratios $r_i/R_n$ and $r_j/R_n$ ($x$- and $y$-axes) for the value a = 2 of the ratio $R_c/R_n$.

## S1.4 Calculation of the weighting function $\varpi(r_i)$

For calculation of the weighting function $w(r_i)$, it is convenient to simplify the notation by choosing $R_n$ as the unit of length and introducing the reduced quantities

$$a = \frac{R_c}{R_n} \; ; \quad x = \frac{r}{R_n}; \quad x_i = \frac{r_i}{R_n}; \quad x_p = \frac{r_p}{R_n} \quad . \tag{26}$$

This transforms the expressions in Eqs. (12) and (13) into Eq. (27):

$$\overline{d_1} = \int_{x_p}^{a+1} D_1(x_i) x_i{}^2 \times 4\pi\varpi(x_i)\, dx_i \, ; \qquad \varpi(x_i) = 3\int_0^1 \frac{\Omega(x_i, x)}{4\pi} x^2 dx \tag{27}$$

The expression for $\Omega(x_i, x)$ is obtained from Eq. (10) by substituting $a$ for $R_c$, $x$ for $r$, and $x_i$ for $r_i$.

$$\frac{\Omega(x_i, x)}{4\pi} = \begin{cases} 0 & x < x_i - a \\ 1 & x < a - x_i \\ \frac{1}{2} + \frac{1}{4x_i}\left(\frac{a^2 - x_i{}^2}{x} - x\right) & \text{else} \end{cases} \tag{28}$$

The calculation of $\varpi(x_i)$ requires three cases to be distinguished.

### *S1.4.1 Case 1: $r_i \leq R_c - R_n$, i.e. $x_i \leq a - 1$*

For this case, the second case of Eqs. (10) and (28) applies to all points inside the nucleus (Fig. 4) so that

$$\varpi(r_i) = 3\int_0^1 x^2 dx = 1 \tag{29}$$

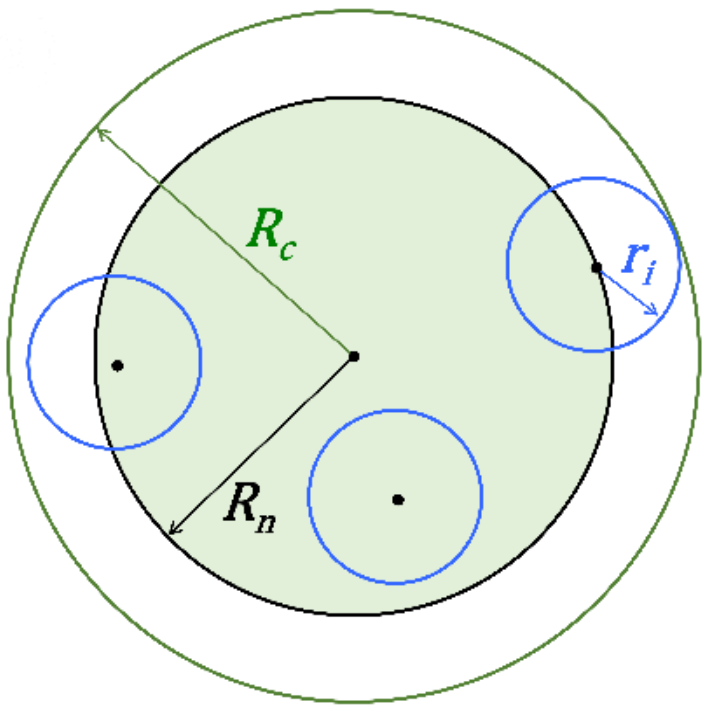

Fig. 4: Schematic illustration of case 1. The drawing shows a central cut through the spherical nucleus (black circle, radius $R_n$), the concentric sphere containing NPs (green circle, radius $R_c$), and several spheres (blue circles, radius $r_i$) of the maximum radius that are completely inside the region containing NPs when their center is within the nucleus.

### *S1.4.2 Case 2a: $R_c - R_n \leq r_i \leq R_c$, i.e., $a - 1 \leq x_i \leq a$*

In this case, the second case of Eq. (10) applies all points inside a sphere of radius $(R_c - r_i)$, while the third case applies to larger radial distances from the center of the nucleus (Fig. 5). This leads to Eq. (30):

$$\varpi(r_i) = 3\int_0^{a-x_i} x^2\, dx + \frac{3}{2}\int_{a-x_i}^1 x^2\, dx + \frac{3}{4x_i}\int_{a-x_i}^1 [(a^2 - x_i{}^2)x - x^3]\, dx \tag{30}$$

Performing the integrals gives Eq. (31):

$$\varpi(x_i) = \frac{1}{2} + \frac{1}{2}(a - x_i)^3 + \frac{3}{4x_i}\left[\frac{1}{2}(a^2 - x_i{}^2)(1 - (a - x_i)^2) - \frac{1}{4}[1 - (a - x_i)^4]\right] \tag{31}$$

Expanding the terms to obtain a polynomial in $x_i$ gives

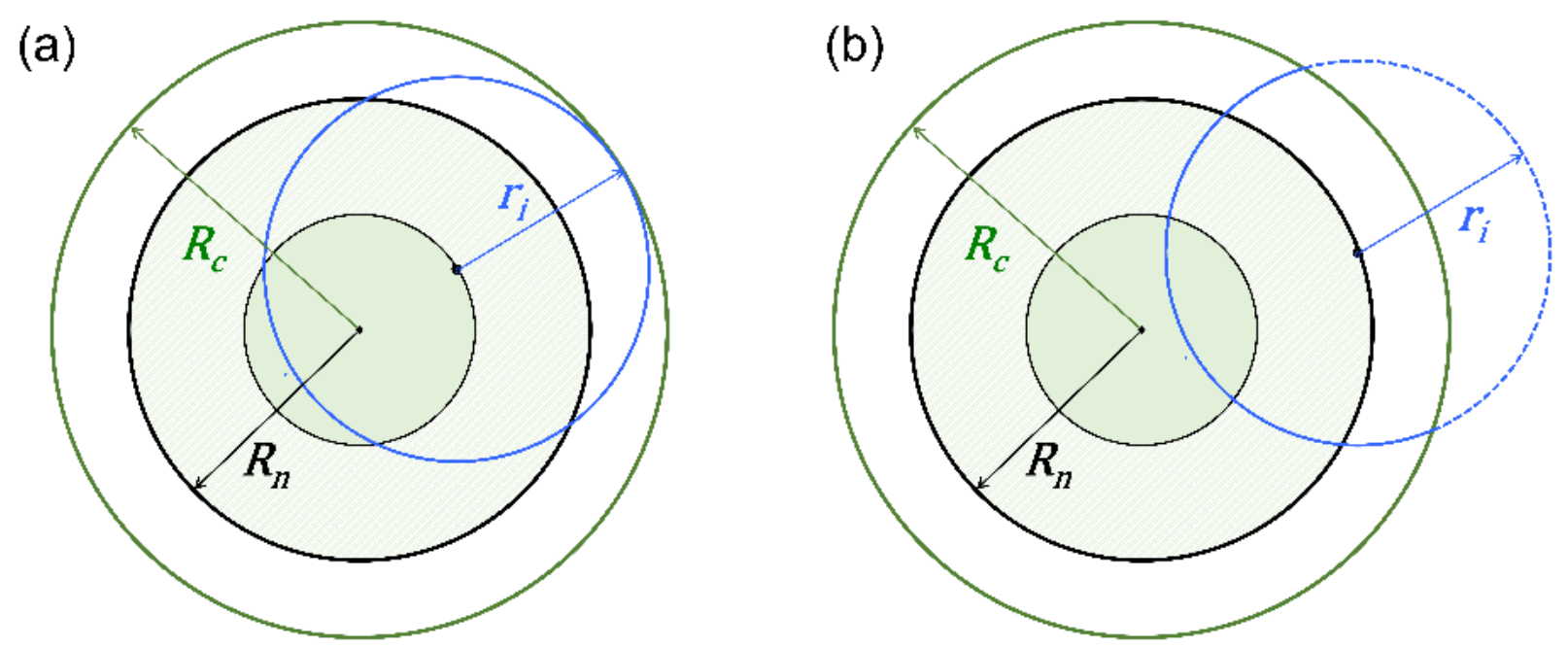

Fig. 5: Schematic illustration of case 2a: (a) For center points inside the solid green area, the sphere of radius $r_i$ (blue circle) is entirely within the sphere containing NPs (green circle, radius $R_c$). (b) For center points in the green stroked area, a part of the sphere is outside this region (dashed blue arc).

$$\begin{aligned}\overline{\omega}(x_i) = &\frac{1}{2}(1+a^3) - \frac{3}{2}a^2x_i + \frac{3}{2}a{x_i}^2 - \frac{3}{2}{x_i}^3 \\ &+ \frac{3}{4x_i}\left[\left(\frac{1}{2}a^2(1-a^2) - \frac{1}{4}(1-a^4)\right) + (a^3-a^3)x_i \right. \\ &\left. + \left(-\frac{1}{2}(a^2+1-a^2) + 2a^2 + \frac{3}{2}a^2\right){x_i}^2 + \frac{3}{4}(-a-2a-a){x_i}^3 + \left(\frac{1}{2}+\frac{2}{3}+\frac{1}{4}\right){x_i}^4\right]\end{aligned} \tag{32}$$

and finally results in Eq. (33):

$$\overline{\omega}(x_i) = \frac{1}{2}(1+a^3) - \frac{3}{8}(1+a^2)x_i + \frac{1}{16}{x_i}^3 - \frac{3}{16x_i}(1-a^2)^2 \tag{33}$$

For $R_c = R_n$ and reverting to dimensional quantities, this simplifies to Eq. (34):

$$\overline{\omega}(r_i) = 1 - \frac{3}{4}\frac{r_i}{R_n} + \frac{1}{16}\left(\frac{r_i}{R_n}\right)^3 \tag{34}$$

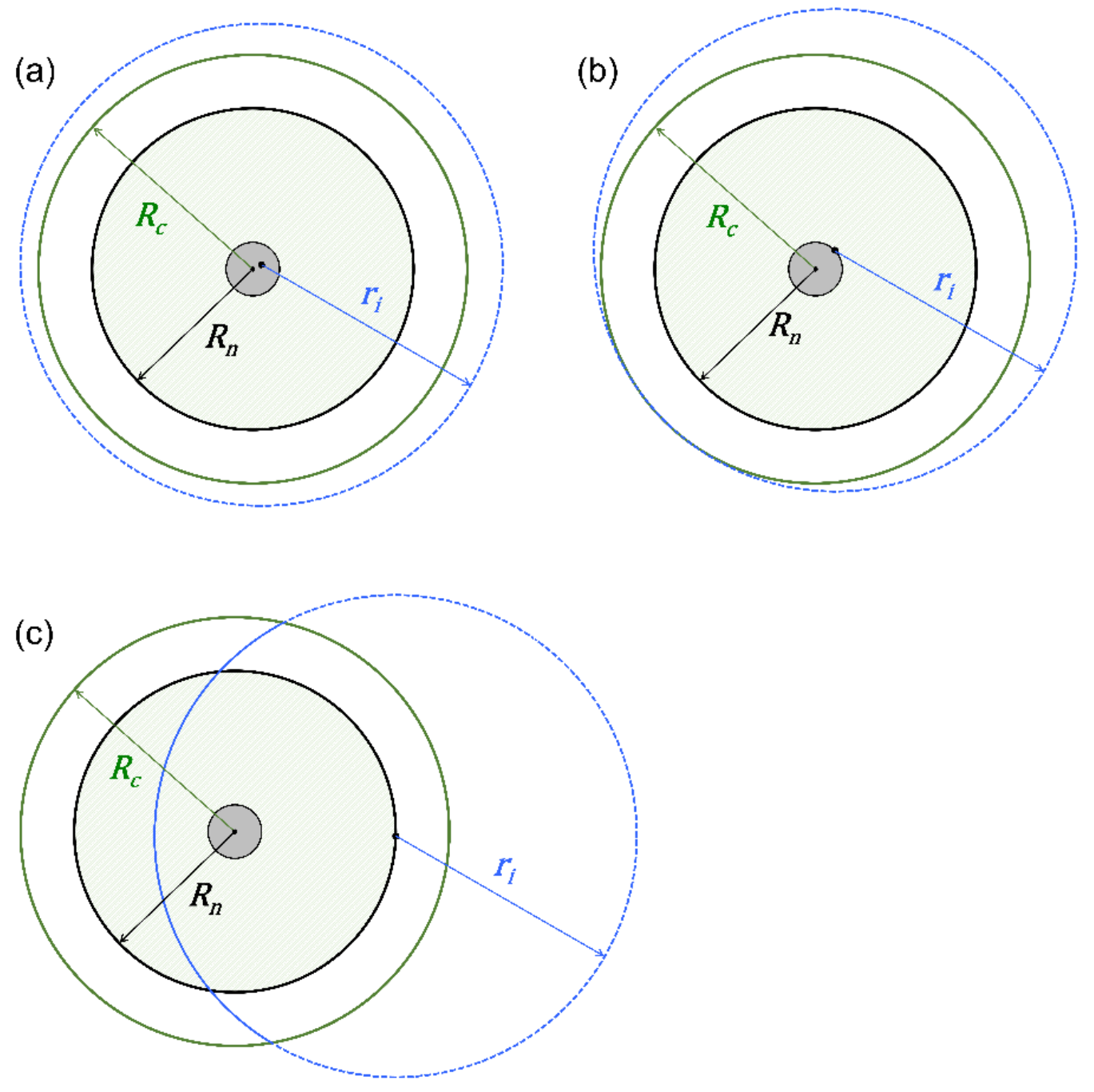

Fig. 6: Schematic illustration of case 2b: (a) For center points inside the solid gray area, the sphere of radius $r_i$ (blue circle) is entirely outside the sphere containing NPs (green circle, radius $R_c$). (b) For center points in the green-stroked area, a part of the sphere is inside this region (solid blue arc).

This expression is equivalent to the known expression for the distribution of the distance between two random points in a sphere[1].

### S1.4.3 Case 2b: $R_c \leq r_i \leq R_c + R_n$

In this case (Fig. 6), the surface of the sphere with radius $r_i$ is outside the sphere containing NPs for all points with a radial distance from the center smaller than $r_i - R_c$ and the weighting function is given by Eq. (35):

$$\varpi(x_i) = \frac{3}{2}\int_{x_i-a}^{1} x^2\,dx + \frac{3}{4x_i}\int_{x_i-a}^{1} [(a^2 - x_i^2)x - x^3]\,dx \tag{35}$$

$$\varpi(x_i) = \frac{1}{2}(1 - (x_i - a)^3) + \frac{3}{4x_i}\left[\frac{1}{2}(a^2 - x_i^2)(1 - (a - x_i)^2) - \frac{1}{4}[1 - (a - x_i)^4]\right] \tag{36}$$

This is the same expression as Eq. (31). This means that the result for this case is also given by Eq. (33).

### S1.4.4 Case 3: $r_i > R_c + R_n$

In this case, the surface of the sphere with radius $r_i$ is outside the sphere containing NPs for all points inside the nucleus, so that trivially $\varpi(r_i) = 0$.

### S1.4.5 Summary

In summary, the dependence of the part of the sphere surface inside the sphere containing NPs as given by Eq. (10) translates into a weighting function for the radial dependence of the excess dose around a NP undergoing an interaction as given by Eq. (37):

$$\varpi(r_i) = \begin{cases} 1 & r_i \leq R_c - R_n \\ \frac{1}{2}\left(1 + \frac{R_c^3}{R_n^3}\right) - \frac{3}{8}\left(1 + \frac{R_c^2}{R_n^2}\right)\frac{r_i}{R_n} + \frac{1}{16}\left(\frac{r_i}{R_n}\right)^3 - \frac{3R_n}{16r_i}\left(1 - \frac{R_c^2}{R_n^2}\right)^2 & R_c - R_n \leq r_i \leq R_c + R_n \\ 0 & r_i > R_c + R_n \end{cases} \tag{37}$$

[1] Parry M and Fischbach E 2000 Probability distribution of distance in a uniform ellipsoid: Theory and applications to physics *Journal of Mathematical Physics* **41** 2417–33

## S1.5 Calculation of the bivariate weighting function $\bar{\omega}_2(x_i, x_j)$

The calculation of the bivariate weighting function requires more cases to be considered separately (see Fig. 7).

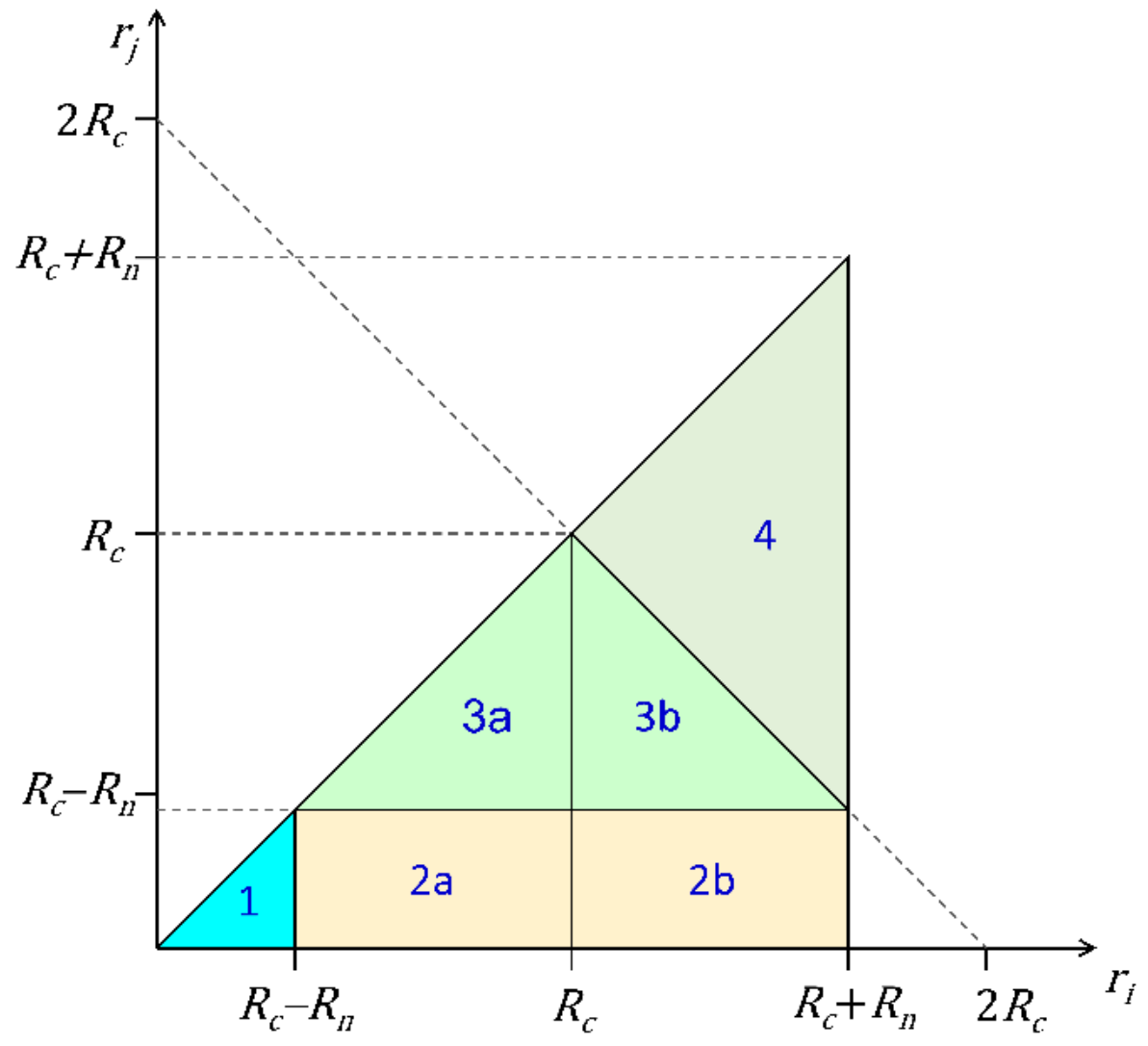

Fig. 7: Illustration of the ranges of the larger radius $r_i$ and the smaller radius $r_j$ in different cases to be distinguished in the calculation of the bivariate weighting function Eq. (21).

To simplify the notation, the substitutions from Eq. (26) are adapted as follows:

$$a = \frac{R_c}{R_n} \;;\quad x = \frac{r}{R_n};\quad x_i = \frac{r_i}{R_n} \;;\quad x_j = \frac{r_j}{R_n} \;;\quad x_p = \frac{r_p}{R_n} \quad . \tag{38}$$

### S1.5.1 Case 1: $r_j \le r_i \le R_c - R_n$

In this case, the second condition of Eq. (10) applies to both NPs such that the bivariate weighting function is unity:

$$\bar{\omega}_2(x_i, x_j) = 1 \quad r_i \le R_c - R_n \tag{39}$$

### S1.5.2 Case 2a: $r_j \le R_c - R_n \le r_i \le R_c$

In this case, the second condition of Eq. (10) applies to the closer NP such that the bivariate weighting function is identical to the weighting function for the larger radial distance of the two NPs:

$$\bar{\omega}_2(x_i, x_j) = \bar{\omega}(x_i) = \frac{1}{2}(1 + a^3) - \frac{3}{8}(1 + a^2)x_i + \frac{1}{16}{x_i}^3 - \frac{3}{16x_i}(1 - a^2)^2 \tag{40}$$

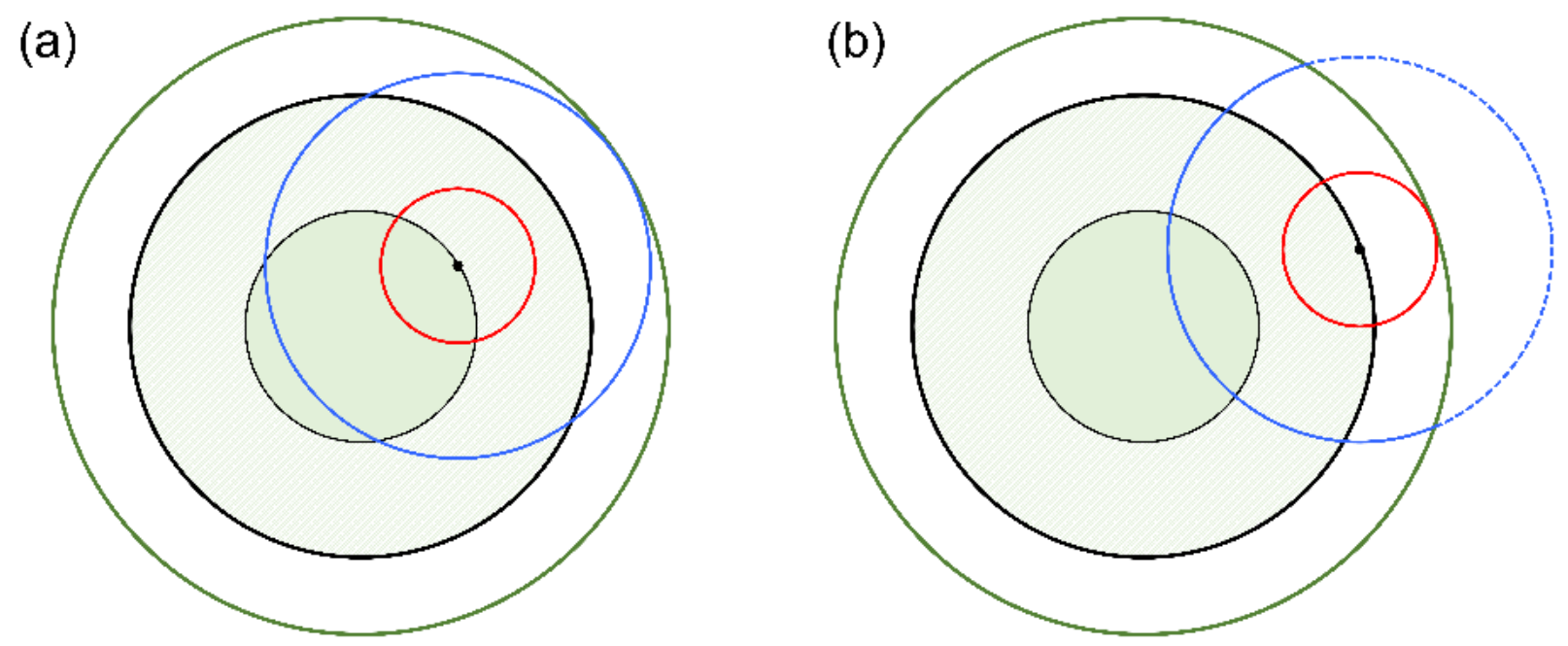

Fig. 8: Schematic illustration of case 2a: (a) For center points inside the solid green area, the sphere of radius $r_i$ (blue circle) is entirely within the sphere containing NPs (green circle, radius $R_c$). (b) For center points in the green stroked area, a part of this sphere is outside this region (dashed blue arc), while the sphere of radius $r_j$ (red circle) remains entirely inside the region containing NPs.

### *S1.5.3 Case 2b:* $r_j \le R_c - R_n \ \wedge \ R_c \le r_i \le R_c + R_n$

Also in this case (Fig. 9) does the second condition of Eq. (10) apply to the closer NPs such that the bvariate weighting function is identical to the weighting function for the larger radial distance of the two NPs.

$$\overline{\omega}_2(x_i, x_j) = \overline{\omega}(x_i) = \frac{1}{2}(1 + a^3) - \frac{3}{8}(1 + a^2)x_i + \frac{1}{16}x_i{}^3 - \frac{3}{16x_i}(1 - a^2)^2 \qquad (41)$$

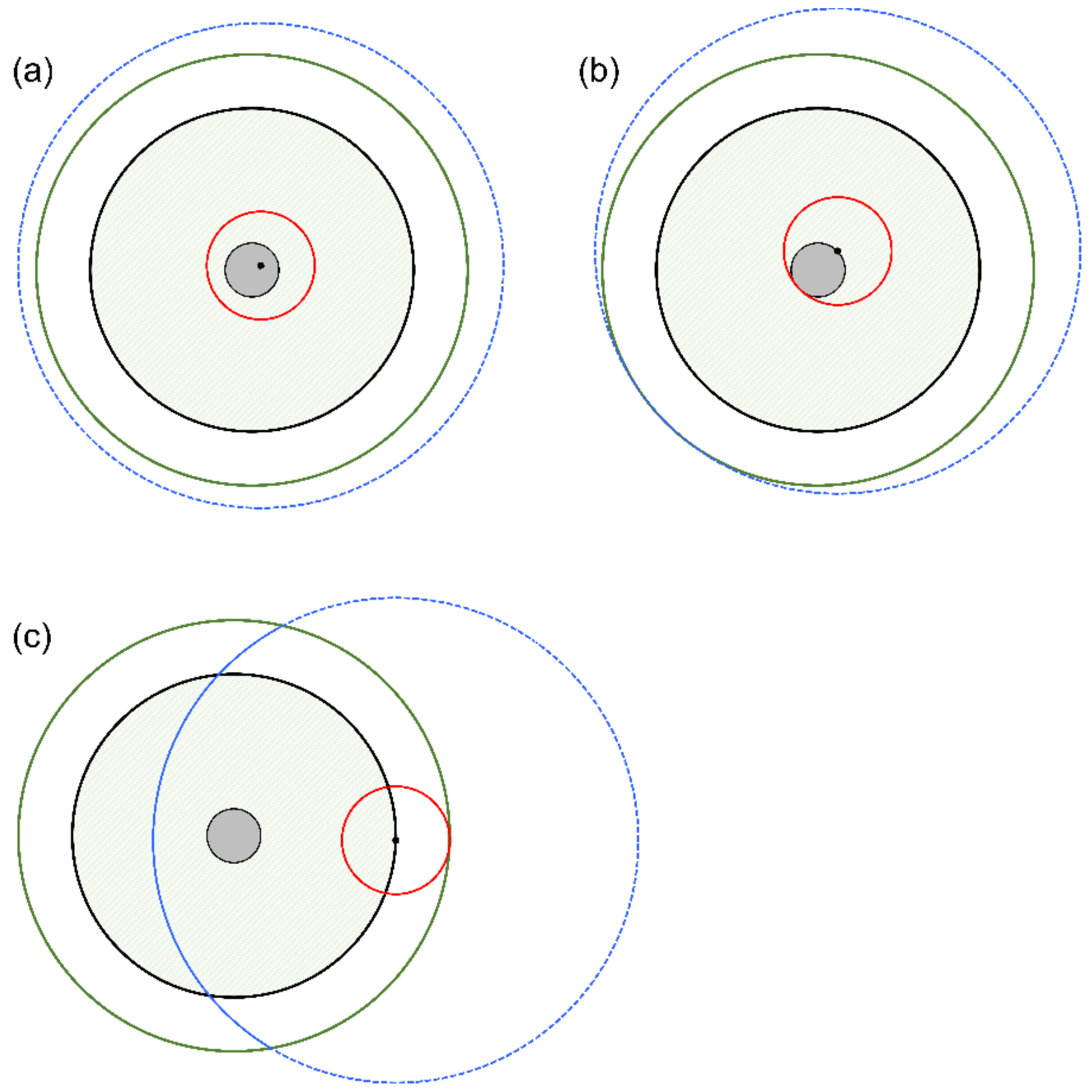

Fig. 9: Schematic illustration of case 2b: (a) For center points inside the solid gray area, the sphere of radius $r_i$ (blue circle) is entirely outside the sphere containing NPs (green circle, radius $R_c$). (b) For center points in the green stroked area, a part of this sphere is inside this region (solid blue arc) and a part is outside (dashed blue arc), while the sphere of radius $r_j$ (red circle) remains entirely inside the region containing NPs.

### *S1.5.4 Case 3a:* $R_c - R_n \le r_j \le r_i \le R_c$

In this case, the second condition of Eq. (10) is fulfilled for both NPs and all points inside a smaller sphere of radius $R_c - r_i$. This region is represented by the solid green circles in Fig. 10. Fig. 10(a) shows the limiting case of a point on the perimeter of this region when the sphere with the larger radius $r_i$ touches the spherical boundary of the region containing NPs. For radial distances between $R_c - r_i$ and $R_c - r_j$, the third condition of Eq. (10) applies to $r_i$ and the second condition to $r_j$. The corresponding region is shown as green stroked areas in Fig. 10, and Fig. 10(b) shows the limiting case of a point on the outer surface of this region with the sphere of the smaller radius $r_j$ touching the spherical boundary of the region containing NPs. For radial distances from the center of the nucleus larger than $R_c - r_j$ (semitransparent green areas in in Fig. 10), condition 3 of Eq. (10) applies to both NPs. This leads to Eq. (42):

$$\bar{\omega}_2(x_i, x_j) = (a - x_i)^3 + \frac{3}{2}\int_{a-x_i}^{a-x_j} x^2\,dx + \frac{3}{4x_i}\int_{a-x_i}^{a-x_j} [(a^2 - x_i{}^2)x - x^3]\,dx + \frac{3}{16x_i x_j}\int_{a-x_j}^{1} (a^2 - x_i{}^2 + 2x_i x - x^2)\,(a^2 - x_j{}^2 + 2x_j x - x^2)dx \tag{42}$$

Considering Eq. (30), Eq. (42) can be rewritten as follows:

$$\begin{aligned}\bar{\omega}_2(x_i, x_j) = \bar{\omega}(x_i) - \frac{3}{2}\int_{a-x_j}^{1} x^2\,dx - \frac{3}{4x_i}\int_{a-x_j}^{1} [(a^2 - x_i{}^2)x - x^3]\,dx \\ + \frac{3}{8x_i}\int_{a-x_j}^{1} [(a^2 - x_i{}^2)x - x^3]dx + \frac{3}{8x_j}\int_{a-x_j}^{1} [(a^2 - x_j{}^2)x - x^3]dx + \frac{3}{4}\int_{a-x_j}^{1} x^2\,dx \\ + \frac{3}{16x_i x_j}\int_{a-x_j}^{1} (a^2 - x_i{}^2 - x^2)\,(a^2 - x_j{}^2 - x^2)dx\end{aligned} \tag{43}$$

Summarizing the related term transforms Eq. (43) into Eq. (44)

$$\begin{aligned}\bar{\omega}_2(x_i, x_j) = \bar{\omega}(x_i) - \frac{1}{4} + \frac{1}{4}(a - x_j)^3 + \frac{3}{8x_j}\int_{a-x_j}^{1} [(a^2 - x_j{}^2)x - x^3]dx \\ - \frac{3}{8x_i}\int_{a-x_j}^{1} [(a^2 - x_i{}^2)x - x^3]dx + \frac{3}{16x_i x_j}\int_{a-x_j}^{1} (a^2 - x_i{}^2 - x^2)\,(a^2 - x_j{}^2 - x^2)dx\end{aligned} \tag{44}$$

Using again Eq. (30) leads to Eq. (45):

$$\bar{\omega}_2(x_i, x_j) = \bar{\omega}(x_i) + \frac{1}{2}\bar{\omega}(x_j) - \frac{1}{2} + \Delta\bar{\omega}_2(x_i, x_j) \tag{45}$$

$$\Delta\bar{\omega}_2(x_i, x_j) = \frac{3}{8}x_i\int_{a-x_j}^{1} x dx - \frac{3}{8x_i}\int_{a-x_j}^{1} [a^2 x - x^3]dx + \frac{3}{16x_i x_j}\int_{a-x_j}^{1} (a^2 - x_i{}^2 - x^2)\,(a^2 - x_j{}^2 - x^2)dx \tag{46}$$

Performing the integrals in Eq. (46) gives Eq. (47):

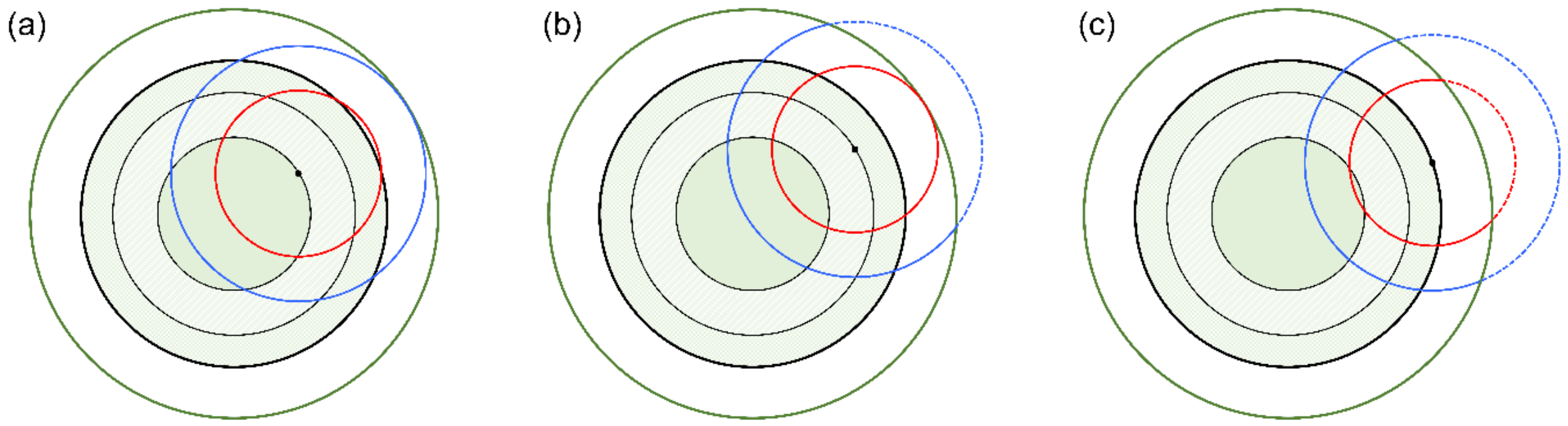

Fig. 10: Schematic illustration of case 3a: (a) For center points inside the solid green area, both the sphere of radius $r_i$ (blue circle) and the sphere of radius $r_j$ (red circle) remain entirely inside the region containing NPs. (b) For center points in the green stroked area, a part of the larger sphere is outside this region (dashed blue arc), while the smaller sphere remains entirely inside the region containing NPs. (c) With center points in the semitransparent green area, both spheres are partially outside the sphere containing NPs (dashed arcs).

$$\Delta\overline{\omega}_2(x_i,x_j) = \frac{3}{16}x_i\left(1-(a-x_j)^2\right) - \frac{1}{x_i}\left[\frac{3}{16}a^2\left(1-(a-x_j)^2\right) - \frac{3}{32}\left[1-(a-x_j)^4\right]\right] + \frac{3}{16x_ix_j}\left[(a^2-x_i^2)(a^2-x_j^2)(1-a+x_j) + \frac{1}{3}(x_i^2+x_j^2-2a^2)\left(1-(a-x_j)^3\right) + \frac{1}{5}\left(1-(a-x_j)^5\right)\right] \quad (47)$$

Spelling out the terms leads to Eq. (48)

$$\begin{aligned}\Delta\overline{\omega}_2(x_i,x_j) = {} & \frac{3}{16}x_i(1-a^2+2ax_j-x_j^2) \\ & - \frac{3}{32}\frac{1}{x_i}\left[2a^2(1-a^2)-(1-a^4)-2(1-a^2)x_i^2+4a^3x_j-4a^3x_j-2a^2x_j^2 + 6a^2x_j^2-4ax_j^3+x_j^4\right] \\ & + \frac{3}{16x_ix_j}\left[a^4(1-a)+a^4x_j-a^2(1-a)x_i^2-a^2(1-a)x_j^2-a^2x_i^2x_j+x_i^2x_j^3 - a^2x_j^3\right] \\ & + \frac{1}{16x_ix_j}\left[-2a^2(1-a^3)+x_i^2(1-a^3+3a^2x_j-3ax_j^2+x_j^3)-6a^4x_j+6a^3x_j^2 - 2a^2x_j^3+(1-a^3)x_j^2+3a^2x_j^3-3ax_j^4+x_j^5\right] \\ & + \frac{3}{80x_ix_j}\left[(1-a^5)+5a^4x_j-10a^3x_j^2+10a^2x_j^3-5ax_j^4+x_j^5\right]\end{aligned} \quad (48)$$

Sorting term as powers of $x_i$ gives:

$$\begin{aligned}\Delta\overline{\omega}_2(x_i,x_j) = {} & +x_i\left[\frac{3}{16}(1-a^2)+\frac{3}{8}ax_j-\frac{3}{16}x_j^2-\frac{3}{16}a^2+\frac{3}{16}x_j^2+\frac{3}{16}a^2-\frac{3}{16}ax_j+\frac{1}{16}x_j^2\right] \\ & + \frac{x_i}{x_j}\left[-\frac{3}{16}a^2(1-a)+\frac{1}{16}(1-a^3)\right] \\ & + \frac{1}{x_i}\left[\frac{3}{32}(a^2-1)^2-\frac{3}{8}a^2x_j^2+\frac{3}{8}ax_j^3-\frac{3}{32}x_j^4+\frac{3}{16}a^4-\frac{3}{16}a^2(1-a)x_j-\frac{3}{16}a^2x_j^2 \right.\\ & -\frac{3}{8}a^4+\frac{3}{8}a^3x_j-\frac{1}{8}a^2x_j^2+\frac{1}{16}(1-a^3)x_j+\frac{3}{16}a^2x_j^2-\frac{3}{16}ax_j^3+\frac{1}{16}x_j^4+\frac{3}{16}a^4 \\ & \left. -\frac{3}{8}a^3x_j+\frac{3}{8}a^2x_j^2-\frac{3}{16}ax_j^3+\frac{3}{80}x_j^4\right] \\ & + \frac{1}{x_ix_j}\left(\frac{3}{16}a^4(1-a)-\frac{1}{16}2a^2(1-a^3)+\frac{3}{80}(1-a^5)\right)\end{aligned} \quad (49)$$

This finally gives for $\Delta\overline{\omega}_2(r_i,r_j)$:

$$\begin{aligned}\Delta\overline{\omega}_2(x_i,x_j) = {} & \frac{3}{16}(1-a^2)x_i+\frac{3}{16}x_ix_j+\frac{1}{16}x_ix_j^2+\frac{3}{32}\frac{(a^2-1)^2}{x_i}+\frac{1}{16}\frac{(a-1)^2(2a+1)x_j}{x_i}-\frac{1}{8}\frac{x_j^2}{x_i} \\ & +\frac{1}{160}\frac{x_j^4}{x_i}+\frac{1}{16}\frac{(a-1)^2(2a+1)x_i}{x_j}-\frac{1}{80}\frac{(a-1)^3(8a^2+9a+3)}{x_ix_j}\end{aligned} \quad (50)$$

And for $\overline{\omega}_2(x_i,x_j)$:

$$\begin{aligned}\overline{\omega}_2(x_i,x_j) = {} & \overline{\omega}(x_i)+\frac{1}{2}\overline{\omega}(x_j)-\frac{1}{2}+\frac{3}{16}(1-a^2)x_i+\frac{3}{16}x_ix_j+\frac{1}{16}x_ix_j^2+\frac{3}{32}\frac{(a^2-1)^2}{x_i} \\ & +\frac{1}{16}\frac{(a-1)^2(2a+1)x_j}{x_i}-\frac{1}{8}\frac{x_j^2}{x_i}+\frac{1}{160}\frac{x_j^4}{x_i}+\frac{1}{16}\frac{(a-1)^2(2a+1)x_i}{x_j} \\ & -\frac{1}{80}\frac{(a-1)^3(8a^2+9a+3)}{x_ix_j}\end{aligned} \quad (51)$$

Substituting Eq. (38) leads to the explicit expression for $\overline{\omega}_2(x_i,x_j)$:

$$\overline{\omega}_2(x_i, x_j) = \frac{1}{4}(1+3a^3) - \frac{3}{16}(1+3a^2)x_i + \frac{1}{16}x_i^3 - \frac{3}{32x_i}(1-a^2)^2 - \frac{3}{16}(1+a^2)x_j + \frac{1}{32}x_j^3 - \frac{3}{32x_j}(1-a^2)^2 + \frac{3}{16}x_i x_j + \frac{1}{16}x_i x_j^2 + \frac{1}{16}\frac{(a-1)^2(2a+1)x_j}{x_i} - \frac{1}{8}\frac{x_j^2}{x_i} + \frac{1}{160}\frac{x_j^4}{x_i} + \frac{1}{16}\frac{(a-1)^2(2a+1)x_i}{x_j} - \frac{1}{80}\frac{(a-1)^3(8a^2+9a+3)}{x_i x_j} \quad (52)$$

For $R_c = R_n$, this simplifies to

$$\overline{\omega}_2(x_i, x_j) = 1 - \frac{3}{4}x_i + \frac{1}{16}x_i^3 - \frac{3}{8}x_j + \frac{1}{32}x_j^3 + \frac{3}{16}x_i x_j + \frac{1}{16}x_i x_j^2 - \frac{1}{8}\frac{x_j^2}{x_i} + \frac{1}{160}\frac{x_j^4}{x_i} \quad (53)$$

### S1.5.5 *Case 3b:* $R_c - R_n \leq r_j \leq 2R_c - r_i \;\wedge\; R_c \leq r_i \leq R_c + R_n$

In this case, the first condition of Eq. (10) is fulfilled for the outer NP at all points inside a smaller sphere of radius $(r_i - R_c)$. For radial distances between $(r_i - R_c)$ and $(R_c - r_j)$, the third condition of Eq. (10) applies to $r_i$ and the second condition of Eq. (10) applies to $r_j$. For radial distances from the center of the nucleus larger than $(R_c - r_j)$, condition 3 of Eq. (10) applies to both NPs. This leads to Eq. (54).

$$\overline{\omega}_2(x_i, x_j) = \frac{3}{2}\int_{x_i-a}^{a-x_j} x^2\,dx + \frac{3}{4x_i}\int_{x_i-a}^{a-x_j}[(a^2 - x_i^2)x - x^3]\,dx + \frac{3}{16x_i x_j}\int_{a-x_j}^{1}(a^2 - x_i^2 + 2x_i x - x^2)(a^2 - x_j^2 + 2x_j x - x^2)dx \quad (54)$$

Except for the missing first term and the change in the lower boundary of the first two integrals, this equation is identical to Eq. (42). The primitive of the second integral is an even function, so that $x_i - a$ and $a - x_i$ give the same result. Similar to case 2b for the calculation of $\overline{\omega}$, the result is therefore the same expression as in case 3a.

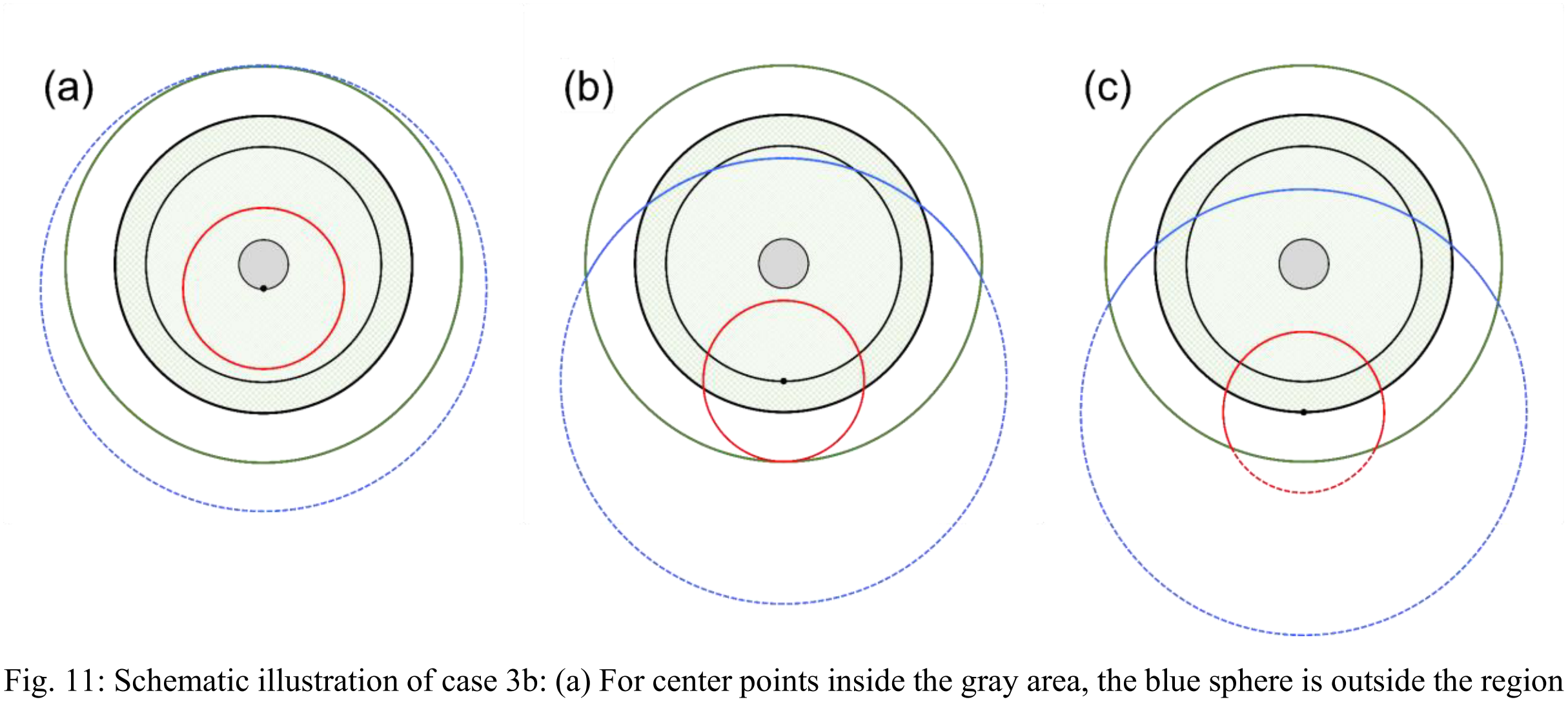

Fig. 11: Schematic illustration of case 3b: (a) For center points inside the gray area, the blue sphere is outside the region containing NPs. For center points in the solid green area, both the sphere of radius $r_i$ (blue circle) and the sphere of radius $r_j$ (red circle) are only partly inside the region containing NPs. (b) For center points in the green stroked area, a part of the larger sphere is outside this region (dashed blue arc), while the smaller sphere remains entirely inside the region containing NPs. (c) With center points in the semitransparent green area, both spheres are partially outside the sphere containing NPs (dashed arcs).

### S1.5.6 *Case 4:* $2R_c - r_i \leq r_j \leq r_i \leq R_c + R_n$

In this case, the first condition of Eq. (10) is fulfilled for the more distant NPs and all points inside a smaller sphere of radius $r_i - R_c$ (see Fig. 12). For radial distances between $r_i - R_c$ and $R_n$, the third condition applies to both $r_i$ and $r_j$. This leads to Eq. (55):

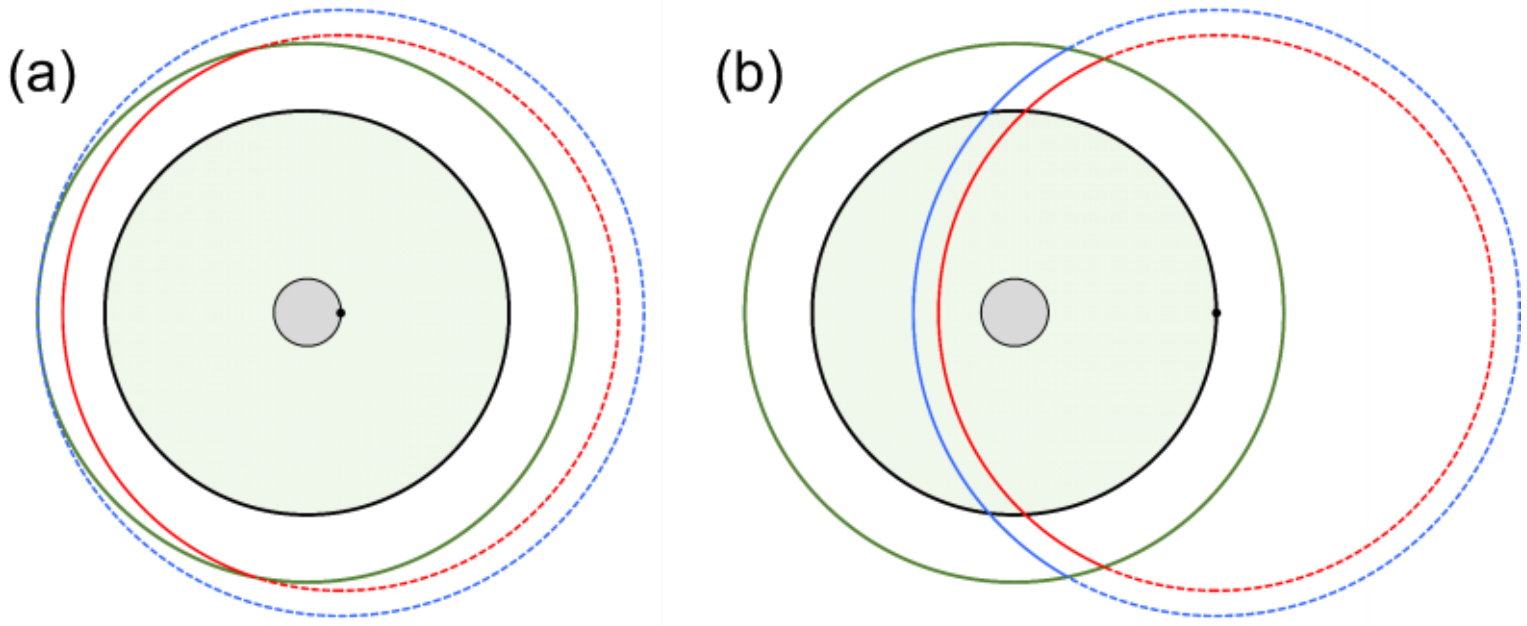

Fig. 12: Schematic illustration of case 4: For center points inside the solid green area, both the sphere of radius $r_i$ (blue circle) and the sphere of radius $r_j$ (red circle) are entirely outside the region containing NPs. (b) For center points in the semi-transparent green area, both spheres are partially outside the sphere containing NPs (dashed arcs).

$$\varpi_2(x_i, x_j) = \frac{3}{16x_i x_j}\int_{x_i-a}^{1}(a^2 - x_i^2 + 2x_i x - x^2)\left(a^2 - x_j^2 + 2x_j x - x^2\right)dx \tag{55}$$

Splitting into powers of $x_i$ and $x_j$ leads to Eq. (56):

$$\begin{aligned}\varpi_2(x_i, x_j) = \frac{3}{4}\int_{x_i-a}^{1} x^2\,dx - \frac{3}{8}(x_i + x_j)\int_{x_i-a}^{1} xdx + \frac{3}{8}\left(\frac{1}{x_i}+\frac{1}{x_j}\right)\int_{x_i-a}^{1}[a^2x - x^3]dx + \\ -\frac{3}{16}\left(\frac{x_i}{x_j}+\frac{x_j}{x_i}\right)\int_{x_i-a}^{1}(a^2 - x^2)dx + \frac{3}{16x_i x_j}\int_{x_i-a}^{1}(a^4 - 2a^2x^2 + x^4)\,dx\end{aligned} \tag{56}$$

Performing the integral gives Eq. (57):

$$\begin{aligned}\varpi_2(x_i, x_j) &= \frac{1}{4}(1 - (x_i - a)^3) - \frac{3}{16}(x_i + x_j)(1 - (x_i - a)^2) \\ &+ \frac{3}{32}\left(\frac{1}{x_i}+\frac{1}{x_j}\right)[2a^2(1 - (x_i - a)^2) - 1 + (x_i - a)^4] \\ &- \frac{1}{16}\left(\frac{x_i}{x_j}+\frac{x_j}{x_i}\right)[3a^2(1 + a - x_i) - 1 + (x_i - a)^3] \\ &+ \frac{3}{16x_i x_j}\left[a^4(1 + a - x_i) - \frac{2}{3}a^2(1 - (x_i - a)^3) + \frac{1}{5}[1 - (x_i - a)^5]\right]\end{aligned} \tag{57}$$

Spelling out the terms gives Eq. (58):

$$\begin{aligned}\varpi_2(x_i, x_j) &= \frac{1}{4}\left(1 + a^3 - 3a^2x_i + 3ax_i^2 - x_i^3\right) - \frac{3}{16}(x_i + x_j)(1 - a^2 + 2ax_i - x_i^2) \\ &+ \frac{3}{32}\left(\frac{1}{x_i}+\frac{1}{x_j}\right)[2a^2 - 2a^4 + 4a^2x_i - 2a^2x_i^2 \\ &- 1 + a^4 - 4a^3x_i + 6a^2x_i^2 - 4ax_i^3 + x_i^4] \\ &- \frac{1}{16}\left(\frac{x_i}{x_j}+\frac{x_j}{x_i}\right)[3a^2 + 3a^3 - 3a^2x_i - 1 + a^3 - 3a^2x_i + 3ax_i^2 - x_i^3] \\ &+ \frac{3}{16x_i x_j}[a^4(1 + a) - a^4x_j\,] - \frac{1}{8x_i x_j}[a^2(1 + a^3) - 3a^4x_i + 3a^3x_i^2 - a^2x_i^3] \\ &+ \frac{3}{80x_i x_j}[(1 - a^5) + 5a^4x_j - 10a^3x_j^2 + 10a^2x_j^3 - 5ax_j^4 + x_j^5]\end{aligned} \tag{58}$$

Sorting terms as powers of $x_i$ gives Eq. (59):

$$\begin{aligned}\bar{\omega}_2(x_i,x_j) &= \frac{1}{4}(1+a^3) + \left(-\frac{3}{4}a^2\right)x_i + \left(\frac{3}{4}a\right)x_i^2 + \left(-\frac{1}{4}\right)x_i^3 \\ &+ \frac{1}{x_i}\left[\frac{3}{32}(a^2-1)^2 - \frac{3}{8}a^2x_j^2 + \frac{3}{8}ax_j^3 - \frac{3}{32}x_j^4 + \frac{3}{16}a^4 - \frac{3}{16}a^2(1-a)x_j - \frac{3}{16}a^2x_j^2\right. \\ &- \frac{3}{8}a^4 + \frac{3}{8}a^3x_j - \frac{1}{8}a^2x_j^2 + \frac{1}{16}(1-a^3)x_j + \frac{3}{16}a^2x_j^2 - \frac{3}{16}ax_j^3 + \frac{1}{16}x_j^4 + \frac{3}{16}a^4 \\ &\left.- \frac{3}{8}a^3x_j + \frac{3}{8}a^2x_j^2 - \frac{3}{16}ax_j^3 + \frac{3}{80}x_j^4\right] \\ &+ x_i\left[\frac{3}{16}(1-a^2) + \frac{3}{8}ax_j - \frac{3}{16}x_j^2 - \frac{3}{16}a^2 + \frac{3}{16}x_j^2 + \frac{3}{16}a^2 - \frac{3}{16}ax_j + \frac{1}{16}x_j^2\right. \\ &\left.+ \frac{x_i}{x_j}\left(-\frac{3}{16}a^2(1-a) + \frac{1}{16}(1-a^3)\right)\right] \\ &+ \frac{1}{x_ix_j}\left(\frac{3}{16}a^4(1-a) - \frac{1}{16}2a^2(1-a^3) + \frac{3}{80}(1-a^5)\right)\end{aligned} \tag{59}$$

This finally gives Eqs. (60) and (61):

$$\begin{aligned}\bar{\omega}_2(x_i,x_j) &= \frac{1}{4}(1+a^3) - \frac{3}{16}(1+a^2)x_i - \frac{3}{32x_i}(1-a^2)^2 + \frac{1}{32}x_i^3 - \frac{3}{16}(1-a^2)x_j - \frac{3}{32x_j}(1-a^2)^2 \\ &+ \frac{3}{16}x_ix_j - \frac{1}{16}x_i^2x_j + \frac{1}{16}\frac{(1-a^2-a^3)x_j}{x_i} - \frac{1}{8}\frac{x_i^2}{x_j} - \frac{1}{160}\frac{x_i^4}{x_j} + \frac{1}{16}\frac{(1-a^2-a^3)x_i}{x_j} \\ &- \frac{1}{80}\frac{(8a^5+15a^4-a^2+3)}{x_ix_j}\end{aligned} \tag{60}$$

$$\begin{aligned}\bar{\omega}_2(x_i,x_j) &= \frac{1}{2}\bar{\omega}(x_i) - \frac{3}{16}(1-a^2)x_j - \frac{3}{32x_j}(1-a^2)^2 + \frac{3}{16}x_ix_j - \frac{1}{16}x_i^2x_j + \frac{1}{16}\frac{(1-a^2-a^3)x_j}{x_i} \\ &- \frac{1}{8}\frac{x_i^2}{x_j} - \frac{1}{160}\frac{x_i^4}{x_j} + \frac{1}{16}\frac{(1-a^2-a^3)x_i}{x_j} - \frac{1}{80}\frac{(8a^5+15a^4-a^2+3)}{x_ix_j}\end{aligned} \tag{61}$$

For $R_c = R_n$, this simplifies to

$$\bar{\omega}_2(x_i,x_j) = \frac{1}{2} - \frac{3}{8}x_i + \frac{1}{32}x_i^3 + \frac{3}{16}x_ix_j - \frac{1}{16}x_i^2x_j - \frac{1}{16}\left(\frac{x_i}{x_j} + \frac{x_j}{x_i}\right) - \frac{1}{8}\frac{x_j^2}{x_i} - \frac{1}{160}\frac{x_j^4}{x_i} - \frac{5}{16}\frac{1}{x_ix_j} \tag{62}$$

### *S1.5.7 Case 5: $r_j > R_c + R_n$*

In this case, the surface of the sphere with radius $r_j$ is outside the sphere containing NPs for all points inside the nucleus, so that trivially $\bar{\omega}_2(x_i,x_j) = 0$.

### *S1.5.8 Summary*

In summary, the bivariate weighting function $\bar{\omega}_2(x_i,x_j)$ is given by Eq. (63).

$$\bar{\omega}_2(x_i,x_j) = \begin{cases} \bar{\omega}(r_i) & x_j \le a-1 \le x_i \le a+1 \\ \bar{\omega}_2^{(3)}(x_i,x_j) & a-1 \le x_j \le 2a-x_i \ \wedge\ x_j \le x_i \le a+1 \\ \bar{\omega}_2^{(4)}(x_i,x_j) & 2a-x_i \le x_j \le x_i \le a+1 \\ 0 & x_i \ge x_j > a+1 \end{cases} \tag{63}$$

| | |
|---|---|
| $$\varpi(x_i) = \begin{cases} 1 & x_i \le a-1 \\ \frac{1}{2}\left(1+\frac{R_c{}^3}{R_n{}^3}\right) - \frac{3}{8}\left(1+\frac{R_c{}^2}{R_n{}^2}\right)x_i + \frac{1}{16}x_i{}^3 - \frac{3}{16x_i}\left(1-\frac{R_c{}^2}{R_n{}^2}\right)^2 & a-1 \le x_i \le a+1 \\ 0 & x_i > a+1 \end{cases}$$ | (64) |
| $$\varpi_2{}^{(3)}(x_i,x_j) = \varpi(x_i) + \frac{1}{2}\varpi(x_j) - \frac{1}{2} + \frac{3}{16}(1-a^2)x_i + \frac{3}{16}x_ix_j + \frac{1}{16}x_ix_j{}^2 + \frac{3}{32}\frac{(a^2-1)^2}{x_i} + \frac{1}{16}\frac{(a-1)^2(2a+1)x_j}{x_i} - \frac{1}{8}\frac{x_j{}^2}{x_i} + \frac{1}{160}\frac{x_j{}^4}{x_i} + \frac{1}{16}\frac{(a-1)^2(2a+1)x_i}{x_j} - \frac{1}{80}\frac{(a-1)^3(8a^2+9a+3)}{x_ix_j}$$ | (65) |
| $$\varpi_2{}^{(4)}(x_i,x_j) = \frac{1}{2}\varpi(x_i) - \frac{3}{16}(1-a^2)x_j - \frac{3}{32x_j}(1-a^2)^2 + \frac{3}{16}x_ix_j - \frac{1}{16}x_i{}^2x_j + \frac{1}{16}\frac{(1-a^2-a^3)x_j}{x_i} - \frac{1}{8}\frac{x_i{}^2}{x_j} - \frac{1}{160}\frac{x_i{}^4}{x_j} + \frac{1}{16}\frac{(1-a^2-a^3)x_i}{x_j} - \frac{1}{80}\frac{(8a^5+15a^4-a^2+3)}{x_ix_j}$$ | (66) |

## S1.6 Validation of the transformations in sections S1.4 and S1.5

The evaluation of the integrals in sections S1.4 and S1.5 to construct polynomials in the variables $x_i$ and $x_j$ was performed by using custom-built scripts in the GNU data language (GDL).

### *S1.6.1 General approach.*

The procedure for verification of the polynomials was such that the expressions obtained by analytical evaluation of the integrals were implemented in corresponding GDL functions. Using these functions, the values of the weighting functions for the different cases were calculated for a set of values for the parameter *a* and for values of the variables $x_i$ and $x_j$ covering the corresponding domains. Then, in a first step, for each value of parameter *a*, the data were fitted to polynomials of $x_i$ and $x_j$, where the polynomial degree was taken from the analytical integrals. Each of the corresponding coefficients was then fit to polynomials of ($a$-1), where again the degree of the polynomial was inferred from the analytical expressions. Finally, the coefficients were rounded to the nearest multiple of the inverse of a common denominator that was also inferred from the analytical expressions. (The largest such denominator was 160.) Eventually, the polynomial formulas were evaluated and the absolute deviations from the values obtained with the original expressions were determined to verify that they were within what is expected from computational accuracy.

The whole procedure was performed interactively case by case. The code of the scripts used is listed in the next subsection.

*S1.6.2 GDL code used for verifying the polynomials.*

```
pro transform_args,xiv,xjv,xi,xj
  if n_elements(xiv) ne n_elements(xjv) then begin
    xi=xiv#replicate(1d,n_elements(xjv))
    xj=replicate(1d,n_elements(xiv))#xjv
  endif else begin
    xi=xiv
    xj=xjv
  endelse
end
```

```
function w2c2_poly,a,xi
  w1= 0.5d0*xi * (1.0d0+(a-xi)^3) $ ; for xi <= a, the xi-terms are (a-xi)^3-0.5*(a-xi)^3, for xi>a, -0.5*(xi-a)^3
     +0.375d0  * (a^2-xi^2)*(1.0d0-(a-xi)^2) $ ; 3./8.
     -0.1875d0 * (1.0d0-(a-xi)^4)              ; 3./16.
return,w1
end
```

```
function w2c3_poly1,a,xiv,xjv
  transform_args,xiv,xjv,xi,xj
  ;
  w2= 0.5d0*xi * ((a-xj)^3+(a-xi)^3) $ ; for xi <= a, the xi-terms are (a-xi)^3-0.5*(a-xi)^3, for xi>a, -0.5*(xi-a)^3
     +0.375d0  * (a^2-xi^2)*((a-xj)^2-(a-xi)^2) $ ; 3./8.
     -0.1875d0 * ((a-xj)^4-(a-xi)^4)              ; 3./16.
return,w2
end
```

```
function w2c3_poly1diff,a,xiv,xjv
  transform_args,xiv,xjv,xi,xj
  ;
  w2= -0.1875d0*xj*( (a^2-xi^2)*(1.0d0-(a-xj)^2) $
                   - 0.5d0*(1.0d0-(a-xj)^4))
return,w2
```

```
end
```

```
function w2c3_poly2,a,xiv,xjv
  transform_args,xiv,xjv,xi,xj
  w2=(  (a^2-xi^2)*(a^2-xj^2)*(1-a+xj) $
        +((a^2-xi^2)*xj+(a^2-xj^2)*xi)*(1-(a-xj)^2) $
        +1./3.*(xi^2+xj^2+4*xi*xj-2*a^2)*(1-(a-xj)^3) $
        -0.5*(xi+xj)*(1-(a-xj)^4) +0.2*(1-(a-xj)^5))*3./16.
return,w2
end
```

```
function w2c3_poly2diff,a,xiv,xjv
  transform_args,xiv,xjv,xi,xj
  ;
  w2 = 0.1875d0 * (a^2-xi^2)*(a^2-xj^2)*(1-a+xj) $
      +0.0625d0 * (xi^2+xj^2-2*a^2)*(1-(a-xj)^3) $
      +0.0375d0 * (1-(a-xj)^5)
return,w2
end
```

```
function w2c4_poly,a,xiv,xjv
  transform_args,xiv,xjv,xi,xj
  w2=(  (a^2-xi^2)*(a^2-xj^2)*(1+a-xi) $
        +((a^2-xi^2)*xj+(a^2-xj^2)*xi)*(1-(xi-a)^2) $
        +1./3.*(xi^2+xj^2+4*xi*xj-2*a^2)*(1-(xi-a)^3) $
        -0.5*(xi+xj)*(1-(xi-a)^4) +0.2*(1-(xi-a)^5))*3./16.
return,w2
end
```

```
function lemmix_calc_w2,a,xiv,xjv
  ; Process input and generate output array
  npts=[n_elements(xiv),n_elements(xjv)]
  if npts[0] eq 1 or npts[1] eq 1 then begin
    xi=xiv & xj=xjv & w2=dblarr(max(npts))
  endif else begin
    xi=xiv#replicate(1d,npts[1])
```

```
    xj=replicate(1d,npts[0])#xjv
    w2=fltarr(npts[0],npts[1])
  endelse
  ;
; ----------------------------------------------------------------------
; Case 1: both spheres always inside NP region ---------------------------------------------------------------
  qq=where(xi le a-1 and xj le a-1,nq)
  if nq gt 0 then w2[qq]=1.
  ;
; ----------------------------------------------------------------------
; Case 2: smaller sphere always inside NP region ---------------------------------------------------------------
  qq=where(xj le a-1 and xi ge a-1 and xi le a+1,nq) ; xj <= xi
  if nq gt 0 then w2[qq]=w2c2_poly(a,xi[qq])/xi[qq]
  ;
  qq=where(xi le a-1 and xj gt a-1 and xj le a+1,nq) ; xj > xi
  if nq gt 0 then w2[qq]=w2c2_poly(a,xj[qq])/xj[qq]
  ;
; ----------------------------------------------------------------------
; Case 3: Both can intersect the surface of the NP region but there is
;         a region where both are inside and a region where the larger
;         one  intersects while the smaller sphere is inside ----------------------------------------------------------
-----
  qq=where(xj gt a-1 and xj le xi and xj le 2*a-xi and xi le a+1,nq) ; xj<=xi
  if nq gt 0 then w2[qq]=(w2c3_poly1(a,xi[qq],xj[qq])+w2c3_poly2(a,xi[qq],xj[qq])/xj[qq])/xi[qq]
  ;
  qq=where(xi gt a-1 and xi le xj and xi le 2*a-xj and xj le a+1,nq) ; xj>xi
  if nq gt 0 then w2[qq]=(w2c3_poly1(a,xj[qq],xi[qq])+w2c3_poly2(a,xj[qq],xi[qq])/xi[qq])/xj[qq]
  ;
; ----------------------------------------------------------------------
; Case 4: Both spheres always intersect the suface of the NP region
  qq=where(xj gt 2*a-xi and xj le xi and xi le a+1,nq) ; xj<=xi
  if nq gt 0 then w2[qq]=w2c4_poly(a,xi[qq],xj[qq])/xj[qq]/xi[qq]
  ;
  qq=where(xi gt 2*a-xj and xi le xj and xj le a+1,nq) ; xj>xi
  if nq gt 0 then w2[qq]=w2c4_poly(a,xj[qq],xi[qq])/xi[qq]/xj[qq]
  ;
; ----------------------------------------------------------------------
```

```
; Case 1: both spheres always inside NP region ------------------------------------------------------------
  qq=where(xi le a-1 and xj le a-1,nq)
  if nq gt 0 then w2[qq]=1.
  ;
; ----------------------------------------------------------------------
return,w2
end
```

```
pro add,was1, was2
  common dieformel,formel
  formel=formel+strtrim(was1,2)
  if n_params() eq 2 then formel=formel+strtrim(was2,2)
end
```

```
function clean,input_value, denim
  if n_params() eq 1 then den=160 else den=denim
  value = input_value
  if den gt 1 then begin
    if 5*long(value/5) eq value and 5*long(den/5) eq den then begin
      value /= 5
      den /= 5
    endif
    while 2*long(value/2) eq value and 2*long(den/2) eq den do begin
        value /= 2
        den /= 2
    endwhile
  endif
  if den gt 1 then suffix='/'+strtrim(den,2) else suffix=''
return,strtrim(value,2)+suffix
end
```

```
function poli_a_1_to_a, coeff
  common pascal,n_over_k
  ncoeff=n_elements(coeff)
  if n_elements(n_over_k) lt ncoeff^2 then begin
    n_over_k=intarr(ncoeff,ncoeff)
```

```
    n_over_k[0,0]=1
    for i=1,ncoeff-1 do begin
      n_over_k[i,i]=1
      n_over_k[i,0]=-n_over_k[i-1,0]
      for j=1,i-1 do n_over_k[i,j]=n_over_k[i-1,j-1]-n_over_k[i-1,j]
    endfor
    n_over_k=transpose(n_over_k)
  endif
  res=coeff
  for i=0,ncoeff-2 do for j=i+1,ncoeff-1 do res[i] += n_over_k[i,j]*coeff[j]
  ;print,'coeff',coeff,f='(a,7i10)'
  ;print,'res  ',res,f='(a,7i10)'
return,res
end
```

```
function fit_poly_xi_xj,ndeg,w2,xi,xj
  npts=[n_elements(xi),n_elements(xj)]
  ;
  if n_params() eq 3 then qq=w2 else begin
  ; first fit xj-dependence
    xx=dblarr(npts[1],ndeg[1]+1)
    for i=0,ndeg[1] do xx[*,i]=xj^i
    ;regression for x_j
    minv=invert(transpose(xx)#xx)
    qq=(w2#xx)#minv
  endelse
  ;
  ; then the xi-dependence
  xx=dblarr(n_elements(xi),ndeg[0]+1)
  for i=0,ndeg[0] do xx[*,i]=xi^i
  ;regression for x_i
  minv=invert(transpose(xx)#xx)
  ;
  ; this is the matrix of parameters
  pp=minv#(transpose(xx)#qq)
  ;
```

```
return,pp
end
```

```
function xPolyCoeff, a, theCase, npts=npts
  common polgrade,ndeg
  IF not keyword_set(npts) then npts=[100,110]
  IF n_elements(theCase) eq 0 then theCase=2
  If n_elements(npts) eq 1 then npts=[npts,npts+10]
  npts=long(npts)
  ;
  xmax=double(a+1) & xmin=0
  ;
  steps=(dindgen(npts[0])+0.5)/double(npts[0])
  xi=(xmax+steps*(xmin-xmax)) & sort,xi
  ;
  steps=dindgen(npts[1])/double(npts[1])
  xj=(xmax+steps*(xmin-xmax)) & sort,xj
  xxj=replicate(1d,n_elements(xi))#xj
  ;
  case theCase of
    2: begin
          w2=w2c2_poly(a,xi)
          ndeg=[4,0]
          pp=fit_poly_xi_xj(ndeg,w2,xi)
       end
    3: begin
          w2=xxj*w2c3_poly1(a,xi,xj) & ndeg=[4,5]
          pp=fit_poly_xi_xj(ndeg,w2,xi,xj)
          ;
          w2=w2c3_poly2(a,xi,xj) & ndeg=[2,5]
          pp2=fit_poly_xi_xj(ndeg,w2,xi,xj)
          for i=0,2 do pp[i,*]=pp[i,*]+pp2[i,*]
       end
    -3: begin
          w2=w2c3_poly1diff(a,xi,xj)+w2c3_poly2diff(a,xi,xj) & ndeg=[2,5]
          ndeg=[2,5]
```

```
          pp=fit_poly_xi_xj(ndeg,w2,xi,xj)
       end
    4: begin
          w2=w2c4_poly(a,xi,xj)
          ndeg=[5,2]
          pp=fit_poly_xi_xj(ndeg,w2,xi,xj)
       end
    else: stop
  endcase
return,pp
end
```

```
function lemmix_PolyCoeff,npts=npts,theCase,terme=terme
  common dieformel,formel
  common polgrade,ndeg
  IF not keyword_set(theCase) then theCase=2
  IF not keyword_set(npts) then npts=[128,512]
  IF n_elements(npts) eq 1 then npts=npts*[1,1]
  IF n_elements(npts) lt 3 then npts=[npts,100]
  ;
  case theCase of
    2: ndeg=[4,0]
    3: ndeg=[4,5]
    -3: ndeg=[2,5]
    4: ndeg=[5,2]
    else: stop
  endcase

  ncombi=(ndeg[0]+1)*(ndeg[1]+1)
  pwr1=indgen(ncombi) mod (ndeg[0]+1)
  pwr2=indgen(ncombi)/(ndeg[0]+1)+(theCase eq 2) ; make sure to have something to divide
  ;
  a=1.+findgen(npts[2])/npts[2]
  dd=dblarr(npts[2],ncombi)
  for i=0,npts[2]-1 do begin
    print,f='(a,$)','.'
```

```
  dd[i,*]=reform(xPolyCoeff(a[i],npts=npts[1],theCase),ncombi)
endfor
;
ndeg=[ndeg,max(ndeg)]
pp=dblarr(ncombi,ndeg[2]+1)
xx=replicate(1.d,npts[2])
for i=1,ndeg[2] do xx=[[xx],[(a-1)^i]]
minv=invert(transpose(xx)#xx)
pp=minv#(transpose(xx)#dd)
ip=transpose(round(pp*160))
print,ncombi
;print,ip
;stop
;
;so=sort(pwr1+pwr2) & pp=pp[so,*] & ip=ip[so,*] & pwndeg,5]r1=pwr1[so] & pwr2=pwr2[so]
;
for i=0,ncombi-1 do begin
  formel=''
  qq=where(ip[i,*] ne 0,nq)
  if nq gt 0 then begin
    if nq eq 1 then begin
      if ip[i,qq] gt 0 then add,'+'
      add,clean(ip[i,qq]),'.'
      if qq[0] ge 1 then add,'*(a-1)'
      if qq[0] gt 1 then add,'^',qq[0]
    endif else begin
      pwr0=qq[0]
      ip[i,qq[0]:*]=poli_a_1_to_a(ip[i,qq[0]:*])
      qq=where(ip[i,*] ne 0,nq)
      if nq eq 1 then begin
        if ip[i,qq[0]] gt 0 then add,'+'
        add,clean(ip[i,qq[0]]),'.'
        if pwr0 gt 0 then add,'*(a-1)'
        if pwr0 gt 1 then add,'^',pwr0
        add,'*a'
        if qq[0]-pwr0 gt 1 then add,'^',qq[0]-pwr0
      endif else begin
```

```
          if ip[i,qq[0]] lt 0 then begin
            ip[i,qq] *= -1
            add,'-'
          endif else add,'+'
          ;
          last=ip[i,qq[nq-1]]
          last=min(abs(ip[i,qq]))
          if theCase eq 2 then den=16 else den=160
          ;stop
          if total(last*(ip[i,qq]/last) eq ip[i,qq]) eq nq then begin
            add,clean(last),'.'
            ip[i,qq] /= last
            den=1
            ;stop
          endif else begin
            if total(5*(ip[i,qq]/5) eq ip[i,qq]) eq nq then last=5 else last=1
            while total(last*(ip[i,qq]/last) eq ip[i,qq]) eq nq do last=last*2
            last=last/2
            add,clean(last),'.'
            ip[i,qq] /= last
            den=1
          endelse
          ;
          if pwr0 ge 1 then add,'*(a-1)'
          if pwr0 gt 1 then add,'^',pwr0
          add,'*(',clean(ip[i,qq[0]],den)
          for j=1,nq-1 do begin
            if ip[i,qq[j]] gt 0 then add,'+'
            if den gt 1 or abs(ip[i,qq[j]]) gt 1 then add,clean(ip[i,qq[j]],den),'*' else if ip[i,qq[j]] lt 0 then add,'-
'
            add,'a'
            if qq[j]-pwr0 gt 1 then add,'^',qq[j]-pwr0
          endfor
          add,')'
        endelse
      endelse
      if nq gt 0 then begin
```

```
        if pwr1[i] eq 0 then add,'/xi'
        if pwr1[i] gt 1 then add,'*xi'
        if pwr1[i] gt 2 then add,'^',pwr1[i]-1
        if pwr2[i] eq 0 then add,'/xj'
        if pwr2[i] gt 1 then add,'*xj'
        if pwr2[i] gt 2 then add,'^',pwr2[i]-1
      endif
      print,formel
      if n_elements(res) eq 0 then begin
        terme=formel
        res=pp[*,i]
      endif else begin
        terme=[terme,formel]
        res=[[res],[pp[*,i]]]
      endelse
    endif
  endfor

  formel=terme[0]
  for i=1,n_elements(terme)-1 do formel=formel+terme[i]
  ;
  return,formel
end
```

```
pro lemmix_PolyCoeff,npts=npts,theCase,aarr
  print
  formel=lemmix_PolyCoeff(npts=npts,theCase,terme=terme)
  ;print,terme,f='(a,$)'
  print
  print,formel
  ;
  for i=0, n_elements(aarr)-1 do begin
    a=aarr[i]
    xi=a-1+(dindgen(40)+0.5)/20
    xj=xi
    xi=xi#replicate(1.d0,40)
```

```
    xj=transpose(xi)
    case theCase of
      2: z=w2c2_poly(a,xi)/xi
      3: z=(w2c3_poly1(a,xi,xj)+w2c3_poly2(a,xi,xj)/xj)/xi
      -3: z=(w2c3_poly1diff(a,xi,xj)+w2c3_poly2diff(a,xi,xj)/xj)/xi
      4: z=w2c4_poly(a,xi,xj)/xi/xj
      else: stop
    endcase

    dummy=execute('y='+formel)
    case theCase of
      2: qq=where(xi)
      3: qq=where(xi+xj  le 2*a)
      -3: qq=where(xi+xj le 2*a)
      4: qq=where(xi+xj gt 2*a)
      else: stop
    endcase
    print,'case:',theCase,'   a=',a,'   maxabsdiff=',max(abs(z[qq]-y[qq]))
    ;stop
  endfor
end
```